\documentclass[12pt]{article}
\begin{document}
\input epsf
\def\be{\begin{equation}}
\def\bea{\begin{eqnarray}}
\def\ee{\end{equation}}
\def\eea{\end{eqnarray}}
\def\d{\partial}
\def\db{{\bar\partial}}
\def\la{\lambda}
\def\eps{\epsilon}
\def\gam{\gamma}
\def\vareps{\varepsilon}
\def\zb{\bar z}
\def\wb{\bar w}
\def\Zb{\bar Z}
\def\Wb{\bar W}

\begin{titlepage}
\begin{center}
% \vspace*{-3cm}
\hfill EFI-08-02\\

\vskip 1cm

{\LARGE {\bf Brane webs and 1/4--BPS geometries\\ \ \  \\
  }}

\vskip 1.5cm

{\large  Oleg Lunin }

\vskip 1cm

{\it Enrico Fermi Institute, University of Chicago,
Chicago, IL 60637
}

\vskip 3.5cm

\vspace{5mm}

\noindent

{\bf Abstract}

\end{center}

We discuss brane webs preserving eight supercharges and derive geometries produced by them. 
Consistency conditions of supergravity are shown to impose certain requirements on the 
locations of the sources, and these restrictions are found to be in a perfect agreement with 
results of the probe analysis. In particular, solutions of IIB SUGRA describing $(p,q)$ stings 
are inconsistent, unless the web consists of straight line segments whose orientation is 
correlated with charges of the string. The geometries produced by membranes and D3 branes are 
only consistent if brane profiles are holomorphic. Using perturbation theory, we show that a 
unique gravity solution exists for any allowed distribution of sources. We also revisit 
1/4--BPS geometries with $AdS_p\times S^q$ asymptotics and derive the boundary conditions 
leading to regular geometries. All degenerate limits of regular solutions are shown to agree 
with expectations from the brane probe analysis. 

\vskip 4.5cm

\end{titlepage}

\newpage

\setcounter{tocdepth}{2}
%{\footnotesize\tableofcontents}
{\small\tableofcontents}

\newpage

%=======================================

\section{Introduction and summary.}

\label{SectIntro}

\renewcommand{\theequation}{1.\arabic{equation}}
\setcounter{equation}{0}

%========================================

Brane intersections have been instrumental in improving our understanding of black holes \cite{StromVafa} and in constructing string 
realization of interesting field-theoretic phenomena \cite{HanWitten}. While 
the vast majority of work on intersecting branes has been devoted to 
orthogonal intersections, studies of branes intersecting at angles, 
initiated in \cite{BerDougLeih}\footnote{See also \cite{Ohta} for the 
discussion of intersections in M theory.}, led to significant progress in classifying supersymmetric objects in string theory and to applications to model building \cite{uranga}. The situation becomes especially interesting if intersecting branes have the same dimensionality of the worldvolume: in this case one can get complicated systems known as 
"brane webs" \cite{StrWeb,D5Webs}. In particular, string webs have emerged in the descriptions of the dyonic black holes \cite{BHWebs}, so 
by finding a geometry produced by an individual web, one can provide
a gravitational representation of a microscopic state contributing to 
the black hole entropy. String webs also provide a nice geometric interpretation of various field--theoretic phenomena \cite{FTBion}. 
The webs of higher--dimensional branes are even more interesting: unlike the $(p,q)$ strings, which follow straight lines, these objects can 
have very complicated shapes. Brane webs occupy a very special niche among curved branes since no gauge field is turned on 
along their worldvolume (usually such gauge field acts as a cause of 
a nontrivial shape, see \cite{CalMald} for a discussion of a prototypical example), and they clearly deserve better understanding. 

Most of the work on curved D--branes has been performed 
using an open string picture for the branes \cite{Leigh}. In 
particular, this description
has been used to determine the shapes of the branes and to analyze the supersymmetries preserved by them. However, D--branes have also emerged as solitons in the closed string theory \cite{HorowStrom},
which gives an alternative way of studying their dynamics. 
Moreover, since the open-- and closed--string pictures describe the same object in different corners of the parameter space \cite{polch}, a comparison of brane dynamics derived from these complementary descriptions should provide a test of an open/closed
string duality proposed in \cite{polch}. In the last decade many such checks have been performed, but mainly they were done in the decoupling limit, where open string description reduces to field theory 
\cite{malda,gkp,WittAdS}. It would be very nice to extend this agreement to the full--fledged duality between open and closed strings.

Unfortunately, in the presence of Ramond--Ramond fluxes, genuine string computations are very hard, if not impossible, so often one 
has to rely on low-energy effective actions: supergravity for closed strings, and DBI action for the open ones. 
For orthogonal brane intersections preserving eight supercharges, 
a perfect agreement between SUGRA and DBI analyses was found in 
\cite{myCM}, and  one of the goals of this paper is to extend the results of that work to the brane webs. To perform a comparison, we will 
construct the geometries produced by various webs, and these metrics 
might have numerous applications even apart from testing DBI/SUGRA
duality. Once the agreement between two descriptions is demonstrated in the asymptotically--flat case, one is naturally led to a correspondence between various quantities in the near--horizon limit. 

In this article we will study the  webs constructed from branes with the same dimensionality, and three such systems admit regular near--horizon limits: the webs of D3, M2 and M5 branes. Even for a single stack of branes, a near--horizon limit of the geometry is not 
geodesically--complete, and, to recover the entire $AdS_p\times S^q$, 
some continuation of the metric is required \cite{WittAdS}. In the case of 1/4--BPS brane webs, 
an analogous extension leads to the geometries whose local structure has been determined in \cite{GMSW,donos}. Such continuation corresponds to formulating field theory on $R\times S^{p-2}$ rather than on $R^{1,p-2}$, while on the gravity side it removes infinite 
throats and replaces them by smooth "bubbles". For example, the vacuum of the field theory on $R\times S^{p-2}$ corresponds to the 
$AdS_p\times S^q$ with global coordinates on AdS space, and 
1/2--BPS states in this theory correspond to the smooth geometries discovered in \cite{LLM}. Similarly, 1/4--BPS states in field theory correspond to some of the solutions constructed in 
\cite{GMSW,donos}, but, according to the rules of AdS/CFT correspondence, only regular geometries (and their degenerate limits) 
are allowed.
This implies that the local analysis of \cite{GMSW,donos} should be supplemented by some boundary conditions, and we will derive them in this paper. As in the 1/2--BPS case \cite{LLM}, the boundary
conditions can be formulated in terms of droplets in some Kahler space, but, unlike their 1/2--BPS cousins, the droplets corresponding to 1/4--BPS solutions cannot have arbitrary shapes. We will derive the 
conditions which should be obeyed by the boundaries of droplets, and 
these restrictions will be shown to agree with expectations from the analysis of brane probes. This will serve as one of the checks of the 
open/closed duality in the near--horizon region. 

This article has the following organization. In section \ref{SectProbe} we 
review the description of brane webs in terms of open strings. In particular, 
we classify the nontrivial 1/4--BPS webs in IIB string and M theories and show that they either form planar networks built from straight lines,
or follow holomorphic profiles.
While most of the results presented in that section are well--known, it is useful to 
write them in the uniform notation for later comparison with gravity analysis.

In section \ref{SectStrWeb} we derive the geometries produced by 
the webs of 
$(p,q)$--strings, and we demonstrate that, for consistency of SUGRA, 
such webs must be built from straight line segments, and orientation of the 
segments must be correlated with charges $p$ and $q$. Although this conclusion comes from supergravity (and it does not rely on any 
information about branes), it agrees perfectly with outcome of the probe 
(or open--string) analysis, and this agreement can be viewed 
as a nontrivial check of the open/closed string duality. 
Section \ref{SectMembr} extends such agreement to the webs of M2 
branes, where situation is even more interesting: worldvolume analysis suggests that the membranes must follow holomorphic profiles, and we confirm this result by an independent 
computation in supergravity. In section \ref{Sect35Brn} we use various 
dualities to construct geometries produced by other brane webs, and we compare 
with earlier results of \cite{myCM}. Again, a perfect agreement between probe and SUGRA descriptions is found. 

Notice that the metrics discussed in section 
\ref{SectMembr} have been written before \cite{marolf}, but they have never 
been derived from the first principles. The authors of \cite{marolf}  
proposed an ansatz 
which contained a Kahler space fibered over ${\bf R}^6$, enforced the projectors which were exported from the probe analysis, and checked that all 
SUGRA equations were satisfied. While this approach yielded a solution of 
eleven--dimensional supergravity, it was not clear whether there were any generalizations, moreover, since the open--string projectors were introduced by hand, the construction of \cite{marolf} could not provide an independent check of the open/closed string duality.  In contrast to the approach of \cite{marolf}, our derivation is based only on supergravity, so an agreement with probe analysis appears as a nontrivial agreement.  

The authors of \cite{marolf} also argued that a solution produced by a 
curved M2 brane did not exist. In particular, perturbative arguments were  
used to conclude that gravity solution breaks down {\it everywhere} 
in space. However, this statement is somewhat counter--intuitive, since, starting with probe approximation, one should be able to turn on gravity without creating problems sufficiently far away from the branes. In section 
\ref{SectMembrPert} we show that the divergences encountered in 
\cite{marolf} originate from the perturbation theory in charges which was introduced in that paper, but,
performing a more natural multipole expansion instead, one arrives at a 
well--defined perturbation series which converges away from the sources. We view this fact as a strong evidence pointing to the existence of the supergravity solution, and we argue that the "improved" perturbation 
theory produces a unique geometry for any allowed distribution of membranes. 

The second part of this paper is devoted to studying 1/4--BPS geometries with $AdS_p\times S^q$ asymptotics. While such solutions can be constructed by taking decoupling limits of the brane webs discussed in 
sections \ref{SectMembr}--\ref{Sect35Brn}, the resulting space would 
not be 
geodesically complete. This situation has also been encountered for a
single stack of D3 branes, and in that case the space can be continued to produce a global $AdS_5\times S^5$. As a result of such continuation, one finds a string dual of a field theory on a sphere \cite{WittAdS}. 
Unfortunately, it is not clear 
how to perform a similar continuation for the near--horizon geometry of a 
brane web, but, fortunately, the local structure of the desired solutions have 
been derived from the first principles \cite{GMSW,myUnp,donos}. 
In section \ref{SectIIBbl} we analyze the global properties of the geometries constructed in \cite{myUnp,donos} and determine the restrictions imposed
by regularity. 

The solutions of \cite{myUnp,donos} have 
$SO(4)\times U(1)\times U(1)_t$ symmetry,
and the metric contains a coordinate $y$ which goes to zero when either $S^3$ or $U(1)$--direction collapses to zero size. Then, requiring regularity of the solution at $y=0$, one arrives at some boundary conditions on this hyperplane. This situation looks similar to the one encountered in the 
1/2--BPS case \cite{LLM}, where regularity implied that one--dimensional Kahler space was divided into droplets by assigning one of two values 
($Z=\pm\frac{1}{2}$) to a certain function. This analogy led the authors of \cite{vaman} to propose a similar picture for the 1/4--BPS droplets: it was 
argued that, by introducing droplets with $Z=\pm\frac{1}{2}$ in 2D Kahler 
space, one produces a regular geometry. It turns out that the 1/4--BPS case is more subtle. First, in addition to requiring certain value for 
function $Z$, one needs to impose a restriction on Kahler potential in 
$Z=-\frac{1}{2}$ regions (see equation (\ref{Jan31BndCnd})):
\bea\label{IntroZ}
y=0:\qquad\begin{array}{l}
Z=-\frac{1}{2},~ 
\d_a \db_b K(z,\zb,y=0)=0,\\
~\\
Z=+\frac{1}{2}.
\end{array}
\eea
The second subtlety is associated with the shapes of the droplets: while 
they could be arbitrary in the 1/2--BPS case \cite{LLM}, now they are
restricted by the regularity conditions. To be more precise, describing the wall between droplets by an equation $v(z_a,\zb_a)=0$, one finds a 
restriction (\ref{RegDgMtr1Cp}) on a real function $v$:
\bea\label{IntroV}
\d_a{\bar\d}_b v+\la \d_a v{\bar\d}_b v=g \d_a w \db_b\wb+O(v),\quad
\mbox{det}(\d_a v\d_b w)|_{v=0}\ne 0.
\eea
Function $w$ defined here must be holomorphic. The conditions 
(\ref{IntroZ}) and (\ref{IntroV}) are derived in section \ref{SectIIBDrpl}. 

While the relation 
(\ref{IntroV}) was derived from SUGRA analysis, it is also crucial for 
ensuring an agreement with probe calculations. As was shown in \cite{mikhail}, the profiles of supersymmetric D3 branes in $AdS_5\times S^5$ are described 
by holomorphic surfaces. In section \ref{SectIIBPrb} we extend this result to branes on an arbitrary 1/4--BPS background\footnote{We also provide a simple geometric interpretation of the radial coordinate introduced in 
\cite{mikhail}.}, but, starting from an arbitrary droplet and contracting it, one can potentially arrive at the sources which do not have holomorphic profiles. In section \ref{SectIIBDrpl} we show that it is the restriction 
(\ref{IntroV}) which prevents this from happening, and that regular 
droplets can collapse 
only to holomorphic cycles (which can be arbitrary). Moreover, in sections 
\ref{SectIIBDbr}, \ref{SectIIBPrb} we also show that holomorphicity of the 
brane embeddings can be independently derived both from probe analysis and from consistency of SUGRA, so, in some sense, a lack of the restriction (\ref{IntroV}) would have implied an internal inconsistency of supergravity.

In section \ref{SectIIBPert} we demonstrate that, for the fixed asymptotic behavior, any distribution of droplets with boundary conditions (\ref{IntroZ}) leads to the unique geometry, and the restriction on the Kahler potential 
appearing in (\ref{IntroZ}) is instrumental in ensuring the uniqueness. Of course, to avoid extra sources on the domain walls, the restrictions 
(\ref{IntroV}) should also be imposed. 

In section \ref{SectIIBTop} we discuss the topology of the 1/4--BPS solutions: we construct two types of non--contractible five--manifolds, and find very simple expressions for the flux of $F_5$ through these cycles. This is very similar to the picture encountered in the 1/2--BPS case 
\cite{LLM}: rather than having sources, the fluxes are supported by non-trivial topology.

Section \ref{SectM4geom} is devoted to the discussion of 1/4--BPS 
geometries 
in M theory, which are obtained by an analytic continuation of the metrics found in \cite{GMSW}. Since the resulting geometries share many qualitative properties with their ten--dimensional cousins, section 
\ref{SectM4geom} is rather brief: we only underline the differences. While we were not able to construct new explicit solutions in ten or eleven dimensions, sections \ref{SectIIBEx} and \ref{SectM4Ex} discuss several 
examples which are obtained by embedding some old solutions into new ansatze. In particular, we embed all geometries constructed in \cite{LLM}.  
While doing this embedding, one notices that there are striking similarities 
between ten-- and eleven--dimensional solutions, which are not obvious at first sight. In section \ref{SectUnif} we rewrite type IIB solutions (both in 
1/2-- and 1/4--BPS case) in a way which makes an analogy with M theory more transparent. 

%=====================================

\section{Webs in the probe approximation}

\label{SectProbe}

\renewcommand{\theequation}{2.\arabic{equation}}
\setcounter{equation}{0}

%=====================================

We begin with recalling some well-known facts about supersymmetric brane webs in IIB string theory. Supersymmetry transformations in this 
theory are parameterized by two Majorana--Weyl spinors which have the same chirality, and it is convenient to combine them into a 32--component real object $\eps$ which satisfies a chirality projection:
\bea
\eps=\left(\begin{array}{c} \eps_1\\ \eps_2\end{array}\right),\quad 
{\bf 1}_2\otimes \Gamma_{11}\eps=-\eps:\qquad \Gamma_{11}\eps_{1,2}=
-\eps_{1,2}.
\eea
Ten--dimensional flat space preserves $32$ supersymmetries corresponding to 
arbitrary constant values 
of $\eps_1$ and $\eps_2$ (modulo the chiral projection). By adding a brane 
to $R^{9,1}$ one breaks half of the supersymmetries and the appropriate 
projections are 
\cite{ScanRef} (see also \cite{smithRev} for a review):
\bea\label{BraneSUSY}
{\rm F1}:&\quad \Gamma=\sigma_3\otimes \Gamma_{(2)},\quad & 
\Gamma\eps=\eps,\nonumber\\
{\rm NS5}:&\quad \Gamma=\sigma_3\otimes \Gamma_{(6)},\quad &
\Gamma\eps=\eps,\\
{\rm D}(2p-1):&\quad \Gamma=i\sigma^p_3\sigma_2\otimes \Gamma_{(2p)},\quad &
\Gamma\eps=\eps.\nonumber
\eea
Here $\Gamma_{(2p)}$ is a product of gamma matrices with indices pointing 
along the worldvolume of the brane. 

Each of the branes appearing in 
(\ref{BraneSUSY}) preserves $16$ real supercharges and there are two other 
interesting objects which have the same amount of SUSY --- a plane wave and a KK monopole:
\bea\label{GeomBranes}
{\rm P}:&\quad \Gamma={\bf 1}_2\otimes \Gamma_{(2)},\quad & 
\Gamma\eps=\eps,\nonumber\\
{\rm KK}:&\quad \Gamma={\bf 1}_2\otimes \Gamma_{(6)},\quad &
\Gamma\eps=\eps.
\eea
These configurations have a pure geometric nature and they do not involve fluxes. 

Once the building blocks preserving half of the supersymmetries are 
specified, one can construct supersymmetric intersections by combining the ingredients with commuting projectors. 
In particular, it is interesting to look at the  
orthogonal intersections of branes which have the same number of worldvolume directions:
\bea\label{IntersBranes}
\left(\begin{array}{c}
F1_1\\ D1_2\end{array}\right),
\left(\begin{array}{c}
D3_{123}\\ D3_{145}\end{array}\right),
\left(\begin{array}{c}D5_{12345}\\ D5_{12367}\end{array}\right),
\left(\begin{array}{c}D5_{12345}\\ D5_{16789}\end{array}\right),
\left(\begin{array}{c}
NS5_{12345}\\ D5_{12678}\end{array}\right),
\left(\begin{array}{c}
NS5_{12345}\\ D5_{12346}\end{array}\right),
\eea
All these configurations preserve eight real supercharges. It turns out that the same supercharges are preserved by more general configurations which are known as brane webs \cite{StrWeb,D5Webs}. Let us 
analyze this in more detail.

{\bf String web.} 

We begin with the first system in 
(\ref{IntersBranes}). Any spinor satisfying the $F1_1$ and $D1_2$ projectors also obeys the following relations:
\bea\label{PQSproj}
\Gamma_{(p,q)}\eps&=&\eps,\nonumber\\
\Gamma_{(p,q)}&=&
\left[\frac{p}{\sqrt{p^2+\frac{q^2}{g_s^2}}}\sigma_3+
\frac{g_s^{-1}q}{\sqrt{p^2+\frac{q^2}{g_s^2}}}\sigma_1\right]
\otimes\Gamma_{0}\left[
\frac{p}{\sqrt{p^2+\frac{q^2}{g_s^2}}}\Gamma_1+
\frac{g_s^{-1}q}{\sqrt{p^2+\frac{q^2}{g_s^2}}}\Gamma_2\right].
\eea
The $\Gamma_{(p,q)}$ projector corresponds to a $(p,q)$--string  which carries $p$ units of string charge and $q$ units of D1 
charge\footnote{For simplicity 
we consider the case of 
vanishing axion: $\tau\equiv \frac{i}{g_s}+a=\frac{i}{g_s}$.} 
\cite{StrWeb} (this can be read off from the first bracket in 
$\Gamma_{(p,q)}$). The second bracket in (\ref{PQSproj}) indicates 
that the 
string stretches along the line
\bea\label{StraightPQ0}
qx_1-g_spx_2=\mbox{const},\quad x_3,\dots x_9 \mbox{ --- fixed}.
\eea
The $(p,q)$ strings which are mutually BPS can be joined to produce 
complicated string webs \cite{StrWeb}, which are constrained only by 
charge conservation at the junctions. 

%==================

\begin{figure}[tb]
\begin{center}
\epsfysize=2.5in \epsffile{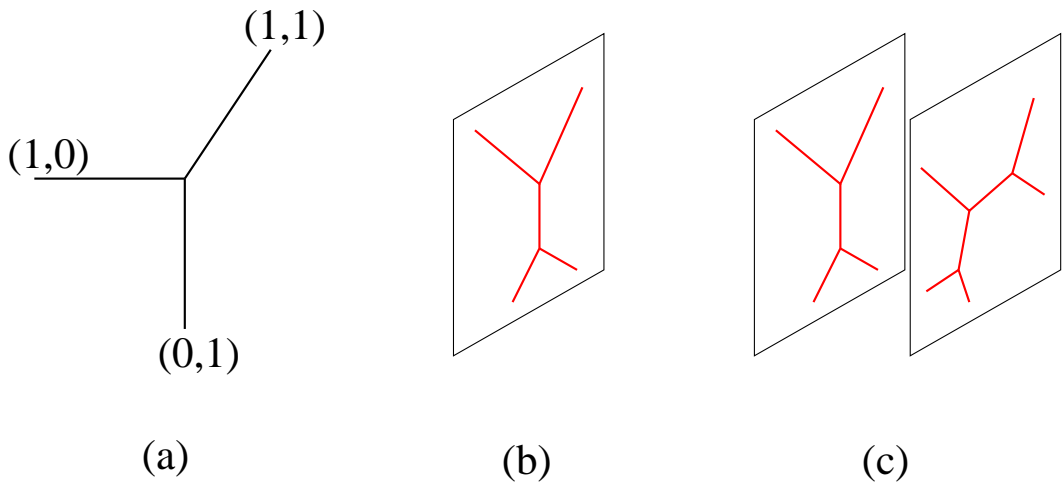}
\end{center}
\caption{
String webs: (a) an elementary junction, (b) a connected web, (c)  
a generic supersymmetric web.
} \label{FigWeb}
\end{figure}

%======================

In particular, it is interesting to consider a junction with incoming fundamental string and D1 brane (see figure \ref{FigWeb}a) and compare it with a general picture for strings ending on branes \cite{CalMald}. 
By charge conservation, the outgoing leg should be a $(1,1)$ string which has a trajectory
\bea\label{StraightPQ}
x_1=g_s x_2,
\eea
and, as expected, for small string coupling this is a small perturbation 
of a vertical D1 brane. The general analysis of \cite{CalMald} suggests 
that the resulting $(1,1)$ string can also be viewed as a D1 brane with some electric flux on its worldvolume and a nontrivial profile 
$x_1(x_2)$:
\bea\label{HarmPQ}
E=\d_2 x_1,\quad \d_2^2 x_1=0
\eea
Clearly the profile (\ref{StraightPQ}) solves the Laplace equation, then 
(\ref{HarmPQ}) gives an expression for the electric field. By analyzing the coupling between the electric field and the Kalb--Ramond tensor in the bulk, one can see that the proper string charge is also 
reproduced\footnote{A similar analysis for a string ending on D3 brane was performed in \cite{myCM}.}. Thus we see that the string junction can be viewed as a simplest example of a bion. 

Starting from any junction on the string web, one can draw a plane through the strings which pass through it, and define Cartesian coordinates 
$(x_1,x_2)$ in this plane. Then the projection (\ref{PQSproj}) implies that 
all legs of the supersymmetric web must be parallel to this plane, so any connected string web must be planar (see figure \ref{FigWeb}b). However it is also possible to have disconnected webs which can be constructed by combining elementary webs oriented in the same directions (see 
figure \ref{FigWeb}c). To construct the geometries produced by the $(p,q)$ systems, it is convenient to begin with metric produced by an elementary block depicted in figure \ref{FigWeb}b, and then generalize it to the case of disconnected webs. This logic will be implemented in section 
\ref{SectStrWeb}. 

{\bf D3--web.}

Let us now consider the D3--D3 system which appears in 
(\ref{IntersBranes}). The preserved supersymmetries satisfy 
two independent relations, and it is useful to combine them in the 
following way:
\bea\label{D3Proj1}
i\sigma_2\otimes \Gamma_{0123}\eps=
i\sigma_2\otimes \Gamma_{0145}\eps=\eps:\quad
\Gamma_{2345}\eps=-\eps.
\eea
The last projector allows us to write two more conditions:
\bea\label{D3Proj2}
i\sigma_2\otimes \Gamma_{0124}\eps=
-i\sigma_2\otimes \Gamma_{0135}\eps,\quad
i\sigma_2\otimes \Gamma_{0125}\eps=
i\sigma_2\otimes \Gamma_{0134}\eps,
\eea 
and the relations listed in (\ref{D3Proj1}), (\ref{D3Proj2}) can be summarized in a compact form:
\bea\label{D3CmplPrj}
i\sigma_2\Gamma_{01a{\bar b}}\eps=
\frac{i}{2}\delta_{a{\bar b}}\eps,\quad
z^1=x_2+ix_3,\quad z^2=x_4+ix_5.
\eea
As in the case of F1--D1 intersections, we find that this Killing spinor is preserved by a larger family of branes. To see this, we recall the 
kappa--symmetry projection associated with a D brane\footnote{We assumed that the branes have no fluxes on their worldvolume, but a 
more general case can be considered as well.} \cite{ScanRef,Kappa}:
\bea\label{DBIProj}
{\rm D}(2p-1):&&
\Gamma=i\sigma^p_3\sigma_2\otimes \left[
{\cal L}^{-1}
\left(\prod_{m=0}^{2p-1}\frac{\d X^{\mu_m}}{\d\xi^m}\right)
\Gamma_{\mu_0\dots \mu_{2p-1}}\right],\quad 
\Gamma\eps=\eps,\nonumber\\
&&{\cal L}=\sqrt{\mbox{det}(G_{\mu\nu}\d_m X^\mu\d_n X^\nu)}.
\eea
In the case of D3 branes it is convenient to introduce complex coordinates both in spacetime and on the worldvolume:
\bea
Z^1=X_2+iX_3,\quad  Z^2=X_4+iX_5,\quad w=\xi_2+i\xi_3
\eea
Then, assuming that $Z^a$ are functions of $w$, ${\bar w}$ and 
imposing the static gauge in the remaining two directions (
$X^0=\xi^0$, $X^1=\xi^1$), we find
\bea\label{HolProj1}
\Gamma\eps&=&i\sigma_2\Gamma_{01}
\frac{1}{\sqrt{\mbox{det}(\d_m Z^a\d_n {\bar Z}^a)}}
\left[
(\d Z^b~{\bar\d}{\bar Z}^{\bar c}-{\bar\d} Z^b~\d{\bar Z}^{\bar c})
\Gamma_{b{\bar c}}+
\d Z^b~{\bar\d}{Z}^{c}\Gamma_{b{c}}+
\d {\bar Z}^{\bar b}~{\bar\d}{\bar Z}^{\bar c}\Gamma_{{\bar b}{\bar c}}\right]\eps\nonumber\\
\\
&=&
\frac{2i}{\sqrt{\mbox{det}(\d_m Z^a\d_n {\bar Z}^a)}}
\left[\frac{1}{4}
(\d Z^b~{\bar\d}{\bar Z}^{b}-{\bar\d} Z^b~{\d}{\bar Z}^{b})
+
\d Z^b~{\bar\d}{Z}^{c}\Gamma_{b{c}2{\bar 2}}+
\d {\bar Z}^{\bar b}~{\bar\d}{\bar Z}^{\bar c}
\Gamma_{{\bar b}{\bar c}2{\bar 2}}\right]\eps.\nonumber
\eea 
To arrive at the second line we used the projector (\ref{D3CmplPrj}). 
The right--hand side of the last expression is proportional to $\eps$ if and only if
\bea
\d Z^a{\bar\d}{Z}^b=0
\eea
for all values of $a$ and $b$. This implies that $Z^1$ and $Z^2$ must be holomorphic functions\footnote{Alternatively, one could have  assumed that these functions are anti--holomorphic, but this would lead to $\Gamma\eps=-\eps$, i.e. such configurations preserve the wrong spinor.}, then the relation (\ref{HolProj1}) simplifies:
\bea
\Gamma\eps=\frac{1}{\sqrt{\frac{i}{2}\d Z^a{\bar\d}{\bar Z}^a}}
\left[\frac{i}{2}\d Z^b~{\bar\d}{\bar Z}^{b}\right]\eps=\eps
\eea 
Thus we learned that a spinor satisfying the projection (\ref{D3CmplPrj}) is preserved by any D3 brane which follows a holomorphic profile in 
$(Z^1,Z^2)$ directions. In particular, we can consider a straightforward 
generalization of (\ref{StraightPQ0}):
\bea
Z_1-aZ_2=\mbox{const},
\eea
which corresponds to a D3 brane with flat worldvolume, and one can form "D3--webs" by constructing the junctions of such objects. Of course, 
more general holomorphic profiles:
\bea\label{Jan31HolProf}
f(Z_1,Z_2)=0
\eea
can also be considered, but  we will refer to them as D3--webs as well. 

{\bf Webs of D5--branes.} 

The third configuration in (\ref{IntersBranes})
has eight supercharges which are also preserved by an 
arbitrary holomorphic web $x_4+ix_5=x_6+ix_7$. The simplest way to see this in to notice that two T dualities relate this configuration of five--branes to the D3--D3 system discussed above. 

The spinors preserved by the fourth configuration in (\ref{IntersBranes})
 obey two independent projections:
\bea
\sigma_1\otimes \Gamma_{012345}\eps=\eps,\quad
\sigma_1\otimes \Gamma_{016789}\eps=\eps.
\eea
To analyze whether these projections can be satisfied by any 
nontrivial 1/4--BPS brane webs, it is convenient to introduce 
complex variables:
\bea
Z^1=x_2+ix_3,\quad Z^2=x_4+ix_5,\quad Z^3=x_6+ix_7,\quad
Z^4=x_8+ix_9.
\eea
In terms of these coordinates the projectors become
\bea\label{QtrnnProj1}
\sigma_1\otimes\Gamma_{01}
\Gamma_{Z_1{\bar Z}_1Z_2{\bar Z}_2}\eps=-\frac{1}{4}\eps,\quad
\sigma_1\otimes\Gamma_{01}
\Gamma_{Z_3{\bar Z}_3Z_4{\bar Z}_4}\eps=-\frac{1}{4}\eps.
\eea
Notice that, since we are looking for 1/4--BPS configurations, these relations must be satisfied by both 
$\eps_-=\frac{1}{2}(1-\sigma_1\otimes\Gamma_{01})\eps$ and 
$\eps_+=\eps-\eps_-$, so we can concentrate on $\eps_-$. If one introduces a basis corresponding to each one of the 
complex coordinates:
\bea\label{CmplGmmStr}
\Gamma_Z|\downarrow\rangle=0,\quad 
\Gamma_Z|\uparrow\rangle=|\downarrow\rangle,\quad
\Gamma_{\bar Z}|\uparrow\rangle=0,\quad
\Gamma_{\bar Z}|\downarrow\rangle=|\uparrow\rangle,
\eea
then an arbitrary spinor $\eps_-$ satisfying projection 
(\ref{QtrnnProj1}) can be written as 
\bea
\eps_-=e_1|\uparrow\uparrow\uparrow\uparrow\rangle+
e_2|\uparrow\uparrow\downarrow\downarrow\rangle+
e_3|\downarrow\downarrow\uparrow\uparrow\rangle+
e_4|\downarrow\downarrow\downarrow\downarrow\rangle.
\eea
To preserve a quarter of supersymmetries, a brane should admit Killing spinors with independent coefficients $e_i$. Application of 
the projector (\ref{DBIProj}) to the spinor $\eps_-$ leads to the relation
\bea
\Gamma\eps_-={\cal L}^{-1}
\left(\prod_{m=1}^{4}\frac{\d X^{\mu_m}}{\d\xi^m}\right)
\Gamma_{\mu_1\dots \mu_4}\eps_-\equiv 
P[X^{\mu_1},X^{\mu_2},X^{\mu_3},X^{\mu_4}]
\Gamma_{\mu_1\mu_2\mu_3 \mu_4}\eps_-,
\eea
and the right--hand side of this expression should be equal to 
$\eps_-$. Looking at coefficients in front of various components of the
spinor and requiring them to match for arbitrary values of $e_1$, $e_2$, $e_3$, $e_4$, we find several relations:
\bea\label{QtrnnProj2}
&&P[Z_1,Z_2,Z_3,Z_4]=0,\quad P[Z_a,Z_b,Z_c,{\bar Z}_c]=0,\quad
\sum_{a,b} P[Z_a,{\bar Z}_a,Z_b,{\bar Z}_b]=\frac{1}{4}\nonumber\\
&&P[Z_1,Z_2,\Zb_3,\Zb_4]=
\sum_a P[Z_1,Z_2,\Zb_a,Z_a]=\sum_a P[Z_3,Z_4,\Zb_a,Z_a]=0
\nonumber\\
&&\sum_{a,b} s_a s_b P[Z_a,{\bar Z}_a,Z_b,{\bar Z}_b]=\frac{1}{4},
\quad s_1=s_2=1,\quad s_3=s_4=-1
\eea
The first equation implies that $(Z_1,Z_2,Z_3,Z_4)$ are 
functionally--dependent, i.e. we can choose a patch where 
$Z_4=f(Z_1,Z_2,Z_3)$. The second equation in (\ref{QtrnnProj2}) implies a functional dependence between $(Z_1,Z_2,Z_3)$ and any one of the three functions $\Zb_1$, $\Zb_2$, $\Zb_3$, 
in particular we find
\bea
Z_3=f_3(Z_1,Z_2,\Zb_1)={\tilde f}_3(Z_1,Z_2,\Zb_2).
\eea
On the chosen patch we can also impose a static
gauge by identifying $(Z_1,Z_2,\Zb_1,\Zb_2)$ with coordinates 
along the brane. 
Since these four coordinates are functionally independent, the last equation can only be satisfied if $f_3$ is a holomorphic function, then we find
\bea
Z_3=f_3(Z_1,Z_2),\quad Z_4=f_4(Z_1,Z_2).
\eea
The two non--homogeneous equations from 
(\ref{QtrnnProj2}) can be combined produce a simple relation: 
\bea
0=\sum_{a=1}^2\sum_{b=3}^4 P[Z_a,{\bar Z}_a,Z_b,{\bar Z}_b]=
\sum_{a=1}^2\sum_{b=3}^4\left|\frac{\d Z_b}{\d Z_a}\right|^2 
P[Z_1,\Zb_1,Z_2,\Zb_2],
\eea
then, since $P[Z_1,\Zb_1,Z_2,\Zb_2]\ne 0$, we conclude that both $Z_3$ and $Z_4$ must be constants on our patch. This brings us 
back to the flat brane $D5_{12345}$ which appeared in 
(\ref{IntersBranes}). Thus we see that, unlike the first three systems in 
(\ref{IntersBranes}), the $1+1$--dimensional orthogonal D5--D5 intersection does not have any interesting generalization analogous 
to the string webs. 

Since D5--D5 intersection and the fifth configuration in 
(\ref{IntersBranes}) can be obtained from the same M theory system:
\bea\label{Jan31RedM5}
\left(\begin{array}{c}
D5_{12345}\\
D5_{16789}
\end{array}\right)\stackrel{R_M,T_1}{\longleftarrow} 
\left(\begin{array}{c}
M5_{M2345}\\
M5_{M6789}
\end{array}\right)\stackrel{R_6,T_2}{\longrightarrow} 
\left(\begin{array}{c}
NS5_{M2345}\\
D5_{M26789}
\end{array}\right),
\eea
we conclude that NS5 and D5 branes intersecting along 
$2+1$--dimensional subspace do not have interesting 
generalizations.   

{\bf (p,q)--fivebranes.}

The last configurations in (\ref{IntersBranes}) is very similar to the 
F1--D1 system: in this case 
an analog of the relation (\ref{PQSproj}) can be written as:
\bea
\Gamma_{(p,q)}\eps&=&\eps,\nonumber\\
\Gamma_{(p,q)}&=&
\left[\frac{g_s^{-1}p}{\sqrt{q^2+\frac{p^2}{g_s^2}}}\sigma_3+
\frac{q}{\sqrt{q^2+\frac{p^2}{g_s^2}}}\sigma_1\right]
\otimes\Gamma_{0}\left[
\frac{g_s^{-1}p}{\sqrt{q^2+\frac{p^2}{g_s^2}}}\Gamma_5+
\frac{q}{\sqrt{q^2+\frac{p^2}{g_s^2}}}\Gamma_6\right]\Gamma_{1234}
\nonumber
\eea
This projection corresponds to a $(p,q)$ five--brane 
\cite{D5Webs} stretching 
in the $x_1,x_2,x_3,x_4$ directions and along the line 
\bea
px_6-g_sqx_5=\mbox{const}.
\eea

This completes our discussion of the 1/4--BPS brane webs appearing in IIB string theory and now we will make few comments about similar systems in eleven dimensions. 

{\bf Brane webs in M theory.}

It is easy to classify the orthogonal intersections of M5 (M2) branes which preserve eight supercharges:
\bea\label{IntrsMBrns}
\left(\begin{array}{c}
M5_{12345}\\ M5_{12367}\end{array}\right),\quad
\left(\begin{array}{c}M5_{12345}\\ M5_{16789}\end{array}\right),\quad
\left(\begin{array}{c}M2_{12}\\ M2_{34}\end{array}\right).
\eea
The first intersection is related by U--duality to the D3--D3 system which was discussed above, so it preserves the same supercharges as a web of the M5--branes with harmonic profiles: 
\bea
x_4+ix_5=f(x_6+ix_7),\quad 
x_8=x_8^{(0)},~x_9=x_9^{(0)},~x_{10}=x_{10}^{(0)}.
\eea

The second intersection in (\ref{IntrsMBrns}) is related to the D5--D5 system (see equation (\ref{Jan31RedM5})), and thus it cannot be generalized in any interesting way. 

The intersecting membranes from (\ref{IntrsMBrns}) can be mapped either into the D1--F1 or into D3--D3 system (depending on the direction of smearing):
\bea
\left(\begin{array}{c}D3_{1256}\\ D3_{3456}\end{array}\right)
\stackrel{R_9,T_{56}}{\longleftarrow}
\left(\begin{array}{c}M2_{12}\\ M2_{34}\end{array}\right)
\stackrel{R_2,T_3}{\longrightarrow}
\left(\begin{array}{c}F1_{1}\\ D1_{4}\end{array}\right).
\eea
The map to D3--D3 demonstrates an existence of the holomorphic web of membranes, and the map to F1--D1 will be useful for the discussion in section \ref{SectStrWeb}. 

{\bf Summary.} 

Let us summarize the results of this section. 
We demonstrated that nontrivial brane webs preserving eight supercharges 
fall into two categories: they are either described by planar networks built from straight lines (this happens for (p,q) strings and five--branes), or they have profiles which are governed by holomorphic functions of two variables.
After this brief overview of brane webs, we turn to construction of the geometries produced by them. 

%=====================================

\section{Geometries produced by string webs}

\label{SectStrWeb}

\renewcommand{\theequation}{3.\arabic{equation}}
\setcounter{equation}{0}

%=====================================

In the previous section we reviewed the open string picture for the brane webs and such description is applicable for the webs which carry small 
charge. As the number of branes becomes larger, the probe approximation breaks down, but one finds another semiclassical description in terms of closed strings, and often a good quantitative description of the dynamics is given by supergravity.
The next two sections will be devoted to finding the gravity solutions produced by webs of branes, and in this section we will describe the geometries corresponding to the webs of $(p,q)$--strings.

In the probe approximation, a generic web preserving eight supercharges contains various independent "elementary webs" located in parallel planes (see figure \ref{FigWeb}c). To construct the relevant supergravity solutions, we begin with describing a geometry 
produced by an "elementary web" and then find its generalization. 
The main advantage of this approach relies on the fact that an 
"elementary web" (see figure \ref{FigWeb}b) has a large bosonic 
symmetry, and one can derive the most general solution of IIB SUGRA 
which has corresponding isometries.
Indeed, if an "elementary web" is stretched in $x_1,x_2$ directions, then it is invariant under $SO(7)$ rotations in the orthogonal directions. To preserve these isometries, one is only allowed to excite a metric, an electric three--form, and an
axion--dilaton in IIB supergravity.
While it is possible to solve the SUSY variations and to find the most general geometry with $SO(7)$ symmetry, we will pursue a different path which will also lead to geometries produced by other brane webs. 
Namely we begin with smearing the web in one of the orthogonal
directions, thus reducing the symmetry to 
$SO(6)\times U(1)$. The advantage of this procedure is that we now have a translational isometry (the Killing spinor is not charged under 
the $U(1)$ transformations), so a T duality along this direction leads to a supersymmetric configuration in IIA theory. A further lift to eleven dimensions yields a configuration of M2 branes. For example, a
fundamental string $F1_1$ and a $D1_2$ brane map into the membranes with following orientations:
\bea\label{MembrScan}
\begin{tabular}{c|cccccccccc}
&M&1&2&3&4&5&6&7&8&9\\
\hline
M2$_F$&$\bullet$&$\bullet$&&$\sim$&&&&&&\\
M2$_D$&$\sim$&&$\bullet$&$\bullet$&&&&&&
\end{tabular}
\eea
Due to the smearings, we have translational invariance in $x_3$ and in 
the M theory direction, there is also an $SO(6)$ rotational symmetry. 
Given these isometries, we find that the most general static configuration of the membranes is described by the following ansatz:
\bea
ds^2&=&-e^{2A}dt^2+e^{2C}(dw_1-\chi dw_2)^2+e^{2D}dw_2^2+g_{MN}dX^MdX^N+
e^{2B}d\Omega_5^2\nonumber\\
G_4&=&dt\wedge W_\alpha\wedge dw^\alpha.\nonumber
\eea
In the appendix \ref{SectAp1} we found the most general 
supersymmetric solution which has this form and, upon translating it back to the IIB theory, we find the geometry which describes string webs:
\bea
ds_{E}^2&=&-e^{9A/4}dt^2+
e^{9A/4}h_{ab}dx^adx^b+e^{-3A/4}(dz^2+dy^2+y^2d\Omega_5^2),\\
e^{2\Phi}&=&e^{3A}h_{11}^2,\quad
C^{(0)}=-\frac{h_{12}}{h_{11}},\quad 
B=e^{3A}h_{1a}~dt\wedge dx^a,\quad
C^{(2)}=e^{3A}h_{2a}~dt\wedge dx^a,\nonumber\\
&&h_{ab}=\frac{1}{2}\d_a\d_b K(x_c,y),\quad e^{-3A}=\mbox{det}~h,\quad
\frac{1}{y^5}\d_y(y^5\d_y K)=-2\mbox{det}~h.\nonumber
\eea
This metric was derived under assumption of $SO(6)\times U(1)_z$ 
symmetry, which can be easily relaxed: 
to describe an 
"elementary web" with $SO(7)$ symmetry, the expression in the parenthesis should be replaced by $dy^2+y^2 d\Omega_6^2$, and for 
the geometry corresponding to the most general web one expects to 
have an arbitrary dependence upon the seven--vector ${\bf y}$:
\bea\label{StrWebGeom}
ds^2_{E}&=&-e^{9A/4}dt^2+e^{9A/4}h_{ab}dx^a dx^b+
e^{-3A/4}d{\bf y}^2\\
\tau&=&\frac{1}{h_{11}}(ie^{-3A/2}-h_{12}),\quad
B+iC^{(2)}=e^{3A}dt\wedge (h_{1a}dx^a+ih_{2a}dx^a)\nonumber\\
\label{StrWebMA}
&&h_{ab}=\frac{1}{2}\d_a\d_b K(x_c,y),\quad e^{-3A}=\mbox{det}~h,\quad
\Delta_{\bf y} K=-2\mbox{det}~h.
\eea
These relations give a local description of the solution away form the sources and now we explain how to incorporate charged objects into
this picture. 
In supergravity one can account for branes in two different ways: they
can either be introduced as sources in the equations of motion, or their
presence can be reflected in boundary conditions. We will pursue the second option. 

Looking at the structure of the two--form potentials, we conclude that 
the $y$--directions should be orthogonal to the branes, so the web breaks into "elementary planes" located at specific values of ${\bf y}$. 
Moreover, in each plane the branes go along certain curves 
$f(x_1,x_2)=0$. Let us introduce $w_1=w_1(x_1,x_2)$ as a 
coordinate along the curve and define $w_2=w_2(x_1,x_2)$ to be 
orthogonal to it. As we approach a string or a D brane, the metric component $g_{tt}$ should go to zero, so $e^{-3A}$ should diverge at the profile of the brane. Moreover, for consistency, the leading order of the metric should split into the longitudinal and transverse parts:
\bea
ds^2_{E}&=&e^{9A/4}(-dt^2+h_{w_1w_1}dw_1^2)+
e^{-3A/4}(e^{3A}h_{w_2w_2}dw_2^2+d{\bf y}^2),
\eea
and the expressions in parenthesis should give regular metrics. 
This implies that functions $h_{w_1w_1}$ and 
${\tilde h}_{w_2w_2}=e^{3A}h_{w_2w_2}$ should remain finite and non--vanishing in the vicinity of the brane, so in the
leading order they must be constant. Moreover, by rescaling $w_a$,
we can set $h_{w_1w_1}={\tilde h}_{w_2w_2}=1$.
This yields the leading contributions to the 
Kahler potential:
\bea
K=w_1^2+2\int^{w_2} d\tilde w_2
\int^{\tilde w_2} d{\hat w}_2 e^{-3A}.
\eea
Taking derivatives of this expression, we find the leading terms in 
the complex two--form:
\bea
B+iC^{(2)}=e^{3A}dt\wedge\left[dw_1
\left(\frac{\d w_1}{\d x_1}+i\frac{\d w_1}{\d x_2}\right)+
e^{-3A}dw_2
\left(\frac{\d w_2}{\d x_1}+i\frac{\d w_2}{\d x_2}\right)\right].\nonumber
\eea
By definition, coordinate $w_2$ is orthogonal to the brane, so the leading 
contribution to the two--form potential should not have any components along $w_2$. To be more precise, one should require that, being rewritten in the orthonormal frame, the $w_2$ component of $B+iC^{(2)}$ should either vanish or be a pure gauge. This implies that
\bea
\left(\frac{\d w_2}{\d x_1}+i\frac{\d w_2}{\d x_2}\right)=A
\eea
is a complex constant. In other words, we see that $w_2$ is a linear function of $x_1$ and $x_2$. Then coordinate $w_1$, which was defined as an orthogonal complement of $w_2$, must be linear as well. 
The rotation from $(x_1,x_2)$ to $(w_1,w_2)$ can be parameterized by one angle $\phi$:
\bea\label{GeomrcWeb}
&&w_1=\cos\phi~x_1+\sin\phi~x_2,\quad 
w_2=-\sin\phi~x_1+\cos\phi~x_2,\\
&&B+iC^{(2)}=dt\wedge\left[e^{3A+i\phi}dw_1+
ie^{i\phi}dw_2\right].\nonumber
\eea

We conclude that the string web must consist of straight elements 
($x_2=\tan\phi~x_1$) and, in a perfect agreement with 
(\ref{StraightPQ0}), the ratio of the D1 and F1 charges is correlated with orientation of these elements. This agreement between the probe picture and supergravity solution is a manifestation of an open/closed string duality for the string webs. 

The most general solution (\ref{StrWebGeom}) should satisfy the equations 
(\ref{StrWebMA}) away from the sources which, for consistency, should be placed along straight lines in ${\bf y}={\bf y}_i$ planes. Such string webs, along with total charge at infinity, uniquely specify the solution: 
since each segment of the string must satisfy (\ref{GeomrcWeb}), 
the distribution of string/brane charge at each junction is uniquely determined by geometric angles. Once proper sources are specified, 
one can use perturbation theory to demonstrate that solution exists and it is unique.
To avoid repetition, we postpone the discussion of perturbative 
expansion until section \ref{SectMembrPert}, where 
geometries produced by membranes are analyzed. To describe
perturbation theory for the string webs, one would need to pick
particular profiles of the M2 branes.

%=====================================

\section{Webs of membranes}

\label{SectMembr}

\renewcommand{\theequation}{4.\arabic{equation}}
\setcounter{equation}{0}

%=====================================

In the previous section we discussed the geometries produced by the webs of $(p,q)$ strings and we found a perfect agreement between the profiles of the brane probes and the allowed sources in supergravity. While providing a very nice check of the duality between open and closed strings, this agreement was somewhat mundane since the webs were constructed out of straight lines. In this section we will explore a more interesting correspondence between the membrane probes and gravity solutions in eleven dimensions, and we will see that holomorphic profiles naturally arise in both descriptions.

%=====================================

\subsection{Structure of the solution}

\label{SectMembrStr}

%=====================================

As reviewed in section \ref{SectProbe}, a membrane 
intersection preserving eight supercharges can be generalized to an M2 brane following an arbitrary 
holomorphic profile. The resulting configuration is still 1/4--BPS. In 
general we do not expect to have additional bosonic symmetries, but, just 
as in the case of string webs, one can consider an "elementary web of holomorphic membranes" which is located at fixed values of the transverse 
coordinates. To be more precise, we begin with an intersection 
(\ref{MembrScan}) and remove the smearing in $x_M$ and $x_3$. The resulting intersection would preserve the same supercharges as a membrane which follows an arbitrary holomorphic profile in $x_M,x_1,x_2,x_3$. Assuming that all such membranes are located at the same values of 
$x_4$,\dots,$x_9$, we can impose a rotational symmetry in these directions. Thus we are interested in configurations which preserve 
$U(1)_t\times SO(6)$ bosonic isometries in addition to eight supercharges. As the membrane charge becomes large, the probe analysis would break down, but at some point a geometric description would becomes semiclassical, so we need to construct supersymmetric metrics with $U(1)_t\times SO(6)$ isometry. While such solutions have been written before \cite{marolf}, that paper started with a guess inspired by \cite{FaySmth} and checked that supersymmetry conditions were satisfied. This approach 
would not serve our purpose of demonstrating the duality between probe and gravitational description since we need to start with the 
{\it most general} gravity solution with given symmetries and 
{\it prove} that it has the same degrees of freedom as the 
probes\footnote{Even solution \cite{FaySmth}, which was only partially derived, would not give an independent check of the duality: while 
constructing that geometry, the authors assumed that there exist complex coordinates in which the projections imposed on the Killing spinor are identical to those satisfied by the probe branes in flat space. 
Thus the agreement between the brane probes and gravity was the basic assumption of that derivation.}. In the appendix \ref{SectAp1} we construct 
the most general supersymmetric solution of eleven dimensional SUGRA which has $U(1)_t\times SO(6)$ 
isometry\footnote{To be precise, we also assumed that the field strength is purely electric.}:
\bea\label{MembrGeomSym}
ds^2&=&-e^{2A}dt^2+2e^{2A}g_{a{\bar b}}dz^ad{\bar z}^b+
e^{-A}(d{y}^2+y^2 d\Omega_5^2)\\
G_4&=&idt\wedge d(e^{3A}g_{a{\bar b}}dz^a\wedge d{\bar z}^b),\quad
g_{a{\bar b}}=\d_a{\bar \d}_b K.\nonumber
\eea
This geometry corresponds to an "elementary web" of the membranes, and on the probe side such elements can be freely superposed to produce a most general $1/4$--BPS web. Such web consists of individual pieces located at different values of ${\bf y}$, and the gravity counterpart of the superposition is obvious: one should relax the requirement of $SO(6)$ symmetry. It is easy to check the the resulting 
geometry,
\bea\label{MembrGeom}
ds^2&=&-e^{2A}dt^2+2e^{2A}g_{a{\bar b}}dz^ad{\bar z}^b+
e^{-A}d{\bf y}^2_6,\\
G_4&=&idt\wedge d(e^{3A}g_{a{\bar b}}dz^a\wedge d{\bar z}^b),\quad
g_{a{\bar b}}=\d_a{\bar \d}_b K,\nonumber
\eea
solves SUSY variations and equations of motion\footnote{Such check 
would essentially follow some of the steps presented in the appendix C 
of \cite{myCM}.}, as long as Kahler potential and warp factor satisfy 
two relations:
\bea
\label{MembrMAH}
\frac{1}{4}e^{-3A}&=&
\d_1{\bar \d}_1K\d_2{\bar \d}_2K-\d_1{\bar \d}_2K\d_2{\bar \d}_1K,\\
\label{MembrMAN}
&&\Delta_{\bf y}K+2e^{-3A}=0,
\eea
but clearly (\ref{MembrGeom}) is not the most general 1/4--BPS metric 
with $U(1)_t$ 
isometry\footnote{For example, a system which consists of KK monopones with worldvolumes along $0123456$, $012789M$ and an 
$M2_{012}$ brane has the same number of (super)symmetries as 
(\ref{MembrGeom}), but it does not fit in that ansatz.}. Nevertheless, based on the physical picture of superposition principle and on the fact that 
(\ref{MembrGeomSym}) is the most general solution with 
$U(1)_t\times SO(6)$ 
isometries, we propose that (\ref{MembrGeom}) is the most 
general solution describing the webs of membranes. For future reference, we also write an equation which does not contain $e^A$:
\bea\label{MembrMA}
\d_1{\bar \d}_1K\d_2{\bar \d}_2K-\d_1{\bar \d}_2K\d_2{\bar \d}_1K=
-\frac{1}{8}\Delta_{\bf y}K.
\eea

As in the case of the string webs, a local solution 
(\ref{MembrGeom})--(\ref{MembrMAN})
is valid away from the sources, and equations (\ref{MembrMAH}), 
(\ref{MembrMAN}) should be 
modified at the locations of the membranes. Let us demonstrate that the sources cannot be placed at arbitrary points, but rather they should follow holomorphic profiles. 

Let us consider the solution in the vicinity of the sources. 
Looking at the three--form potential 
\bea
C_3&=&-ie^{3A}dt\wedge  {h}_{a{\bar b}}dz^a\wedge d\zb^b,
\eea
we conclude that ${\bf y}$--directions should be orthogonal to the membranes. 
In the four dimensional space spanned by $(z_a,{\bar z}_a)$, two directions (we call them $v_1$ and $v_2$) should be transverse to
the branes as well, and two remaining directions ($w_1$, $w_2$) should parameterize the worldvolume.  
By definition of the longitudinal direction, in the vicinity of the membrane the warp factor $e^A$ 
should not depend on $w_1$, $w_2$, i.e.
\bea
\d_w \Delta_{\bf y}K|_{brane}\rightarrow 0.
\eea
This allows us to decompose the Kahler potential into four pieces:
\bea\label{KbreakA}
K=K_1(w_1,w_2,v_1,v_2)+K_2(v_1,v_2,{\bf y})+
K_{harm}+K_0,
\eea
so that $\Delta_{\bf y}K_{harm}$ and all second derivatives of $K_0$ 
vanish in the vicinity of the brane. As we argued above, the brane should be located at a particular value of vector ${\bf y}$, so the Kahler potential can only diverge at a point in eight dimensional space 
$(v_1,v_2,{\bf y})$. Thus we conclude that $K_1(w_1,w_2,v_1,v_2)$ must remain finite in the vicinity of the brane, so it can be replaced by the value on a brane: 
$K_1(w_1,w_2)\equiv K_1(w_1,w_2,v^{(0)}_1,v^{(0)}_2)$. For the same reason the $v_i$ and ${\bf y}$ dependence can be ignored in 
$K_{harm}$ as well. This leads to conclusion that the leading contribution to Kahler potential has a separated form
\bea\label{KbreakB}
K=K_1(w_1,w_2)+K_2(v_1,v_2,{\bf y})
\eea
with divergent $K_2$ and finite $K_1$. This separation splits equation 
(\ref{MembrMA}) into three independent pieces\footnote{The second 
and third equations in this system have different degrees of the singular function 
$K_2$ and thus they have to be considered separately. The first equation is completely regular.}:
\bea\label{EqnForK1}
&&\d_1{\bar \d}_1K_1~\d_2{\bar \d}_2K_1-
\d_1{\bar \d}_2K_1~\d_2{\bar \d}_1K_1=0\\
\label{EqnForK1a}
&&\d_1{\bar \d}_1K_2~\d_2{\bar \d}_2K_2-
\d_1{\bar \d}_2K_2~\d_2{\bar \d}_1K_2=0\nonumber\\
\label{EqnForK1b}
&&e^{-3A}=
(\sigma_1)_{ab}\left[\d_1{\bar \d}_1K_a~\d_2{\bar \d}_2K_b-
\d_1{\bar \d}_2K_a~\d_2{\bar \d}_1K_b\right]=
-\frac{1}{8}\Delta_{\bf y}K_2
\eea
By introducing a new function $\alpha$, the equation 
(\ref{EqnForK1}) can be rewritten as a linear system for $K$:
\bea\label{KzeroEqn}
\d_1{\bar \d}_1K_1=\alpha \d_2{\bar \d}_1K_1,\quad
\d_1{\bar \d}_2K_1=\alpha\d_2{\bar \d}_2K_1.
\eea
To proceed we need to solve an equation
\bea\label{CharacEqn}
(\d_1-\alpha\d_2)\Phi=0,
\eea
which is satisfied by both ${\bar\d}_1 K_1$ and ${\bar\d}_2 K_1$.
Method of characteristics reduces this problem to a system of ODEs:
\bea
\frac{d\Phi}{ds}=0:\qquad \frac{dz_1}{ds}=1,\quad 
\frac{dz_2}{ds}=-\alpha(z_1,z_2,\zb_1,\zb_2). 
\eea
which should be solved for $z_1(s)$, $z_2(s)$, while values of 
$\zb_1$ and $\zb_2$ are kept fixed. The solution is parameterized by 
an integration constant ${\tilde z}_2$ for the second equation:
\bea
z_1=s,\quad z_2=\beta(s,\zb_1,{\bar z}_2,{{\tilde z}}_2):\qquad 
{\tilde z_2}=\gamma(z_1,z_2,\zb_1,\zb_2).
\eea
Since the derivative $\frac{d\Phi}{ds}$ must vanish, function $\Phi$ can depend on $z_1$ and $z_2$ only through the combination 
$\gamma(z_1,z_2,\zb_1,\zb_2)$. Then, after redefining arguments of 
$\Phi$, we
find that the general solution of equation (\ref{CharacEqn}) is
\bea
\Phi=\Phi[z_2+\Psi(z_1,\zb_1,\zb_2),\zb_1,\zb_2)].
\eea
This allows us to find the first integrals of the equations 
(\ref{KzeroEqn}):
\bea\label{Nov14PreK1}
{\bar \d}_1K_1=f_1(z_2+\Psi(z_1,\zb_1,\zb_2),\zb_1,\zb_2),\qquad
{\bar \d}_2K_1=f_2(z_2+\Psi(z_1,\zb_1,\zb_2),\zb_1,\zb_2)
\eea
The mixed derivative ${\bar\d}_1{\bar\d}_2 K_1$ can be computed in two different ways and the consistency condition implies that\footnote{
In the exceptional case where
$f_2=c_1f_1+f_3(\zb_1,\zb_2)$ with constant $c_1$, we have a weaker condition ${\bar\d}_2\Psi=c_1{\bar\d}_1\Psi$. Comparing other mixed derivatives, we find that, up to (anti)holomorphic functions,
$K_1=f[z_2+\Psi(z_1,c_1\zb_2+\zb_1),c_1\zb_2+\zb_1]$. Then reality of $K_1$ implies that $K_1=f[z_2+{\bar c}_1z_1,c_1\zb_2+\zb_1]$. This is a particular case of (\ref{Nov14K1}).}
\bea
\d_1{\bar\d}_1\Psi(z_1,\zb_1,\zb_2)=0,\qquad 
\d_1{\bar\d}_2\Psi(z_1,\zb_1,\zb_2)=0.
\eea
Substitution of these relations in (\ref{Nov14PreK1}) leads to a restriction 
on the form of the Kahler potential $K_1$:
\bea
\begin{array}{c}
{\bar \d}_1K_1={\tilde f}_1(z_2+\Psi(z_1),\zb_1,\zb_2),\\
{\bar \d}_2K_1={\tilde f}_2(z_2+\Psi(z_1),\zb_1,\zb_2)
\end{array}\quad
\rightarrow\quad
K_1=f(z_2+\Psi(z_1),\zb_1,\zb_2).
\eea
Since $K_1$ must be real (up to irrelevant (anti)holomorphic
contributions), we arrive at the final expression:
\bea\label{Nov14K1}
K_1=f(w,{\bar w}),\quad w=z_2+\Psi(z_1).
\eea
Similarly, equation (\ref{EqnForK1a}) implies that $K_2$ depends on 
holomorphic coordinate $v$, its conjugate ${\bar v}$, and ${\bf y}$: 
\bea
K_2=g(v,{\bar v},{\bf y}).
\eea
To check the last equation (\ref{EqnForK1b}),
we pass to coordinates $(w,v)$ and introduce a holomorphic Jacobian
\bea
J=\frac{D(w,v)}{D(z_1,z_2)}.
\eea
Then (\ref{EqnForK1b}) simplifies:
\bea
|J|^2\d_w{\bar\d}_w K_1~\d_v{\bar\d}_v K_2=
-\frac{1}{8}\Delta_{\bf y}K_2
\eea
A consistency condition requires that $J=J_1(w)J_2(v)$, then, introducing reparameterizations of $v$ and $w$, one can remove the Jacobians 
completely. Thus we demonstrated that near the brane there always exists the {\it unique} set of holomorphic coordinates
$(v,w)$ which brings Kahler potential to the form
\bea
K=\frac{1}{2}w{\bar w}+K_2(v,{\bar v},{\bf y}),\qquad
\d_v{\bar\d}_v K_2+\frac{1}{4}\Delta_{\bf y}K_2=0.
\eea
Moreover, we showed that the location of the membrane is determined by the relations ${\bf y}={\bf y}^{(0)}$, $v=v^{(0)}$, which implies that in the original coordinates $(z_a,\zb_a)$ the brane follows a holomorphic profile:
\bea
v(z_1,z_2)=0.
\eea
Thus we find a perfect agreement with results of the probe analysis 
(\ref{Jan31HolProf}). 

Once the allowed brane profiles are specified, it is clear how to introduce sources in the system (\ref{MembrGeom}): the equation for the 
field strength (\ref{MembrMAN}) should be replaced by
\bea\label{M2SrsEqn}
\d_a{\bar\d}_b\left[\Delta_{\bf y}K+8\mbox{det}(\d{\bar\d}K)\right]=
-\sum_{i=1}^k Q_{(i)}
\d_a v{\bar\d}_b{\bar v}\delta({\bf y}-{\bf y}_{(i)})
\delta_{(2)}(v-v_{(i)}).
\eea
Here $\{ {\bf y}_{(i)},v_{(i)}(z)\}$ give the positions of the membranes 
and $Q_{(i)}$ specify their charges. To arrive at (\ref{M2SrsEqn}), one starts with an assumption that the behavior of the left hand side near the brane is not sensitive to the effects of the curvature on the worldvolume, i.e. it can be extracted from the "closeup" limit in which the membrane is flat. In this limit, the holomorphic function $v$ can be chosen to be linear in $z_1$, $z_2$, the determinant in the left hand side of (\ref{M2SrsEqn}) disappears, and the entire equation (\ref{M2SrsEqn}) reduces to the standard Poisson equation for the "harmonic" function $e^{-3A}$. 
Making holomorphic reparameterization of $v$ in this Poisson equation, one arrives at (\ref{M2SrsEqn}).

Equation similar to 
(\ref{M2SrsEqn}) was discussed in \cite{marolf}, where it was argued that geometries corresponding to curved membranes did not exist. 
This conclusion was reached by demonstrating that equation 
(\ref{M2SrsEqn}) did not admit perturbative expansion:
even the second term in the series could not be defined anywhere: formally it contained infinite multiplicative constant.
Physically, such divergence seems counter-intuitive: since  
the effects of the branes should disappear at infinity, one should be able to define a good expansion at least far away from the branes.  
In the next subsection we will demonstrate that the perturbation series for (\ref{M2SrsEqn}) indeed does exist and it is convergent, the curved membranes do produce good asymptotically-flat solutions, and erroneous statement of \cite{marolf} 
was based on an unfortunate choice of the expansion parameter. 

%=====================================

\subsection{Solution in perturbation theory}

\label{SectMembrPert}

%=====================================

In this subsection we will argue that any allowed distribution of membranes leads to the unique solution of equation (\ref{MembrMA}). In 
particular, we will demonstrate that far away from the branes one has 
a well-defined perturbation theory, and we will present some evidence that analytic continuation of this perturbation leads to a good solution. 
This conclusion is in a sharp contradiction with the results of 
\cite{marolf}, where it was argued that neither perturbation series nor geometry exists for the curved branes. 

To expand solutions of (\ref{M2SrsEqn}) in the powers of the charges, 
one should begin with specifying the solution at zeroth order in 
$Q_{(i)}$. Of course, the homogeneous version of (\ref{M2SrsEqn}) has many interesting solutions corresponding to various asymptotic behaviors of the metric, but we would be mostly interested in the case 
where the space is flat. Then in some coordinate system 
$({z}_a,{{\zb}}_a)$ the Kahler potential can be written as
\bea
K_0=\frac{1}{2}({z}_1{{\zb}}_1+{z}_2{{\zb}}_2).
\eea
This starting point was also used in \cite{marolf}, where it was 
shown that perturbative expansion in powers of $Q_{(i)}$ breaks 
down at the second order. Before defining an improved perturbative series, let us recall the arguments of \cite{marolf}. For simplicity we consider $k=1$ and set ${\bf y}_{(1)}=0$, $v_{(1)}=0$ in 
(\ref{M2SrsEqn}). 

In the first 
order of perturbation theory, one gets a Poisson equation:
\bea
\d_a{\bar\d}_b\left[\Delta_{\bf y}K_1+
4(\d_1\db_1+\d_2\db_2)K_1\right]=
-Q\d_a v{\bar\d}_b{\bar v}\delta({\bf y})\delta(v),
\eea
which can be easily integrated:
\bea
\d_a\db_b K_1&=&\frac{Q}{8\Omega_9}\int
\frac{d^2 z' d^2\zb' \d'_a v(z')\db'_b{\bar v}(\zb')}{
[{\bf y}^2+(z_c-z'_c)(\zb_c-\zb'_c)]^{9/2}}\delta^{(2)}(v(z')-v_0)\nonumber\\
&=&\left.\frac{Q}{8\Omega_9}\int dz'_{3-a} d\zb'_{3-b}
\frac{1}{
[{\bf y}^2+(z_c-z'_c)(\zb_c-\zb'_c)]^{9/2}}\right|_{
\tiny
\begin{array}{c}
z'_a=z^{(0)}_a(z'_{3-a})\\
\zb'_b=\zb^{(0)}_b(\zb'_{3-b})
\end{array}}
\eea
Here $z^{(0)}_a(z'_{3-a})$ is defined as a solution of the equation
$v[z^{(0)}_a(z'_{3-a}),z'_{3-a}]=0$.

The second order of (\ref{M2SrsEqn}) leads to a Laplace equation
\bea\label{Mrlf2Ord}
\Delta_{\bf y}K_2+
4(\d_1\db_1+\d_2\db_2)K_2=-8\mbox{det}(\d\db K_1).
\eea
In particular, it is useful to extract the behavior of the right--hand side of this equation near the brane  (where its profile was approximated by 
$z'_a=h_a z$):
\bea
\d_a\db_b K_1&\sim& 
\int \frac{h_{3-a}{\bar h}_{3-b}dzd\zb}{
[{\bf y}^2+|z_c-h_c z|^2]^{9/2}}=
\int_0^\infty \frac{2\pi h_{3-a}{\bar h}_{3-b}dr^2}{
[{\bf y}^2+h^{-1}|h_2z_1-h_1 z_2|^2+hr^2]^{9/2}}\nonumber\\
&\sim&\frac{h_{3-a}{\bar h}_{3-b}}{
h[{\bf y}^2+h^{-1}|h_2z_1-h_1 z_2|^2]^{7/2}},\qquad
h\equiv h_1{\bar h}_1+h_2{\bar h}_2\nonumber\\
\mbox{det}(\d\db K_1)&\sim&\frac{\mbox{det}(h_a{\bar h}_b)}{
h^2[{\bf y}^2+h^{-1}|h_2z_1-h_1 z_2|^2]^{7}}
\eea
Then, trying to solve (\ref{Mrlf2Ord})
using Green's function, one finds a nonsensical answer even away from the sources:
\bea\label{Jan21Divrg}
K_2(z,{\bf y})&=&8\int d^4 w d^6 x G(z,{\bf y}|z',{\bf x})
\left[\mbox{det}(\d\db K_1)\right]_{z',{\bf x}}\nonumber\\
&\ge&
G(z,{\bf y}|z'_a=0,{\bf x}=0)\int_{{\bf x}^2+r^2\le \eps}
d^4 z' d^6 x
\left[\mbox{det}(\d\db K_1)\right]_{z',{\bf x}}=\infty.
\eea
Here $r$ is defined as a distance from the membrane profile 
$v(z_1,z_2)=0$ in four--dimensional space spanned by $(z_a,\zb_a)$.

This argument led authors of \cite{marolf} to the conclusion that perturbation theory in charges is not well-defined and that the solutions corresponding to curved membranes do not exist.
However, as we will now explain, the naive perturbation theory in $Q$ can be modified and the resulting series gives a well--defined solution. 

First we observe that the expression (\ref{Jan21Divrg}) gives a divergent result even far away from the sources, but infinity 
comes from the contribution at the location of 
the branes. This divergence appeared since the authors of \cite{marolf} assumed that equation (\ref{Mrlf2Ord}) was not modified near the sources (this was a consequence of making an expansion in powers 
of $Q$). A completely opposite situation is encountered for the multipole expansion: there solution is regular at infinity, but the series cannot be trusted near the sources. Physically it is clear that one is interested in the multipole rather than $Q$--expansion, but these two series can be easily confused. For example, looking at a function
\bea\label{ExGeomProgr}
f=\frac{1}{1+\frac{Q}{r}}=1-\frac{Q}{r}+\frac{Q^2}{r^2}+\dots,
\eea
one may conclude that the large--$r$ and small--$Q$ expansions look the same\footnote{This is expected since $Q/r$ is the only dimensionless parameter in the problem.}, but it is important to keep in mind that the series should not be taken seriously at small values of $r$, and one is allowed to modify any {\it perturbative} equation for $f$ at $r=0$ to ensure the correct large--distance behavior.  

To illustrate such modification, we go back to the equation (\ref{Mrlf2Ord}) and add an extra term to its right--hand side:
\bea\label{Mrlf2OrdMd}
\Delta_{\bf y}K_2+
4(\d_1\db_1+\d_2\db_2)K_2=-8\mbox{det}(\d\db K_1)+
Q_0^{(2)}\delta({\bf y})\delta^{(2)}(v). 
\eea 
The value of $Q_0^{(2)}$ can be fixed by eliminating the leading divergent contribution to $K_2$ at infinity (i.e. by requiring 
the correction to the membrane charge to vanish). 
To show that such $Q_0^{(2)}$ can always 
be chosen, we surround 
the membrane by a shell with radius $\eps$ and 
solve equation (\ref{Mrlf2OrdMd}) by restricting integration in 
(\ref{Jan21Divrg}) to the exterior of the shell and by adding a term proportional to $Q_0^{(2)}$:
\bea\label{Jan21K2eps}
K^{\eps}_2(z,{\bf y})&=&8\int_{r>\eps} d^4 z' d^6 x G(z,{\bf y}|z',{\bf x})
\left[\mbox{det}(\d\db K_1)\right]_{z',{\bf x}}\nonumber\\
&&-Q^{(2)}_0\int d^4 z' G(z,{\bf y}|z',{\bf 0})\delta_{(2)}(v[z'])
\eea
The first term in this equation diverges as $\eps$ goes to zero, and the 
leading pole has the same functional dependence on $(z,{\bf y})$ as the second term. Thus by taking $Q_0^{(2)}=\frac{{\tilde Q}_0^{(2)}}{\eps^p}$, 
one can eliminate the leading divergence in $K^{\eps}_2(z,{\bf y})$. 
Notice that leading order in $1/\eps$ is also the leading contribution in $1/r$ expansion (it gives the charge of the membrane measured 
from infinity), so it is convenient to shift $Q_0^{(2)}$ by an 
$\eps$--independent term and to require
\bea
\frac{K^{\eps}_2(z,{\bf y})}{
\int d^4 z' G(z,{\bf y}|z',{\bf 0})\delta^{(2)}(v[z'])}=O(r^{-1}).
\eea

Going back to equation (\ref{Jan21K2eps}), we observe that the new 
leading term in $\eps$--expansion behaves like a potential of a 
"dipole membrane" far away from the sources, and it can be canceled 
by adding an appropriate local counterterm to the right--hand side
of (\ref{Mrlf2OrdMd}).
Acting in a similar fashion, one can modify 
(\ref{Mrlf2OrdMd}) by adding a series of extra "multipole" sources which are localized on the membrane:
\bea\label{Jan21TotK2}
K^{\eps}_2(z,{\bf y})&=&8\int_{r>\eps} d^4 z' d^6 x G(z,{\bf y}|z',{\bf x})
\left[\mbox{det}(\d\db K_1)\right]_{z',{\bf x}}\nonumber\\
&&-\sum_k Q^{(2)}_k\int d^4 z' \Delta_{\bf y}^kG(z,{\bf y}|z',{\bf 0})\delta_{(2)}(v[z']),
\eea
and which make $K^{\eps}_2$ finite. Of course, once $\eps$ is sent to zero, the coefficients in front of these sources 
would diverge, but the resulting function $K^{\eps=0}_2$ is well-defined and it satisfies equation (\ref{Mrlf2Ord}) away form the branes. Then we 
will {\it define} $K_2\equiv K^{\eps=0}_2$ as a second term in the perturbation series in $Q$:
\bea\label{Jan21Exp}
K=K_0+QK_1+Q^2 K_2+\dots
\eea
The same procedure can be repeated for the higher orders in perturbation series. Moreover, by choosing the finite contributions to $Q^{(m)}_p$, we can also ensure that 
$K_n\ll K_{n-1}$ far away from the branes and that the first $p$ terms in the series (\ref{Jan21Exp}) correctly reproduce the first $p$ multipole moments of the brane configuration (the moments of 
$e^{-3A}$ can be extracted from the probe analysis, then the moments 
of $K$ are found by integrating (\ref{MembrMAN})). 
Notice that the series 
(\ref{Jan21Exp}) should be viewed as $1/r$ rather than 
$Q$--expansion. To make such series possible, we had to modify the sources at the location of the brane, but since the series 
(\ref{Jan21Exp}) breaks down long before these points (e.g. the expansion (\ref{ExGeomProgr}) breaks down at $r=Q$, while the sources are
located at $r=0$), the vicinity of the branes requires special consideration. 

To summarize, we showed that, 
while the series in powers of $Q$ does not make sense \cite{marolf}, 
the large $r$--expansion is well--defined, but it requires introduction of 
new sources at the location of the brane. The $1/r$--expansion is expected to have a nonzero radius of convergence, and the  
Kahler potential  in the entire space can be constructed by analytic continuation (equation (\ref{ExGeomProgr}) gives the simplest example). 
Moreover, by construction, $K$ satisfies a differential equation 
\bea\label{Jan21AvSrc}
\Delta_{\bf y}K+8\mbox{det}(\d\db K)=-\frac{Q}{2\pi}\delta({\bf y})
\log (v{\bar v})
\eea
away from the sources and all multipole moments of $K$ match those of the brane configuration. To prove the relation (\ref{M2SrsEqn}), we need to 
demonstrate that (\ref{Jan21AvSrc}) holds everywhere. Unfortunately, such proof cannot be performed in perturbation theory (since an analytic continuation was involved), so a better understanding of the nonlinear equation (\ref{Jan21AvSrc}) is required. However, it seems plausible that, for any given (infinite) set of multipole moments, there 
exists only one solution of (\ref{Jan21AvSrc}) away from the sources.
Then it seems natural to {\it define} such solution as a geometry produced by the curved membrane, and, if such solution has extra terms on the right--hand side of (\ref{M2SrsEqn})\footnote{Notice that our construction guarantees that such terms may arise only on the surface of the membrane.}, one should take it as a sign of breakdown in 
(\ref{M2SrsEqn}): after all, the source terms in that equation came from the simplest generalization of branes in flat space\footnote{The same sources were derived in \cite{marolf} from the probe analysis, but there it was assumed that geometry did not backreact to deform the profiles. This assumption is essentially equivalent to taking a locally--flat approximation for the branes which led to (\ref{M2SrsEqn}).}.  However, we believe that 
the source terms in (\ref{M2SrsEqn}) are correct, and that analytic continuation of (\ref{Jan21Exp}) satisfies equation (\ref{Jan21AvSrc}) everywhere. The argument was outlined in section 
\ref{SectMembrStr}: near the brane one expects the curvature effects to become irrelevant, then (\ref{M2SrsEqn}) comes from the source term for the flat 
brane. It would be very nice to find a rigorous proof of this proposal. 

To summarize, in this subsection we demonstrated that, while the solutions of equation (\ref{M2SrsEqn}) cannot be written as naive series in powers 
of $Q$ \cite{marolf}, they admit a well--defined multipole expansion which converges far away from the branes and which can be extended to the entire space away from the brane profiles. By construction, our solution shares an infinite set of quantum numbers with probe configuration, so we declared that our geometry should describe an appropriate stack of membranes. We also conjectured 
that an analytic continuation of the series satisfies a nonlinear equation with sources (\ref{M2SrsEqn}). While some heuristic evidence for this 
proposal was given, it would be nice to find a more rigorous proof of the conjecture. It is clear that, while we only discussed a single stack 
of membranes, the results also hold for an arbitrary number of stacks. 

%=====================================

\section{Webs of five-- and three--branes}

\label{Sect35Brn}

\renewcommand{\theequation}{5.\arabic{equation}}
\setcounter{equation}{0}

%=====================================

\subsection{Summary of the solutions}

\label{Sect35BrnSum}

%=====================================

As we discussed in section \ref{SectProbe}, there are two ways of 
constructing interesting webs of five--branes in IIB string theory. Both configurations 
can be obtained by the chain of dualities from the system which we already discussed. For example, starting with a web of membranes, one can perform the following transformations:
\bea\label{Dual5Web}
\left(\begin{array}{c}
M2_{45}\\ M2_{67}
\end{array}\right)\rightarrow&
\left(\begin{array}{c}
D2_{45}\\ D2_{67}
\end{array}\right)&\rightarrow
\left(\begin{array}{c}
D5_{12345}\\ D5_{12367}
\end{array}\right)_1\\
&\downarrow&
\phantom{\frac{e^x}{e^6}}
\nonumber\\
&\left(\begin{array}{c}
D4_{1245}\\ D4_{1267}
\end{array}\right)&\rightarrow
\left(\begin{array}{c}
M5_{(10)1245}\\ M5_{(10)1267}
\end{array}\right)_2
\rightarrow
\left(\begin{array}{c}
NS5_{12345}\\ D4_{1237}
\end{array}\right)\rightarrow
\left(\begin{array}{c}
NS5_{12345}\\ D5_{12347}
\end{array}\right)_3
\nonumber
\eea
Tracing the metrics through the chain of dualities, we find the geometries produced by the webs of five--branes:

{\bf 1. Webs of D5 branes} Performing the first two steps in 
(\ref{Dual5Web}), we find the geometry describing a web of D5 branes:
\bea
ds_{IIB}^2&=&e^{3A/2}\left[-dt^2+d{\bf x}_3^2+
2g_{a{\bar b}}dz^ad{\bar z}^b\right]+
e^{-3A/2}d{\bf y}^2_2\\
F_7&=&idt\wedge d(e^{3A}g_{a{\bar b}}dz^a\wedge d{\bar z}^b)\wedge d^3{\bf x},\quad e^{2\Phi}=e^{3A},\quad
g_{a{\bar b}}=\d_a{\bar \d}_b K.
\eea
For this solution Kahler potential should be a function of $z_a$, 
$\zb_a$ and ${\bf y}_2$ and away for the sources it should satisfy 
differential equation:
\bea\label{D5eqnNS}
\d_a{\bar\d}_b\left[\Delta_{\bf y}K+8\mbox{det}(\d{\bar\d}K)\right]=0.
\eea
Repeating the analysis presented in the previous section, one can demonstrate that D5 branes can be added to this solution, but for consistency of IIB supergravity they should follow holomorphic profiles
$v(z_1,z_2)=0$. This conclusion is in a perfect agrees with results of the probe analysis.
In the presence of sources equation (\ref{D5eqnNS}) should be replaced by (\ref{M2SrsEqn}).

{\bf 2. Webs of M5 branes} Going back to the membrane web and performing dualities outlined in (\ref{Dual5Web}), we arrive at the
geometry produced by a web of M5 branes:
\bea\label{WebM5}
ds^2&=&e^{A}\left[-dt^2+d{\bf x}_3^2+
2g_{a{\bar b}}dz^ad{\bar z}^b\right]+
e^{-2A}d{\bf y}_3^2\\
F_7&=&idt\wedge d(e^{3A}g_{a{\bar b}}dz^a\wedge d{\bar z}^b)\wedge d^3{\bf x}.\nonumber
\eea
The regularity conditions work in the same way as before. This ansatz has been previously discussed in \cite{FaySmth}. 

{\bf 3. (p,q)--fivebranes} The web of $(p,q)$ fivebranes is a 
magnetic counterpart of the string web. To find a geometry produced by it, we need to assume two translational isometries in four dimensional subspace spanned by $(z_a,\zb_a)$ in (\ref{WebM5}), then applying the arguments presented in the appendix \ref{SectApIsom}, one can show 
that $z_a=r_a+iw_a$, and that nothing depends on $w_a$. We can then reduce this system on $w_1$ and perform a T duality along $w_2$:\\
\bea
ds_{IIB,E}^2&=&e^{3A/4}\left[-dt^2+d{\bf x}_3^2+dz^2+
h_{ab}dr^adr^b\right]+
e^{-9A/4}d{\bf y}_3^2,\\
e^{2\Phi}&=&h_{11}^2 e^{3A},\quad C_0=-\frac{h_{12}}{h_{11}},\quad
h_{ab}=\frac{1}{2}\d_a\d_b K,\quad e^{-3A}=\mbox{det}~h.
\eea
To evaluate the two--form potentials, one needs to perform an electric--magnetic duality in (\ref{WebM5}), and we will not do this here.

As in the case of the string webs, one can see that the branes must 
go along straight lines in $(r_1,r_2)$ directions and that the orientation of the elements of the web is correlated with the amount of D5/NS5 charge (see discussion which led to equation (\ref{GeomrcWeb})). 
 
{\bf 4. Webs of D3 branes} The last interesting system in 
(\ref{IntersBranes}) is a web of D3 branes. To construct the appropriate supergravity solution we need to apply one T duality to the D2--D2 system in (\ref{Dual5Web}):
\bea\label{WebD3Metr}
ds_{IIB}^2&=&e^{3A/2}\left[d{\bf x}_{1,1}^2+
2g_{a{\bar b}}dz^ad{\bar z}^b\right]+
e^{-3A/2}d{\bf y}^2_4\\
F_5&=&-id^2 x\wedge d(e^{3A}g_{a{\bar b}}dz^a\wedge d{\bar z}^b),
\quad e^{2\Phi}=1.\nonumber
\eea
The regularity conditions again lead to the holomorphic profiles of the branes ($v(z_a,\zb_a)=0$), and the equations of motion with sources become
\bea
&&e^{-3A}=4\mbox{det}(\d{\bar\d}K),\nonumber\\
&&\d_a{\bar\d}_b\left[\Delta_{\bf y}K+8\mbox{det}(\d{\bar\d}K)\right]=
-\sum_{i=1}^k Q_{(i)}
\d_a v{\bar\d}_b{\bar v}\delta({\bf y}-{\bf y}_{(i)})
\delta(v-v_{(i)}).
\eea
It is interesting to compare this solution with an alternative description of metrics produced by D3 branes which was presented in 
\cite{myCM}. 

%=====================================

\subsection{Alternative description}

\label{Sect35BrnAlt}

%=====================================

Let us briefly review the construction of \cite{myCM}. In that paper it was shown that all metrics produced by 1/4--BPS webs of D3 branes 
can be written in terms of one function 
$F({\bf x},{\bf y},w)$:\footnote{Equation (\ref{AltD3Metr}) was derived 
in \cite{myCM} assuming translational invariance in $u$ direction.
To compare (\ref{AltD3Metr}) with 
\cite{myCM}, one should make two replacements: 
$e^{-H-\phi}\rightarrow H$, $e^{-2\phi}\rightarrow h$ in equation (4.47) of that paper. The expression for $F_5$ can be found in \cite{myCM}.}
\bea\label{AltD3Metr}
ds^2&=&H^{-1}ds_{1,1}^2+Hd{\bf y}_4^2\\
&&+H^{-1}\left\{H^2h^{-1}\left[(\d_w F dw+\d_{\bf y} Fd{\bf y})^2+
(du+\eps_{ij}\d_i Fdx^j)^2\right]+hd{\bf x}_2^2\right\},
\nonumber
\eea
\bea
\d_w h=-\Delta_{\bf x}F|_{y,w},\quad H^{-2}=h^{-1}\d_w F|_{\bf x,y}.
\eea
Function $F$ satisfies a system of differential equations:
\bea
\d_F H^2+(\Delta_{\bf y} w)_{x,F}=0,\quad
\Delta_{\bf y} e^{-2\phi}+\Delta_{\bf x} H^2+
\Delta_{\bf y}(\d_w F \d_{x_i} w\d_{x_i} w)|_{x,F}=0,\nonumber
\eea
which allow to determine $F$ uniquely once the sources are specified (see \cite{myCM}). In particular, supergravity equations are consistent if and only if the branes follow harmonic profiles:
\bea
\{F=p({\bf x}),{\bf y}={\bf y}_{(0)}\},\quad 
\{{\bf x}={\bf x}_{(0)},{\bf y}={\bf y}_{(0)}\}.
\eea
To compare with analysis of the previous subsection, we need to introduce complex coordinates in the metric appearing in the curly brackets in (\ref{AltD3Metr}): 
\bea\label{Jan18BB}
ds_4^2&=&H^2h^{-1}\left[(dF-\d_{\bf x} Fd{\bf x})^2+
(du+\eps_{ij}\d_i Fdx^j)^2\right]+hd{\bf x}_2^2\nonumber\\
&=&H^2h^{-1}\left[(dF+\frac{h}{H^2}\d_{\bf x} wd{\bf x})^2+
(du-\frac{h}{H^2}\eps_{ij}\d_i wdx^j)^2\right]+
hd{\bf x}_2^2\nonumber\\
&=&H^2h^{-1}dWd{\bar W}+2({\bar\d}_z w~dWd{\bar z}+cc)+
h(4H^{-2}|\d_z w|^2+1)dzd{\bar z}.
\eea
The holomorphic coordinates turned out to be $z=x_1+ix_2$, 
$W=F+iu$. In the process of deriving (\ref{Jan18BB}) we used the following relations:
\bea
\d_x F=-\d_w F\d_x w=-H^{-2}h\d_x w,\quad
\eps_{ij}\d_i wdx^j=i(\d_z w dz+\d_{\zb} w d\zb)
\eea
The Kahler potential corresponding to the metric (\ref{Jan18BB}) is
\bea
K=2\int dF w:&& \d_W\db_{W} K=\frac{1}{4}\d_F^2 K=
\frac{1}{2}H^2 h^{-1},\quad 
\d_z\db_W K=\d_z w,\nonumber\\
&&\d_z{\bar\d}_z K=2\int dF \Delta_{\bf x}w.\nonumber
\eea
Thus we conclude that the metric (\ref{AltD3Metr}) can be rewritten as
\bea
ds^2=H^{-1}ds_{1,1}^2+Hd{\bf y}_4^2+
2H^{-1}g_{a{\bar b}}dz^ad\zb^b.
\eea
Moreover, one finds that
\bea
\mbox{det}g_{a{\bar b}}=\frac{1}{4}H^2,
\eea
in a perfect agreement with (\ref{MembrMAH}). 
Thus we have shown that some of the solutions derived in \cite{myCM} 
fit nicely into the metric ansatz (\ref{WebD3Metr}).

%=====================================

\section{1/4--BPS bubbling solutions of IIB SUGRA}

\label{SectIIBbl}

\renewcommand{\theequation}{6.\arabic{equation}}
\setcounter{equation}{0}

%=====================================

In the last three sections we discussed various brane intersections 
preserving eight supercharges. However, while deriving the gravity 
solutions, we made important assumptions that the geometries were static, and that Killing spinor did not depend on the time coordinate. The first assumption was motivated by the probe 
analysis of section \ref{SectProbe}, where it was shown that brane webs 
in {\it flat} space formed static configurations. 
The second assumption originated from the fact
that supersymmetric geometries with flat asymptotics must have a
translational Killing vector. For the solutions with more general asymptotics, supersymmetry only requires an existence of a time--like (or light--like) Killing vector which does not have to be hypersurface--orthogonal (i.e. "rotating geometries" are allowed). Moreover, 
the Killing spinors might be charged under time 
translations\footnote{Such symmetry is usually called "rotational" to distinguish it from the "translational" symmetry where the spinors 
are neutral.}.  The simplest example of the "rotational time--like symmetry" is a translation along global time in AdS space, but such 
Killing vector is still hypersurface--orthogonal. On the other hand, by shifting one of the angular coordinates, one can introduce a new "time" on AdS$_5$ which mixes with other coordinates:
\bea
ds_5^2&=&\cosh^2\rho d\tau^2+d\rho^2+\sinh^2\rho\left[d\theta^2+
\cos^2\theta (d\phi+d\tau)^2+\sin^2\theta d\psi^2\right],\nonumber
\eea
but half of the Killing spinors is neutral under the shift symmetry in 
$\tau$ (while the other half is charged). These simple examples show that we can independently relax the requirements of staticity and neutrality of the Killing spinors. 

While study of general $1/4$ geometries on spaces with arbitrary asymptotics goes beyond the scope of this paper,  
in this section we will discuss the geometries which asymptote to 
$AdS_5\times S^5$. Such configurations are important for constructing the closed-string description of various states in 
${\cal N}=4$ super--Yang--Mills. Let us begin with recalling the situation for the $1/2$--BPS states. On the field theory side, they are described by excitations of matrix harmonic oscillator, 
in particular semiclassical states are represented by the droplets of incompressible Fermi fluid \cite{beren}. On the bulk side, one has three regimes which are well-understood. The operators with small conformal dimension are represented by perturbative gravitons on 
$AdS_5\times S^5$, the operators with $\Delta\sim N$ have semiclassical description in terms of D branes \cite{giant}, and semiclassical states with 
$\Delta\sim N^2$ correspond to regular geometries \cite{LLM}. 
The map presented in \cite{LLM} is very explicit: 
the boundary conditions for the metrics are identified with 
distributions of the fermionic droplets in the phase space. 
It would be very nice to have a similar picture for the states preserving a smaller amount of SUSY.

In the $1/4$ BPS case, the field theory side is understood, and 
states are constructed from two commuting matrices \cite{beren4}. On 
the bulk side, we again expect to have gravitons for small values of $\Delta$ and branes ("giant gravitons") 
for $\Delta\sim N$. In this case the giant gravitons are parameterized by holomorphic surfaces \cite{mikhail} and we review this construction in section \ref{SectIIBPrb}. As dimension of the operator becomes of order 
$N^2$, the geometric description takes over and the local structure of the metric was described in \cite{donos}. However, to compare with field theory, one also needs to specify the allowed boundary conditions 
and, unfortunately, this ingredient has been missing. The main goal of this section is to clarify the admissible boundary conditions for the geometries and to show that they are consistent with expectation coming from both field theory and brane probe analysis. 

%=====================================

\subsection{Local description}

\label{SectIIBLocal}

%=====================================

Let us recall construction of bubbling geometries preserving $8$ supercharges. On the field theory side, $1/4$--BPS states can be represented as "words" constructed out of two commuting 
matrices\footnote{We are using the standard notation for adjoint scalars in ${\cal N}=4$ theory: six matrices $\Phi_1$,\dots $\Phi_6$ are often combined into three holomorphic objects 
$X=\Phi_1+i\Phi_2$, $Y=\Phi_3+i\Phi_4$, $Z=\Phi_5+i\Phi_6$.} 
$X$, $Y$. Starting with elementary building blocks
\bea\label{QrtBldBlk}
\mbox{tr}(X^{n_1}Y^{n_2}),
\eea
one can write the most general state by combining various products of traces. Since matrix $Z$ does not appear in the wavefunction, all 
$1/4$--BPS states are invariant under $U(1)$ rotation 
$Z\rightarrow e^{i\psi}Z$. In the context of AdS/CFT one is interested in the field theory defined on $R_\tau\times S^3$, and it turns out that for a state (\ref{QrtBldBlk}) to preserve SUSY, it can only contain zero modes 
of $X$ and $Y$ on $S^3$. This implies that $1/4$--BPS states have an $SO(4)\times U(1)$ symmetry which should also be preserved by their gravity duals. Moreover, the states constructed from building blocks 
(\ref{QrtBldBlk}) are also symmetric under exchange 
of $Z$ and ${\bar Z}$. This implies that the dual metric should be invariant under $Z_2$ symmetry $\psi\rightarrow -\psi$, and this angle should not mix with the remaining coordinates.

The local supersymmetric geometries with $SO(4)\times U(1)$ isometries have been constructed in 
\cite{myUnp,donos}\footnote{Solution (\ref{DonosSoln}) is written in the notation introduced in \cite{vaman}. More general family of $1/4$--BPS geometries with $SO(4)\times U(1)$ isometries was constructed in 
\cite{donos2}, but due to the mixing between $U(1)$ direction and other coordinates, these solutions break $Z_2$ symmetry, so they are not relevant for describing the states (\ref{QrtBldBlk}), and we will not discuss these metrics further.}:
\bea\label{DonosSoln}
ds_{10}^2&=&-h^{-2}(dt+\omega)^2+h^2\left[\frac{2}{Z+\frac{1}{2}}
\d_a{\bar\d}_b Kdz^ad\zb^b+dy^2\right]+y(e^Gd\Omega_3^2+
e^{-G}d\psi^2)\nonumber\\
F_5&=&\left\{-d[y^2 e^{2G}(dt+\omega)]-y^2 d\omega+2i\d\db K\right\}
\wedge d\Omega_3+\mbox{dual}\nonumber\\
&&h^{-2}=2y\cosh G,\quad Z\equiv 
\frac{1}{2}\tanh G=-\frac{1}{2}y\d_y(y^{-1}\d_y K)\nonumber\\
&&\mbox{det} h_{a{\bar b}}=y(Z+\frac{1}{2})\exp\left[y^{-1}\d_y K\right]
W(z){\bar W}({\bar z})\\
d\omega&=&\frac{i}{y}\left(\d_a{\bar\d}_b\d_y K~dz^ad\zb^b-
\d_a Z~dz^a dy+{\bar\d}_a Z d\zb_a dy\right)=
\frac{i}{2}d\left[\frac{1}{y}{\bar\d}\d_y K-\frac{1}{y}{\d}\d_y K\right] 
\nonumber
\eea
Using reparameterizations, we can impose the gauge $W=\frac{1}{2}$. 
Since $y$ coordinate is equal to the product of two warp--factors (one for $S^3$ and one for $S^1$), we should impose certain boundary conditions 
to avoid singularities at $y=0$ hypersurface. This issue will be addressed in section \ref{SectIIBDrpl}, here we just observe that when 
$S^3$ contracts to zero size (while $g_{\psi\psi}$ remains finite) function $Z$ is necessarily equal to $-\frac{1}{2}$, while 
$Z=\frac{1}{2}$ when $\psi$--circle collapses. This situation is completely analogous to the picture for the $1/2$--BPS states 
\cite{LLM}, but, as we will see later, regularity for the $1/4$--BPS case leads to some additional requirements. 

%=====================================

\subsection{Examples}

\label{SectIIBEx}

%=====================================

Before discussing the boundary conditions for an arbitrary $1/4$--BPS solution, it might be useful to consider some examples. In particular, the simplest example of geometry which fits into the ansatz (\ref{DonosSoln}) 
is $AdS_5\times S^5$, although coordinates used in 
(\ref{DonosSoln}) are not very standard. Another important example is given by a family of 
$1/2$--BPS solutions which have an enhanced $SO(4)\times SO(4)$ symmetry \cite{LLM}. In this subsection we will embed these two solutions into (\ref{DonosSoln}).

\subsubsection{AdS$_5\times$S$^5$ as a 1/4--BPS state in a theory on $R\times S^3$.}

Starting from the standard metric on AdS$_5\times$S$^5$:
\bea\label{NormAdS5S}
ds^2&=&-\cosh^2\rho dt^2+d\rho^2+\sinh^2\rho d\Omega_3^2\nonumber\\
&&+\sin^2\theta d\psi^2+d\theta^2+\cos^2\theta\left[
\cos^2\alpha d\phi_1^2+\sin^2\alpha d\phi_2^2+d\alpha^2\right],
\eea 
it is easy to find the change of coordinates which leads to 
(\ref{DonosSoln}). 
The appropriate map was derived in \cite{vaman}, and here we just 
write down the answer which will be used later on. Following 
\cite{vaman}, we introduce complex coordinates
\bea\label{Jan11Ads5}
z_1=r\cos\alpha e^{i(\phi_1+t)},\quad z_2=r\sin\alpha e^{i(\phi_2+t)},
\quad r=\cosh\rho\cos\theta.
\eea
By construction, the subspace spanned by $(z_a,\zb_a)$ is orthogonal to $y=\sinh\rho\sin\theta$. Then direct computations lead to expressions for the Kahler potential:
\bea\label{KahlerAdS}
K&=&\frac{1}{2}\left[\Psi-\log\Psi-y^2\log(\Psi-r^2)+y^2\log y-
y^2\right],\\
\Psi&\equiv&
\frac{1}{2}(r^2+y^2+1)+\sqrt{\frac{1}{4}(r^2+y^2-1)^2+y^2},\nonumber
\eea
for the one--form $\omega$, and for the function $Z$:
\bea\label{OldEqn99}
\omega=\frac{h^2}{\cosh^2\rho}
\mbox{Im}(\zb_1dz_1+\zb_2dz_2),\quad 
Z=\frac{h^2}{2}(r^2+y^2-1),\quad
h^{-2}=\sinh^2\rho+\sin^2\theta.
\eea
It is interesting to look at the boundary conditions for function $Z$ at $y=0$. Since $y$ is a product of two functions, the hypersurface $y=0$ is divided into two regions:
\bea
\rho=0:&&\quad r=\cos\theta\le 1,\quad Z=-\frac{1}{2},\nonumber\\
\theta=0:&&\quad r=\cosh\rho\ge 1,\quad Z=\frac{1}{2},\nonumber
\eea
and the boundary between the regions is a three--sphere in 
${\bf C}^2$:
\bea\label{StandAdSBdry}
z_1\zb_1+z_2\zb_2=1.
\eea

\subsubsection{AdS$_5\times$S$^5$ as a 1/4--BPS geometry with 
$SO(4)$ R--symmetry.}

While we will mostly be interested in the representation 
(\ref{KahlerAdS}), there is an alternative way of embedding 
$AdS_5\times S^5$ into the general $1/4$--BPS ansatz (\ref{DonosSoln}).
Unlike (\ref{KahlerAdS}) which preserves the sphere from $AdS_5$ 
(this fact makes (\ref{KahlerAdS}) useful for studying normalizable states 
in SYM on 
$R\times S^3$), the other representation breaks space--time rotational invariance, while keeping a large part of the R--symmetry group. 

To find such alternative embedding, 
we begin with rewriting the metric on $AdS_5\times S^5$:
\bea\label{WickAdS}
ds^2&=&-\cosh^2\rho dt^2+d\rho^2+\sinh^2\rho\left[
\sin^2\alpha d\psi^2+\cos^2\alpha d{\tilde\beta}^2+d\alpha^2\right]
\nonumber\\
&&+d\theta^2+\cos^2\theta d{\tilde\phi}^2+\sin^2\theta d\Omega_3^2
\eea
Looking at the warp factors, we can easily extract $y$ and $e^G$:
\bea
y=\sin\theta\sinh\rho\sin\alpha,\quad 
e^G=\frac{\sin\theta}{\sinh\rho\sin\alpha},\quad
h^{-2}=\sinh^2\rho\sin^2\alpha+\sin^2\theta
\eea
To put the metric (\ref{WickAdS}) in the form (\ref{DonosSoln}), one need 
to shift the 
angular variables (${\tilde\beta}=\beta+t$,
${\tilde\phi}=\phi+t$), this will ensure that in the new coordinates 
$g_{tt}=-h^{-2}$.  Such shift also introduces mixings between time and angular coordinates and one can easily read off the relevant one--form: 
\bea\label{AltAdSOm}
\omega=-h^2(\sinh^2\rho\cos^2\alpha d\beta+\cos^2\theta d\phi)
\eea
The Kahler base is parameterized by the two angles $(\beta,\phi)$ and two more coordinates which should be orthogonal to $y$. Starting with three--dimensional space spanned by $(\rho,\alpha,\theta)$, one can use the metric (\ref{WickAdS}) to construct a subspace orthogonal to 
$y$, and to show that it can be parameterized by
\bea
x_1=\cosh\rho\cos\theta,\quad
x_2=\tanh\rho\cos\alpha. 
\eea
It is now easy to invert the relations between $(\rho,\alpha,\theta)$ 
and $(x_1,x_2,y)$:
\bea
\cosh\rho&=&\frac{1}{\sqrt{2(1-x_2^2)}}\left[1+r^2+y^2+
    \sqrt{(1+r^2+y^2)^2-4r^2}\right]^{1/2},\nonumber\\
\cos\theta&=&\frac{1}{\sqrt{2}}\left[1+r^2+y^2-
    \sqrt{(1+r^2+y^2)^2-4r^2}\right]^{1/2},\quad
r=x_1\sqrt{1-x_2^2}\nonumber\\
\sin\alpha&=&\frac{\sqrt{1-x_2^2}}{\sqrt{2}}\left[\frac{-2y^2+
x_2^2(r^2+y^2-1+
    \sqrt{(1+r^2+y^2)^2-4r^2})}{x_2^2(r^2+y^2-1)+x_2^4-y^2}
\right]^{1/2}.
\eea
Using these expressions, we can rewrite $e^G$ and $Z$ as functions 
of $(x_1,x_2,y)$, and the expression for $Z$ turns out to be especially simple:
\bea\label{AldAdSZ}
Z=-\frac{1}{2}\frac{r^2+y^2-1}{\sqrt{(1+r^2+y^2)^2-4r^2}}
\eea
As expected, at the $y=0$ surface this function takes only two values: 
$Z=\pm \frac{1}{2}$. Since $Z$ is related to the $y$--derivatives of the Kahler potential by one of the equations in (\ref{DonosSoln}), we can extract the expression for $K$:
\bea\label{KahlAltAdS}
K&=&\frac{1}{4}\left[-R+(y^2+2)\log(1+r^2+y^2+R)\right.\nonumber\\
&&\quad\left.-y^2\log\left\{2\frac{(r^2-1)R+R^2-
y^2(1+r^2+y^2)}{y^2(r^2-1)^2}
\right\}\right]+K_0+y^2 K_1\\
R&\equiv&\sqrt{(1+r^2+y^2)^2-4r^2}
\eea
The "integration constants" $K_0$ and $K_1$, which may depend upon the coordinates on the base, will be evaluated below. Even though a significant part of the Kahler potential has been determined, 
it cannot be used to calculate the metric unless the proper complex coordinates are found. Comparing the structure of (\ref{AltAdSOm}) and 
(\ref{DonosSoln}) and noticing that Kahler potential does not depend on 
the angular variables, one concludes that such coordinates must have the following form 
\bea\label{Dec18CmplCrd}
z_1=r_1e^{i\phi},\quad z_2=r_2e^{i\beta}:\quad
\omega=\frac{1}{2y}\d_y\left[r_1\d_1 Kd\phi+r_2\d_2 Kd\beta\right]
\eea
Comparing this with (\ref{AltAdSOm}) and performing $y$--integration, 
we find the expression for the derivatives (up to $y$--independent functions):
\bea
r_1\d_1 K=\frac{1}{2}(1+y^2-R)+{\tilde K}_2,\quad
r_2\d_2 K=-\frac{x_2^2(y^2+R)}{2(1-x_2^2)}+{\tilde K}_1
\eea
For these relations to be consistent with (\ref{KahlAltAdS}), we must set 
\bea
r_1=r\sqrt{1-x_2^2},\quad r_2=x_2,\quad 
K_1=\frac{1}{2}\log\frac{r(1-x_2^2)}{r^2-1}
\eea
One can also express ${\tilde K}_1$ and ${\tilde K}_2$
in terms of derivatives of $K_0$, but we will not discuss this further. 

To complete the expression for the Kahler potential we still need to evaluate $K_0$, and the easiest way to do so is to look at the metric 
on the $y=0$ surface. In particular, a restriction of the metric on the base to the two--dimensional subspace spanned by $(t,\beta,\phi)$ is given by
\bea
\frac{\frac{\d}{\d\log r_a}\frac{\d}{\d\log r_b}K}{(2Z+1)}~d\phi_ad\phi_b
=\left[h_\rho^2 c_\alpha^2(sh^2_\rho+s_\theta^2)d\phi_1^2+
c_\theta^2(sh^2_\rho s_\alpha^2+1)d\phi_2^2+
2sh_\rho^2 c_\alpha^2c_\theta^2 d\phi_1 d\phi_2\right]\nonumber
\eea
and looking at this relation at $y=0$, we find the expression
\bea
K_0=\frac{r^2}{4}-\log r.
\eea
One can check that the resulting Kahler potential satisfies the 
Monge--Ampere equation. 

\begin{figure}[tb]
\begin{center}
\epsfysize=2.7in \epsffile{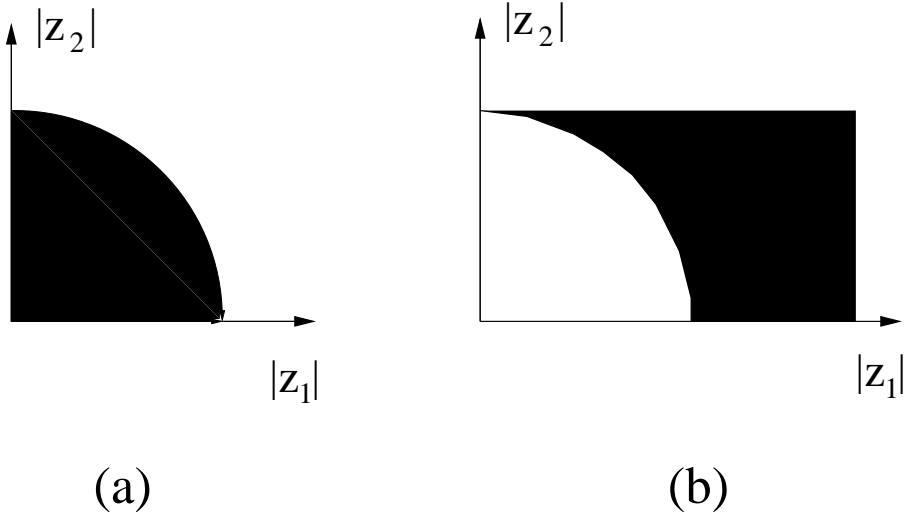}
\end{center}
\caption{
Boundary conditions corresponding to two embeddings of 
$AdS_5\times S^5$: 
(a) spacial sphere $S^3$ is preserved, 
(b) $SO(4)$--part of the R--symmetry group is unbroken. 
} \label{FigAdS5Two}
\end{figure}

Let us summarize the data for the $AdS_5\times S^5$. The 
four--dimensional base is parameterized by the complex coordinates 
(\ref{Dec18CmplCrd}) and the Kahler potential is given by 
(\ref{KahlAltAdS}) with
\bea\label{ExtraInvAdS}
K_0+y^2 K_1=\frac{y^2}{2}\log\frac{r(1-x_2^2)}{r^2-1}+
\frac{r^2}{4}-\log r,\quad r=\frac{r_1}{\sqrt{1-r_2^2}},\quad r_2=x_2
\eea
As already mentioned, function (\ref{AldAdSZ}) takes values 
$\pm\frac{1}{2}$ on the hyperplane $y=0$, and regions with different signs of $Z$ are separated by the surface
\bea\label{AltAdS5Dropl}
r^2=1:\qquad z_1\zb_1+z_2\zb_2=1.
\eea
Although this relation looks the same as (\ref{StandAdSBdry}), the base spaces for two descriptions of $AdS_5\times S^5$ are very different. While (\ref{StandAdSBdry}) represented a sphere carved out of ${\bf C}^4$, in the present case $z$ coordinates cover only a cylinder $|z_2|<1$.  
Notice that the infinity of AdS space is mapped into the region 
$|z_1|=\infty$ and into the boundary of the $z_2$--circle ($|z_2|=1$). 
The boundary conditions for two representations of 
$AdS_5\times S^5$ are depicted in figure \ref{FigAdS5Two}.  

%==================

%======================

\subsubsection{1/2--BPS bubbling solutions.}

While $AdS_5\times S^5$ (along with pp-wave) presents the simplest example of a metric which can be embedded in the ansatz 
(\ref{DonosSoln}), there is also a more general class of known solutions covered by 
(\ref{DonosSoln}). 
Unlike a generic metric (\ref{DonosSoln}), these geometries preserve 16 rather than 8 supercharges, and they have a very explicit description in 
terms of solutions of the Laplace equation \cite{LLM}. Thus it is useful 
to embed these geometries into the general ansatz 
(\ref{DonosSoln}).

\begin{figure}[tb]
\begin{center}
\epsfysize=2.2in \epsffile{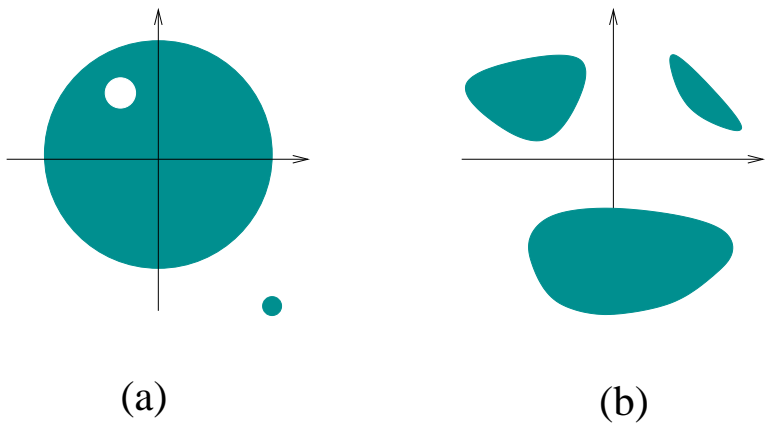}
\end{center}
\caption{
Boundary conditions for the 1/2--BPS geometries of \cite{LLM}: (a) giant graviton and a dual giant, (b) generic distribution of droplets.
} \label{Fig2Bbl}
\end{figure}

We begin with recalling some basic facts about the 1/2--BPS geometries constructed in \cite{LLM}. The metric and fluxes are parameterized by one function which depends on three variables: 
${\tilde Z}(x,z,\zb)$, and, with slight notational modifications, the solution of \cite{LLM} reads:
\bea\label{LLMIIBdy}
ds^2&=&-{\tilde h}^{-2}(dt+V)^2+{\tilde h}^2(dx^2+dzd\zb)+
xe^{H}d\Omega_3^2+xe^{-H}d{\tilde\Omega}_3^2,\nonumber\\
F_5&=&-\frac{1}{4}d\left[x^2 e^{2H}(dt+V)+x^2*_3d\left(
\frac{{\tilde Z}+\frac{1}{2}}{x^2}\right)\right]\wedge d\Omega_3+dual,\\
{\tilde h}^{-2}&=&2x\cosh H,\quad (dV)_{z\zb}=\frac{i}{2x}\d_x {\tilde Z},
\quad x\d_x V=i(dz\d_z-d\zb\d_{\zb}){\tilde Z},
\quad {\tilde Z}=\frac{1}{2}\tanh H,\nonumber
\eea
This construction gives a supersymmetric solution of IIB supergravity as 
long as function ${\tilde Z}$ satisfies a linear differential equation:
\bea\label{LaplLLM}
4\d_z\d_{\zb}{\tilde Z}+x\d_x\left(\frac{\d_x {\tilde Z}}{x}\right)=0.
\eea
Moreover, as shown in \cite{LLM}, the system (\ref{LLMIIBdy}) describes 
a smooth geometry if and only if function ${\tilde Z}$ obeys some special Dirichlet boundary conditions in the plane $x=0$: 
\bea\label{12BPSoldBC}
x=0:\qquad {\tilde Z}=\frac{1}{2}~\mbox{or}~{\tilde Z}=-\frac{1}{2}.
\eea
Thus the entire plane is divided into two types of regions and a typical boundary condition is depicted in figure \ref{Fig2Bbl}b.

%==================

%======================

To compare these 1/2--BPS geometries with (\ref{DonosSoln}), we identify the three dimensional sphere appearing in (\ref{DonosSoln}) with $S^3$ in the metric (\ref{LLMIIBdy}), while embedding the Killing direction 
$\psi$ from 
(\ref{DonosSoln}) into ${\tilde S^3}$:
\bea\label{ThrSphrBdy}
d{\tilde\Omega}_3^2=d\theta^2+\cos^2\theta d\psi^2+
\sin^2\theta d{\tilde\phi}^2.
\eea
Such identification follows from the field theory analysis. Our goal is to 
describe states in SYM on $R\times S^3$ which are constructed from 
elementary blocks (\ref{QrtBldBlk}). To preserve supersymmetry, one 
should look only at zero modes on $S^3$, this leads to the direct 
embedding of $d\Omega_3^2$ into the ansatz (\ref{DonosSoln}). 
The second 
sphere in (\ref{LLMIIBdy}) came from the $SO(4)$ R-symmetry 
preserved by $\mbox{Tr} X^n$, and a generic state constructed out of 
(\ref{QrtBldBlk}) breaks this isometry to $U(1)$. To identify the 
embedding of this $U(1)$ into $SO(4)$, one should notice that the later group can be viewed as a set of rotations in four directions, while 
$U(1)$ corresponds to rotations in one plane. This immediately leads 
to (\ref{ThrSphrBdy}), in particular, it is clear that $\psi$ coordinate, which 
corresponds to the $U(1)$ translations, should be orthogonal to other 
directions. Notice that the $1/2$--BPS geometries (\ref{LLMIIBdy}) can also be embedded into different class of $1/4$--BPS solutions constructed in \cite{donos2} by treating ${\tilde\phi}+\psi$ rather than $\psi$ as a Killing vector preserved by the $1/4$--BPS geometries. The corresponding embedding was discussed in \cite{donos2,vaman}, but 
it appears to be irrelevant for viewing states 
$\prod_i\mbox{Tr} X^{n_i}$ as 
a subset of objects constructed out of (\ref{QrtBldBlk}).

To summarize, we have argued that to embed the geometry 
(\ref{LLMIIBdy}) into the ansatz (\ref{DonosSoln}), one needs to equate
$d\Omega_3^2$ appearing in both expressions and identify $\psi$ direction of (\ref{DonosSoln}) with corresponding term in (\ref{ThrSphrBdy}). Also the time coordinates in (\ref{DonosSoln}) and (\ref{LLMIIBdy}) should be the 
same, but it turns out the coordinate ${\tilde\phi}$ appearing in 
(\ref{LLMIIBdy}) does not belong to the subspace spanned by $y$ and Kahler metric, while the shifted variable $\phi={\tilde\phi}+t$ does\footnote{The simplest way to see this is to compare the coordinate dependence of the Killing spinors in 1/2 and 1/4--BPS cases, but one can also use a purely bosonic argument based on matching $g_{tt}$ in (\ref{DonosSoln}) 
(see Appendix \ref{SectApEmbIIB}).}. Starting with this identification, one can construct the complete map between 1/2--BPS and 1/4--BPS 
variables. The technical details are presented in the Appendix 
\ref{SectApEmbIIB}, and here we just summarize the results. First of all, 
it turns out that, to perform an embedding, one needs to rewrite 
(\ref{LLMIIBdy}) in terms of a new function $D$ which is defined by 
the relations:
\bea\label{Jan7Eqn1}
x\d_x D=\frac{1}{2}-{\tilde Z},\qquad V=-i(dz\d_z-d\zb\d_{\zb}){D}.
\eea
The linear equation for ${\tilde Z}$ implies that $D$ must be harmonic:
\bea\label{Jan7Eqn2}
4\d_z\d_{\zb} D+x^{-1}\d_x(x\d_x D)=0.
\eea
The coordinates $z_a$ and $y$ appearing in (\ref{DonosSoln}) can be expressed through their counterparts from (\ref{LLMIIBdy}):
\bea\label{Jan7Eqn3}
y=x\cos\theta,\quad z_1=z,\quad z_2=x\sin\theta e^{-D+i\phi}.
\eea
The 1/2--BPS metrics are governed by ${\tilde Z}$, the 1/4--BPS ones are parameterized by the Kahler potential $K$, and the relation 
(\ref{Jan5KahlD}):
\bea\label{Jan7Eqn4}
y^{-1}\d_y K=2D-\log y
\eea
maps one description into another.
Equations (\ref{Jan7Eqn1}), (\ref{Jan7Eqn3}) (\ref{Jan7Eqn4}) provide  local embedding of (\ref{LLMIIBdy}) into (\ref{DonosSoln}), and now we will 
relate the boundary conditions in these two cases.

\begin{figure}[tb]
\begin{tabular}{ccc}
\epsfxsize=1.8in \epsffile{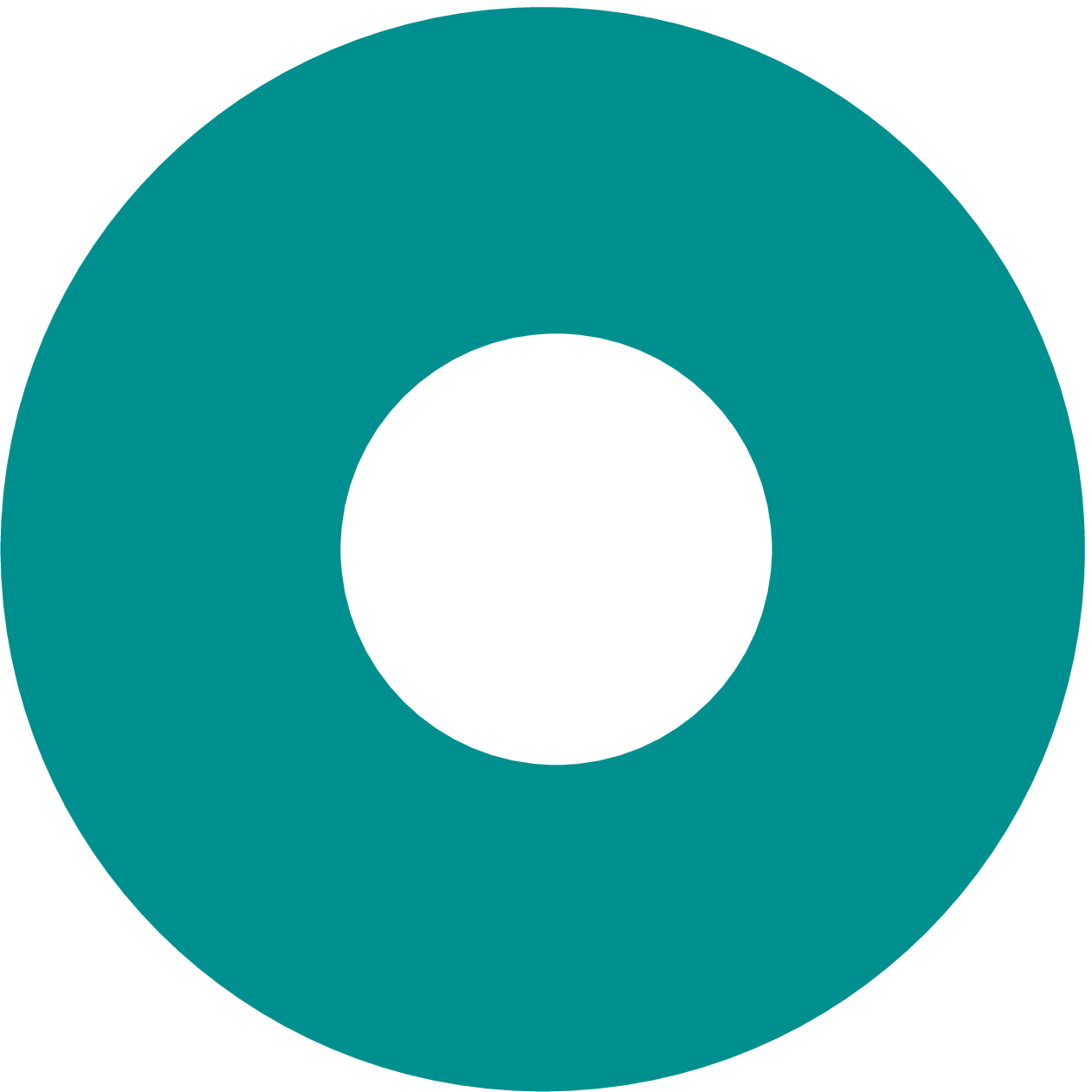}&
\qquad \qquad\qquad&
\epsfysize=2.0in \epsffile{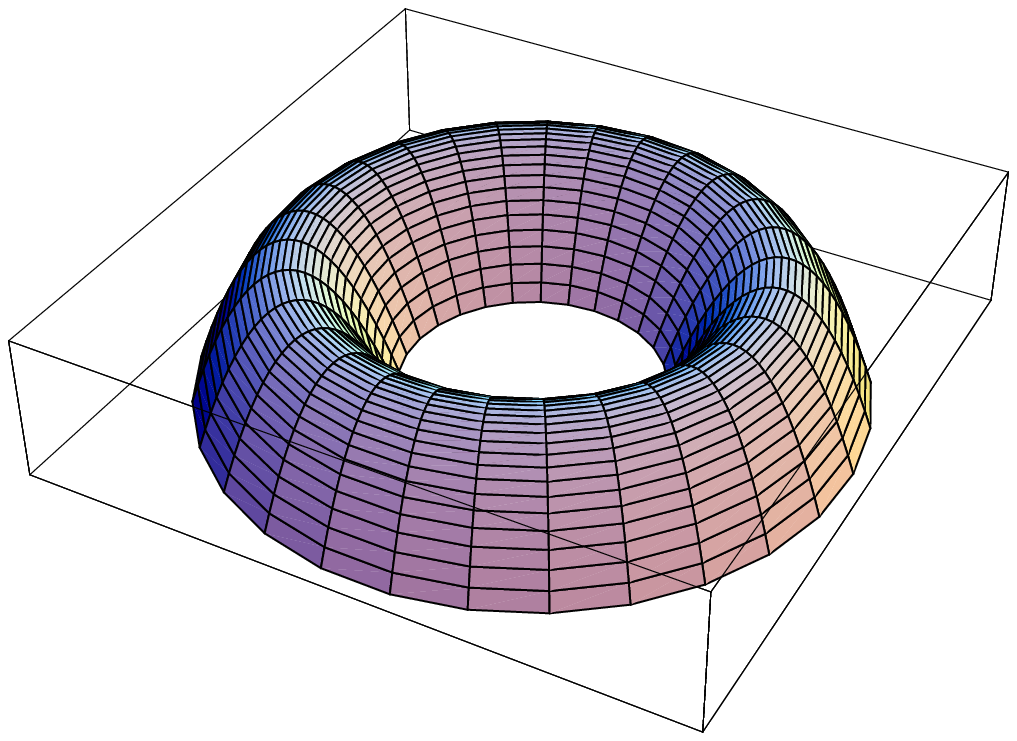}\\
(a)&&(b)\\
\end{tabular}
\caption{
Correspondence between boundary conditions in 1/2--BPS (a) and 
1/4--BPS (b) cases.
} \label{Fig3Dbbl}
\end{figure}

We begin with observing that for the 1/2--BPS states the boundary 
conditions
(\ref{12BPSoldBC}) can be reformulated in terms of $D$:
\bea\label{AltIIBcnd}
x=0:\qquad \d_x D=0\quad\mbox{or}\quad\d_x D=\frac{1}{x}+O(x^0).
\eea
The first relation comes from the regularity condition 
(${\tilde Z}=\pm\frac{1}{2}+O(x^2)$) for the solutions of 
(\ref{LLMIIBdy}).
Notice that 
the boundary conditions (\ref{AltIIBcnd}) are identical to ones found for eleven--dimensional bubbling solutions in \cite{LLM}, and this analogy will be further explored in section \ref{SectUnif}. To interpret (\ref{AltIIBcnd}) in terms of the variables appropriate for the $1/4$--BPS case, we need to 
rewrite the boundary conditions (\ref{AltIIBcnd}) in terms of 
$y$--derivatives. Using relations (\ref{RelXYMIIB}), we find:
\bea
y=0:\qquad \begin{array}{l}
\cos\theta=0:\quad \d_y D=0,\\
\cos\theta\ne 0:\quad \d_y D=0\quad\mbox{or}\quad
\d_y D=\frac{1}{y}+O(y^0).
\end{array}
\eea
Using (\ref{Jan7Eqn4}) and (\ref{DonosSoln}), one can see that this translates into the correct boundary conditions\footnote{As we will see in section \ref{SectIIBDrpl}, regularity also leads to an additional restriction on Kahler potential. This requirement is satisfied by the 1/2--BPS geometries, but we will not present the proof.} $Z=\pm\frac{1}{2}$, and now we will try to extract the shapes of the 1/4--BPS droplets. As in the 
$1/2$--BPS case, the $y=0$ hyperplane is divided into regions with 
$\d_y D=0$ and $\d_y D=\frac{1}{y}$, and pictorially one can use two different colors to distinguish between them.
In the regions with $\d_y D=\frac{1}{y}+O(y^0)$ one has the following relations:
\bea
x=0,\qquad D=\log x+{\hat D}(z,\zb)+O(x),
\eea
which allow us to rewrite (\ref{Jan7Eqn3}) as
\bea
y=0,\quad z_1=z,\quad z_2=\sin\theta e^{-{\hat D}+i\phi}.
\eea
This implies that the bubbles with $\d_y D=\frac{1}{y}$ are defined 
by inequalities involving $|z_2|$ and $z_1$:  
\bea\label{BblBndIeq}
\d_y D=\frac{1}{y}:\qquad |z_2|\le e^{-{\hat D}(z,\zb)}.
\eea
In particular, the coloring of the $y=0$ hyperplane is invariant under phase shift in $z_2$, and an example is presented in figure 
\ref{Fig3Dbbl}b. 

%==================

%======================

To summarize, we demonstrated that 1/2--BPS bubbling solutions of 
\cite{LLM} can be embedded in the more general ansatz 
(\ref{DonosSoln}), and the map between two descriptions is given by 
(\ref{Jan7Eqn3}), (\ref{Jan7Eqn4}). Moreover, the 1/2--BPS geometries correspond to very simple boundary conditions in $y=0$ plane: the walls between regions with $\d_y D=0$ and 
$\d_y D=\frac{1}{y}$ are determined by the equation
\bea\label{Bbl2BcIIB}
z_2\zb_2=e^{-2{\hat D}(z_1,\zb_1)},
\eea
where ${\hat D}=D-\log x$ is a finite part of the harmonic function 
in the ${\tilde Z}=-\frac{1}{2}$ region. 

%=====================================

\subsection{Boundary conditions I: D branes}

\label{SectIIBDbr}

%=====================================

Let us now go back to the solution (\ref{DonosSoln}) and discuss  supersymmetric D3 branes in this geometry. It turns out that there are 
two types of branes preserving eight supercharges, and, just as in the case of the D3--webs discussed in section \ref{SectProbe}, their profiles cannot be arbitrary. In this subsection we will show that a consistency of IIB supergravity leads to some restrictions on the locations 
of D3 branes. In the next subsection we will demonstrate that the allowed profiles are in a perfect agreement with results of the probe analysis.
The $SO(4)$ symmetry 
required by the SUSY algebra leads to two distinguished classes of branes,
and we discuss them one--by--one.  Using terminology of 
\cite{giant}, the first class can be called "giant gravitons", while the second type represents "dual giants".

{\bf 1. Giant gravitons and holomorphic surfaces.}

Let us consider branes which do not wrap the 
three--sphere. To preserve the $SO(4)$ symmetry, these branes must
be located at points where the radius of $S^3$ goes to zero, 
then, to avoid singularities in $g_{\psi\psi}$, $y$ must vanish at the  locations of the branes. Looking at the structure of $F_5$, we conclude that, in addition to being extended in $t$, the brane 
wraps $\psi$--direction. We can introduce two additional coordinates  
$(w_1,w_2)$ on the D3 worldvolume, and they should be functions of 
$(z_a,\zb_a)$. Following notation of section \ref{SectMembrStr}, we use 
$(v_1,v_2)$ to parameterize the complement of $(w_1,w_2)$ in the 
four--dimensional space spanned by $(z_a,\zb_a)$. In the vicinity of the D3 brane the leading contribution to the radius of $S^3$  
should not depend on the longitudinal coordinates, then definition of $Z$ implies a decomposition (\ref{KbreakA}) in the Kahler potential. Then, repeating the same arguments that led to (\ref{KbreakB}), we 
can write the Kahler potential as a sum of two terms
\bea
K=K_1(w_1,w_2)+K_2(v_1,v_2,{\bf x})
\eea
with divergent $K_2$ and finite $K_1$. Substituting this expansion into the 
Monge--Ampere equation appearing in (\ref{DonosSoln}), and 
matching finite $w$--dependent terms, we find a relation 
\bea
\d_1{\bar \d}_1K_1~\d_2{\bar \d}_2K_1-
\d_1{\bar \d}_2K_1~\d_2{\bar \d}_1K_1=0,
\eea
which have been encountered before (see (\ref{EqnForK1})). As was shown in section \ref{SectMembrStr}, this relation implies that 
$w=w_1+iw_2$ is a holomorphic function of $z_1,z_2$. Then one can reparameterize the space transverse to the brane, so that $v=v_1+iv_2$ is also holomorphic. We conclude that consistency of IIB supergravity requires the D3 branes to have a holomorphic profile in $(z_a,\zb_a)$ directions. This conclusion agrees with the results of the probe 
analysis presented in the next subsection. It is also consistent with the fact that the D3 webs are described by holomorphic curves (see section 
\ref{SectProbe}), since asymptotically--flat webs can be obtained from 
the bubbling solutions 
by the procedure outlined in section \ref{SectIIBCmpWb}. 

{\bf 2. Dual giants.}

Let us now consider the D3 branes which wrap $S^3$. To have a Lorentzian worldvolume, the brane should also be stretched along time direction, then six coordinates $(z_a,\zb_a,y,\psi)$ can be used to 
parameterize the directions transverse to the brane. In particular, to preserve $U(1)$ symmetry, the metric component $g_{\psi\psi}$ should go to zero at the location of the brane, then, to avoid a divergence in the warp factor of the sphere, we must require that 
$y=0$. Thus we conclude that the symmetries of the problem require the "dual giant gravitons" to be located at the points
\bea
y=0,\quad z_a=z^{(0)}_a.
\eea

To summarize, we demonstrated that, to have a consistent supergravity solution with eight supercharges, one should only allow two types of 
D--brane sources: they should either follow holomorphic profiles in the Kahler space while wrapping $t$ and $\psi$ directions, or they should wrap $S^3$ and $t$ while being a point in the $(z_a,\zb_a)$ subspace. Both types of branes must be located at $y=0$, but giant gravitons sweep holomorphic surfaces in the subspace where $Z=-\frac{1}{2}$, while dial giants are localized at points in the regions where $Z=\frac{1}{2}$. 

In section \ref{Sect35Brn} it was demonstrated that in 
asymptotically--flat case 
a consistency of supergravity and a probe analysis lead to the same restrictions on the location of sources, this was a manifestation of the 
open/closed string duality. In the next subsection we will show that a similar agreement occurs for the giant gravitons. 

%=====================================

\subsection{Comparison to probe analysis}

\label{SectIIBPrb}

%=====================================

In this subsection we will analyze supersymmetric D--branes in the
geometry (\ref{DonosSoln}). To mimic the discussion of asymptotically flat space presented in section \ref{SectProbe}, one should begin with studying 
branes in $AdS_5\times S^5$ (since this is the closest analog of branes in flat space). Then the DBI action implies that, in a perfect agreement with results of the previous subsection, D3 branes must follow holomorphic profiles \cite{mikhail}. 
We will demonstrate that holomorphicity discovered in 
\cite{mikhail} must be formulated in terms of the complex coordinates 
$z_a$ which were used in (\ref{DonosSoln}). We will also show that  
brane profiles must be holomorphic even in a general $1/4$--BPS metric 
(\ref{DonosSoln}). The last part of this subsection will be devoted to the 
"dual giants", i.e. to branes wrapping $S^3$ in (\ref{DonosSoln}). 

As reviewed in section \ref{SectProbe}, to identify supersymmetric 
branes in an arbitrary background, one needs to solve the kappa-symmetry projections (\ref{DBIProj}). Presently we are interested in D3 branes
in $AdS_5\times S^5$ and we begin with analyzing the "original giant gravitons", which appear as pointlike objects on AdS space \cite{giant}. 
This implies that the brane is located at $\rho=0$, then, recalling the embedding 
(\ref{Jan11Ads5}), one concludes that $y=0$ while $r<1$. 
Kahler potential (\ref{KahlerAdS}) can be expanded in the vicinity of such points:
\bea
\Psi=1+\frac{y^2}{1-r^2}+O(y^4),\quad 
K=\frac{1}{2}\left[1-y^2\log(1-r^2)+y^2\log y\right]+O(y^4)
\eea
To evaluate $Z+\frac{1}{2}$, one can look at subleading terms 
$K$ and use (\ref{DonosSoln}), but equation (\ref{OldEqn99}) gives a 
more direct route to the answer:
\bea
Z+\frac{1}{2}=h^{2}\sinh^2\rho=\frac{y^2}{\sin^4\theta}+O(y^4)=
\frac{y^2}{(1-r^2)^2}+O(y^4),\quad 
e^{G}=\frac{y}{1-r^2}+O(y^2)\nonumber
\eea
Using this data, we can write the leading contributions to the metric appearing in (\ref{DonosSoln}):
\bea\label{MikhailMetr}
ds_{10}^2&=&(1-r^2)\left[-(dt+\omega)^2+
2\d_a{\bar\d}_b {\tilde K}dz^ad\zb^b+d\psi^2\right]+
\frac{1}{1-r^2}\left[dy^2+y^2d\Omega_3^2\right]\nonumber\\
\omega&=&i\left[\d-\db\right]{\tilde K},\quad {\tilde K}=-\frac{1}{2}
\log(1-r^2),\quad r^2=z_a\zb_a.
\eea

We already know that D3 branes are located at $y=0$, let us 
now specify the other coordinates of these objects. Since 1/4--BPS giant gravitons are expected to preserve the $U(1)$ part of the R--symmetry group, they should wrap $\psi$--coordinate, moreover, a position of the giant in the remaining directions should not depend $\psi$. The brane worldvolume extends in the time direction as well, but, since giant graviton is a rotating object, one is tempted to allow for time dependence  
of the transverse coordinates. Indeed, giant graviton moving in the metric (\ref{NormAdS5S}) follows a trajectory with nontrivial $\phi_1(t)$, 
$\phi_2(t)$, but time dependence cancels out in the  complex coordinates (\ref{Jan11Ads5}). Similar situation was encountered in the case of 
1/2--BPS case \cite{LLM}, where giant gravitons turned out to be static in objects $y=0$ plane. In the present case, $z$--coordinates of the 
D3--brane must be time--independent. 
Thus, to analyze the DBI projection 
(\ref{DBIProj}), we can impose a static gauge:
\bea\label{Jan11Gauge}
t=\xi^0,\quad z_a=z_a(\xi^1,\xi^2),\quad \psi=\xi^3. 
\eea
This choice leads to the following induced metric:
\bea\label{Jan11IndMetr}
ds_{ind}^2&=&(1-r^2)\left[-(d\xi^0+i\d_a {\tilde K} Dz^a-
i\db_a {\tilde K}D\zb^a)^2+
2\d_a{\bar\d}_b {\tilde K}Dz^aD\zb^b+d\xi_3^2\right],\nonumber\\
Df&=&\d_{\xi^1}fd\xi^1+\d_{\xi^2}fd\xi^2,
\eea
which allows to compute ${\cal L}$ introduced in (\ref{DBIProj}):
\bea\label{Jan11Lagr}
{\cal L}=(1-r^2)^2~\mbox{det}(
2\d_a{\bar\d}_b {\tilde K}\d_m z^a\d_n\zb^b)
\eea 

The relations (\ref{MikhailMetr}), (\ref{Jan11IndMetr}), (\ref{Jan11Lagr})
can be easily generalized to a case of an arbitrary 1/4--BPS geometry
(\ref{DonosSoln}):
\bea\label{Jan11Gnrl}
ds_{10}^2&=&h^{-2}\left[-(dt+\omega)^2+
2\d_a{\bar\d}_b {\tilde K}dz^ad\zb^b+d\psi^2\right]+
h^2\left[dy^2+y^2d\Omega_3^2\right],\nonumber\\
ds_{ind}^2&=&h^{-2}\left[-(d\xi^0+i\d_a {\tilde K} Dz^a-
i\db_a {\tilde K}D\zb^a)^2+
2\d_a{\bar\d}_b {\tilde K}Dz^aD\zb^b+d\xi_3^2\right],\\
{\cal L}&=&h^{-4}~\mbox{det}(
2\d_a{\bar\d}_b {\tilde K}\d_m z^a\d_n\zb^b),\nonumber
\eea 
and from now on our discussion would refer to this general case. Using an intuition from $AdS_5\times S^5$ solution, we will impose the static gauge (\ref{Jan11Gauge}). 

To evaluate gamma matrices appearing in (\ref{DBIProj}), we 
need expressions for some components of the (reduced) veilbein 
corresponding the ten dimensional metric (\ref{Jan11Gnrl}):
\bea
e^{\bf t}=(dt+\omega),\quad 
e^{\bf \psi}=d\psi,\quad e^{\bf a},\quad e^{\bf{\bar a}}:\quad
\delta_{ab}e^{\bf a}e^{\bf{\bar b}}=
2\d_a{\bar\d}_b {\tilde K}dz^ad\zb^b
\eea

A nontrivial restriction on shape of branes comes from the requirement 
that the projector (\ref{DBIProj}) does not break 
any of the $8$ supercharges which are preserved by 
(\ref{DonosSoln})\footnote{Of course, the $AdS_5\times S^5$ metric has additional supersymmetries which appear accidental from the point of view 
of  1/4--BPS analysis. While giant graviton is allowed break these "extra" supercharges, to be at least $1/4$--BPS, it must preserve the ones which were explicit in (\ref{DonosSoln}).}, so we need to recall the structure of 
the relevant Killing spinors. While these spinors have not been explicitly written down in the literature, some useful information can be extracted from the relations between bilinears found in 
\cite{donos}\footnote{The relations (\ref{ProjFrDns1}) were derived 
in \cite{donos} for a complex spinor in IIB supergravity. 
Translation to the alternative conventions used in (\ref{DBIProj}), is performed in (\ref{ProjFrDns}). 
Notice that relations (\ref{ProjFrDns1}), (\ref{ProjFrDns2}) have 
extra factors of ${\hat\sigma}_1$ in comparison with \cite{donos}, these insertions arise from 
rewriting the reduced spinor of \cite{donos} as a complex spinor in ten dimension (see \cite{donos} for the definition of gamma matrices).}:
\bea\label{ProjFrDns1}
&&{\bar\eps}\Gamma_{\bf ab}\gamma_7{\hat\sigma}_1\eps=
{\bar\eps}\Gamma_{\bf {\bar a}{\bar b}}\gamma_7{\hat\sigma}_1\eps=0,\quad
{\bar\eps}\Gamma_{\bf {a}{\bar b}}\gamma_7{\hat\sigma}_1\eps=
\frac{1}{2}\delta_{\bf ab}{\bar\eps}\gamma_7{\hat\sigma}_1\eps\\
\label{ProjFrDns2}
&&{\bar\eps}\gamma_7{\hat\sigma}_1\eps=
\sqrt{y}e^{-G/2}h{\eps}^\dagger\eps,\quad 
i{\bar\eps}{\hat\sigma}_1\eps=
\sqrt{y}e^{G/2}h{\eps}^\dagger\eps
\eea
It is convenient to rewrite the last two relations in terms of  a rotated spinor $\tilde\eps$:
\bea\label{Jan11e1}
\eps=e^{i\delta\gamma_7}{\tilde\eps}:&&
{\bar\eps}\gamma_7{\hat\sigma}_1\eps=
\cos 2\delta~{\tilde\eps}^\dagger\Gamma^0\gamma_7{\hat\sigma}_1
{\tilde\eps}+
i\sin 2\delta {\tilde\eps}^\dagger\Gamma^0{\hat\sigma}_1{\tilde\eps}
=\frac{e^{-G/2}}{\sqrt{e^G+e^{-G}}}{\tilde\eps}^\dagger{\tilde\eps},
\nonumber\\
&&
i{\bar\eps}{\hat\sigma}_1\eps=
-\sin 2\delta~{\tilde\eps}^\dagger\Gamma^0\gamma_7{\hat\sigma}_1
{\tilde\eps}+
i\cos 2\delta {\tilde\eps}^\dagger\Gamma^0{\hat\sigma}_1{\tilde\eps}
=\frac{e^{G/2}}{\sqrt{e^G+e^{-G}}}{\tilde\eps}^\dagger{\tilde\eps}.
\eea
By setting
\bea
\sin 2\delta=\frac{e^{-G/2}}{\sqrt{e^G+e^{-G}}},\quad
\cos 2\delta=\frac{e^{G/2}}{\sqrt{e^G+e^{-G}}},
\eea
one can reformulate equations (\ref{Jan11e1}) as a projector
\bea
\Gamma^0{\hat\sigma}_1{\tilde\eps}=-i{\tilde\eps}.
\eea
Rewriting (\ref{ProjFrDns1}), (\ref{ProjFrDns2}) in terms of 
${\tilde\eps}$, we find very simple relations:
\bea
{\tilde\eps}^\dagger\Gamma_{\bf ab}{\tilde\eps}=
{\tilde\eps}^\dagger\Gamma_{\bf {\bar a}{\bar b}}{\tilde\eps}=0,\quad
{\tilde\eps}^\dagger\Gamma_{\bf {a}{\bar b}}{\tilde\eps}=
\frac{1}{2}\delta_{\bf ab}{\tilde\eps}^\dagger{\tilde\eps},
\eea
which imply that ${\tilde\eps}$ is annihilated by the holomorphic gamma matrices ($\Gamma_{\bf a}{\tilde\eps}=0$). Recalling 
(\ref{Jan11e1}), we observe that the same relation is satisfied by the original spinor $\eps$:
\bea\label{ProjFrDns}
\Gamma_{\bf a}{\eps}=0.
\eea

After this brief review of the Killing spinor on the geometry 
(\ref{DonosSoln}), we are ready to analyze the DBI projector (\ref{DBIProj}):
\bea\label{MikhProj}
\Gamma=i{\cal L}^{-1}(1-r^2)^2\sigma_2\otimes \Gamma_{t\psi}
\frac{\d X^m}{\d\xi^1}\frac{\d X^n}{\d\xi^2}{\tilde\gamma}_{mn},\quad
\Gamma\eps=\eps
\eea
Here indices $m,n$ go from one to four and coordinates $X^m$ cover the subspace of (\ref{MikhailMetr}) spanned by $z^a,\zb^a$, while 
the action ${\cal L}$ is computed using the metric (\ref{Jan11Gnrl}). 
The matrices ${\tilde\gamma}_{mn}$ are constructed using the veilbein $e^{\bf a}$, $e^{\bf{\bar a}}$ since contribution of 
$e^{\bf t}_m$ disappears from $\gamma_{tm}$:
\bea
\gamma_{tm}=e_t^{\bf t}e_m^{\bf t}\Gamma_{\bf tt}+
e_t^{\bf t}e_m^{\bf m}\Gamma_{\bf tm}=\gamma_t{\tilde\gamma}_m.
\eea

Applying $\Gamma_{\bf ab}$ to the projector (\ref{MikhProj}) and using 
(\ref{ProjFrDns}), we arrive at a relation
\bea
\frac{\d \zb_1}{\d\xi^1}\frac{\d \zb_2}{\d\xi^2}-
\frac{\d \zb_2}{\d\xi^1}\frac{\d \zb_1}{\d\xi^2}=0,
\eea
which implies a functional dependence $\zb_2={\bar f}(\zb_1)$, 
$z_2=f(z_1)$. Thus, by studying the DBI projection (\ref{DBIProj}), 
we showed that the the profiles of supersymmetric D3 branes in the geometry (\ref{DonosSoln}) must be holomorphic: 
\bea\label{Jan11Shape}
f(z_1,z_2)=0.
\eea
This outcome of the open string analysis is in a perfect agreement with closed string picture which was discussed in section \ref{SectIIBDbr}: 
there it was shown that only holomorphic sources are consistent with equations of motion of IIB supergravity.  

\begin{figure}[tb]
\begin{center}
\epsfxsize=4in \epsffile{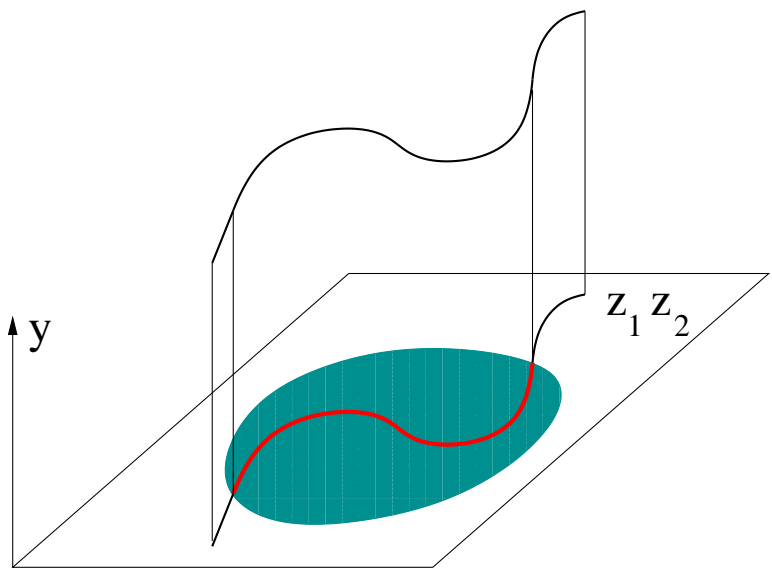}
\end{center}
\caption{
Giant gravitons and holomorphic surfaces. To construct a worldvolume of a 
D3 brane, one should take an intersection of the holomorphic surface with 
$\{y=0,Z=\frac{1}{2}\}$ region (this intersection is shown in red), and fiber 
$\psi$ and $t$ over it. In the case of $AdS_5\times S^5$, this picture gives a geometric interpretation of the radial coordinate introduced in 
\cite{mikhail}
} \label{FigMikhail}
\end{figure}

In the case of $AdS_5\times S^5$ it is interesting to compare our holomorphic curves with analysis of the giant gravitons presented in 
\cite{mikhail}. Mikhailov proposed to "fill in" the five--dimensional sphere by introducing an fictitious radial coordinate $R$ and writing 
a six--dimensional metric of ${\bf C}^3$ as
\bea
ds^2_6=dR^2+R^2d\Omega_5^2=dw_ad\wb_a.
\eea
Then it was shown that supersymmetric D3 branes must be located at 
intersections of two--dimensional holomorphic surfaces in ${\bf C}^3$ (with some additional time dependence) and a sphere $R=1$. A generic surface $f(w_1e^{it},w_2e^{it},w_3e^{it})=0$ gives
rise to a giant graviton preserving four supercharges, while a surface 
$f(w_1e^{it},w_2 e^{it})=0$ leads to a 1/4--BPS object. Matching this 
with equation (\ref{Jan11Shape}), we find a map between $w_a$ and 
$z_a$ coordinates ($z_a=w_ae^{it}$), then definitions 
(\ref{Jan11Ads5}) lead to relations between $w_a$ and standard coordinates on $AdS_5\times S^5$ (see (\ref{NormAdS5S})):
\bea
w_1=\cosh\rho[\cos\theta\cos\alpha e^{i\phi_1}]=\mu_1\cosh\rho,\quad 
w_2=\cosh\rho[\cos\theta\sin\alpha e^{i\phi_2}]=\mu_2\cosh\rho
\eea
Here $\mu_1$ and $\mu_2$ are two of the six coordinates defining 
the five--sphere:
\bea
\sum_{a=1}^3 {\bar\mu}_a\mu_a=1,
\eea
so we arrive at an identification $R=\cosh\rho$. This analysis demonstrates that, rather than being an artificial parameter, Mikhailov's radial coordinate has a very simple meaning for $AdS_5\times S^5$. Moreover, by introducing coordinates $z_a$ and demonstrating holomorphicity, we showed how to define a similar "radial direction" for the most general 
1/4--BPS geometry as well: a holomorphic surface $f(z_a)=0$ is naturally parameterized by one complex coordinate and $y$. The worldvolume of 
the D3 brane is obtained by taking an intersection of this surface with 
any region $\{y=0,Z=\frac{1}{2}\}$, and fibering $\psi$ and $t$ over it. 
A pictorial representation of this construction is given in figure 
\ref{FigMikhail}.
Notice that, since $\psi$ direction shrinks at the boundary of 
$\{y=0,Z=\frac{1}{2}\}$ region, the worldvolume of the resulting D3 brane 
is a manifold without boundaries.  

%==================

%======================

Let us now make a brief comment about the "dual giant graviton".  
This object wraps time direction and three--dimensional sphere which remains unbroken in the ansatz (\ref{DonosSoln}), so, by counting 
dimensions,
one concludes that the dual giant should be located at a point 
$z_a=z^{(0)}_a$ on the $y=0$ surface, and at that point 
$Z=-\frac{1}{2}$. This result trivially agrees with discussion presented in section \ref{SectIIBDbr}.

Let us summarize the results of this subsection. By analyzing supersymmetry conditions for D3 branes on a general geometry 
(\ref{DonosSoln}), we showed that there are two types of 
interesting objects: 
"giant graviton" which wraps $t,\psi$ directions and follows a holomorphic profile $f(z_1,z_2)=0$ in the $Z=-\frac{1}{2}$ subset of $y=0$ space, and "dual giant" which wraps $t$, $S^3$ and occupies a point in 
$Z=\frac{1}{2}$, $y=0$ subspace. No other object can preserve eight supercharges. These results of open string analysis 
agree perfectly with supergravity discussion presented in section 
\ref{SectIIBDbr}.

%=====================================

\subsection{Boundary conditions II: regular droplets.}

\label{SectIIBDrpl}

%=====================================

After reproducing the correct boundary conditions 
corresponding to probe D3 branes, we now study geometries 
produced by the stacks of the branes. While for the small number of 
branes the boundary conditions for (\ref{WebD3Metr}) and 
(\ref{DonosSoln}) are similar, 
the results for multiple branes are very different. This phenomenon has already been seen in the case of 1/2--BPS solutions: for asymptotically--flat geometry one can simply superpose stacks of D3 branes 
and each element in the stack has exactly the same location. On the contrary, the branes in $AdS_5\times S^5$ repel 
each other\footnote{The field theory manifestation of this phenomenon
was discussed in \cite{beren}.} and form non--compressible droplets 
\cite{LLM}. This droplets change the topology of spacetime and, as a result of such bubbling, the geometry remains smooth everywhere. 
A similar phenomenon is expected to take place for the configurations with lower supersymmetry\footnote{See \cite{beren4} for the relevant field 
theory analysis.} and this subsection will be devoted to 
deriving the "bubbling picture" for the 1/4--BPS geometries.

Let us begin with recalling the results pertaining to 1/2--BPS case 
\cite{LLM}. The gravity solutions had $SO(4)\times SO(4)$ isometry and a coordinate $y$ was defined as a product of the warp factors corresponding to the two three--spheres. At $y=0$ one of the spheres had to collapse to zero size, and such contraction would lead to a singularity in the geometry unless some special boundary conditions were imposed. It turned out that the solutions were parameterized by one harmonic function $z$ and regularity led to the requirement that $Z=\pm \frac{1}{2}$ in $y=0$ plane \cite{LLM}. Then the entire plane was separated into two types of regions (see figure \ref{Fig2Bbl}): one of the three--spheres collapsed in the light region and another one did so in the dark region. There were no restrictions on the curves separating the regions. This arbitrariness was in a complete agreement with brane probe analysis: the (dual) giant gravitons corresponded to 
light (dark) points (see figure \ref{Fig2Bbl}a) which could be combined to 
give droplets with arbitrary shapes. As we will see in a moment, in the 
1/4--BPS case the situation is completely different. 

%==================

\begin{figure}[tb]
\begin{tabular}{ccc}
\epsfxsize=2.5in \epsffile{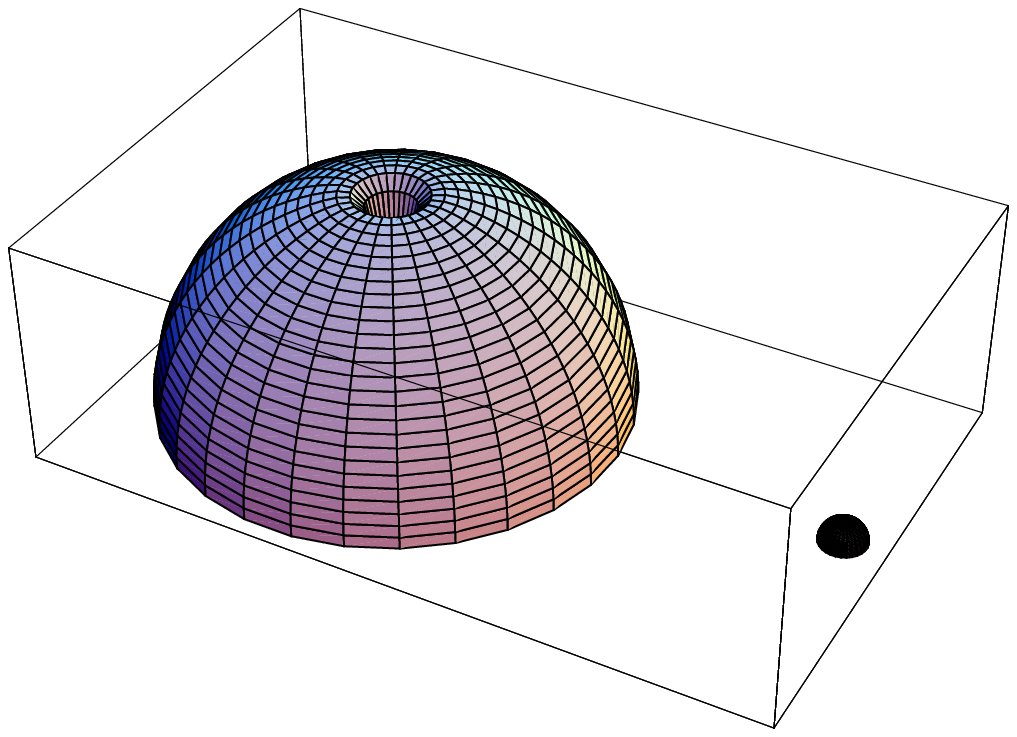}&
\qquad &
\epsfysize=2.0in \epsffile{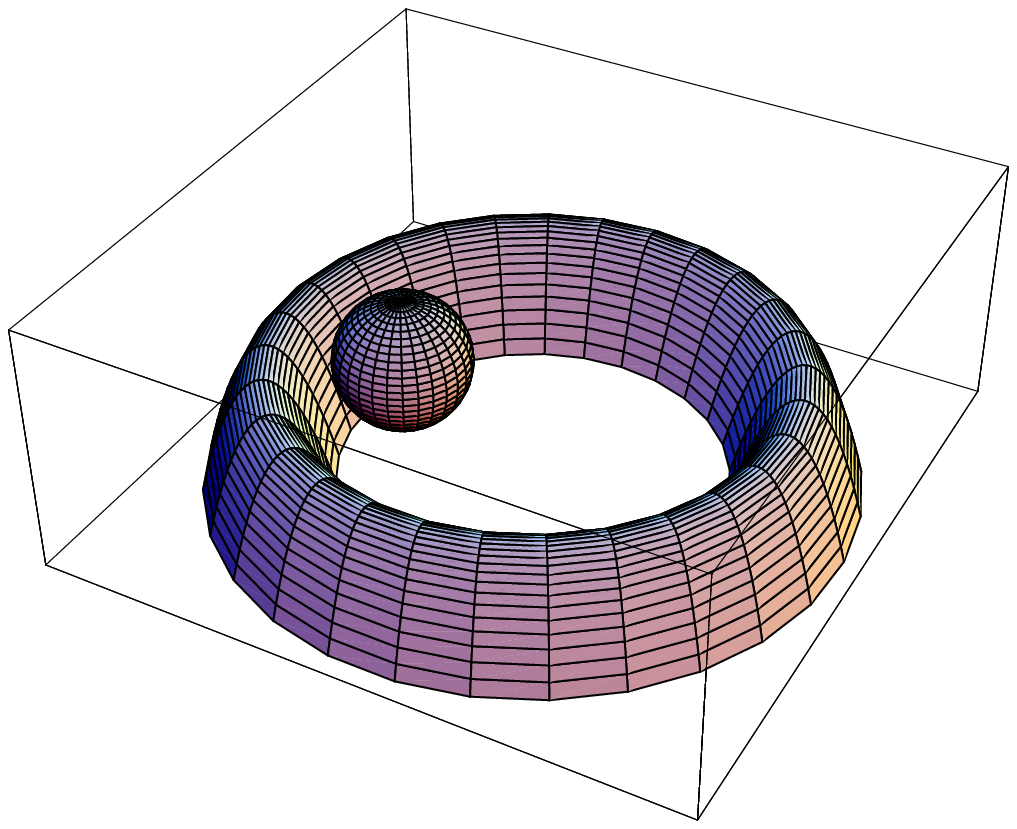}\\
(a)&&(b)\\
\end{tabular}
\caption{
Boundary conditions in the 1/4--BPS case: giant graviton \& dual giant (a) and a generic distribution of droplets (b). To simplify the picture, we use vertical axis for $|z_2|$ and suppress the phase of $z_2$. 
} \label{Fig3DGenu}
\end{figure}

%======================

In a complete analogy with 1/2--BPS case, we observe that 
$y$--coordinate in (\ref{DonosSoln}) is a product of the two warp--factors, 
so on $y=0$ hypersurface either $S^3$ or $S^1$ collapses to zero size. Thus, the entire hypersurface 
is again divided into two types of regions 
(some examples are presented in figure \ref{Fig3DGenu}). We will begin 
with demonstrating that in 
the interior of each droplet the geometry remains regular, once certain restrictions on a Kahler potential are imposed. It turns out that, unlike 1/2--BPS droplets which could have arbitrary shapes, 
their 1/4--BPS counterparts lead to singular solutions unless some additional condition is satisfied by their "walls". One can suspect that 
this should be the case by observing that the 1/4--BPS probe branes 
must follow holomorphic profiles. If arbitrary shapes of the droplets were allowed, one could always take a degenerate limit leading to a source  wrapping a non--holomorphic surface. 
This would imply an existence of some exotic D3 brane which is not allowed in string theory. Fortunately, as we will show below, this situation is ruled out by 
the regularity conditions for the geometries, which imposes certain requirements on the boundaries of the droplets. But before we start analyzing the boundaries, let us demonstrate the regularity at the interior points.

{\bf  Droplets and regularity conditions.}

As already mentioned, one of the spheres collapses at $y=0$ surface,
so one needs to check that the metric remains regular there. 

We will begin with analyzing a vicinity of a point where $y$ goes to zero, while 
$g\equiv ye^{-G}$ remains finite. The definition of $Z$ leads to very simple leading terms in $h$ and in the Kahler potential\footnote{Notice that the Monge-Ampere equation appearing in (\ref{DonosSoln}) can be rewritten 
in terms 
of a new variable $y^2$, an absence of branch cuts in this new description implies that $K$ has expansions in integer powers of 
$y^2$.} :
\bea
Z=-\frac{1}{2}+\frac{y^2}{g^2},\quad
h^{-2}=g
\quad K=\int dy~y\log y+K_0(z,\zb)+
y^2 K_1(z,\zb)+O(y^4).
\eea
Then Monge--Ampere equation implies that 
\bea\label{leadMASph}
\mbox{det}h_{a{\bar b}}=y^4g^{-2}\exp[K_1(z,\zb)]+O(y^6).
\eea
We can now write the leading contribution to the metric
\bea
ds_{10}^2&=&g\left[-(dt+\omega)^2+d\psi^2\right]+g^{-1}
\left[2g^2y^{-2} h_{a{\bar b}}dz^ad\zb^b+dy^2+y^2d\Omega_3^2\right],
\\
\omega&=&i(\db-\d)K_1,\nonumber
\eea
and, to demonstrate regularity, one needs to show that {\it all} components of $h_{a{\bar b}}$ vanish as $y^2$ (the restriction 
(\ref{leadMASph}) on the determinant is not sufficient\footnote{This was 
a loophole in the proof of regularity presented in the Appendix D 
of \cite{vaman}.}).
The leading contribution to (\ref{leadMASph}) leads to an equation for 
$K_0$:
\bea\label{HomMonAmp}
\d_1\db_1 K_0\d_2\db_2 K_0-\d_1\db_2 K_0\d_2\db_1 K_0=0,
\eea
then analysis of section \ref{SectMembr} implies that $K_0=K_0(w,\wb)$, 
where $w$ is a holomorphic function of $z_a$. Unfortunately the Monge--Ampere equation does not impose further restrictions on function $K_0$, so, to ensure regularity, one should supplement the condition 
$Z=-\frac{1}{2}$ by the requirement that $K_0=0$:
\bea\label{Jan23BC}
y=0:&&Z=-\frac{1}{2},\quad  
\d_a \db_b K(z,\zb,y=0)=0.
\eea
Notice that the last condition was missed in \cite{vaman}, but it is crucial
for enforcing regularity. Moreover, as we will show in section 
\ref{SectIIBPert},  both relations in (\ref{Jan23BC}) are needed to 
specify the solution uniquely. As expected, these relations are satisfied 
by the Kahler potential (\ref{KahlerAdS}) is $r<1$ and by the potential 
(\ref{KahlAltAdS}), (\ref{ExtraInvAdS}) if $r>1$. Using the maximum principle for equation (\ref{HomMonAmp}) and fixing the gauge 
freedom in $K$, one can rewrite the last relation in (\ref{Jan23BC}) as two requirements: $K_0=K(z,\zb,y=0)$
must regular in the region where $Z=-\frac{1}{2}$, and $K_0$ should vanish on the boundaries of this region:
\bea\label{Jan24BC}
y=0:&&Z=-\frac{1}{2},\quad  K_0(z,\zb)-\mbox{regular},\quad
K_0(z,\zb)|_{bndry}=0.
\eea
While this condition looks weaker than (\ref{Jan23BC}), equation 
(\ref{HomMonAmp}) makes these two relations equivalent, and each 
of them
is stronger than the requirement $Z=-\frac{1}{2}$. As AdS example 
(\ref{KahlerAdS}) shows, strictly speaking function $K_0$ is not 
well-defined on the wall between different regions, so, while 
(\ref{Jan24BC}) provides a good heuristic picture for the boundary conditions, the relation (\ref{Jan23BC}) is more rigorous, and it will be used in the remaining part of the paper.

Let us now consider a vicinity of a point where $y$ goes to zero, while 
$g\equiv ye^{G}$ remains finite: at such point the warp factor in front of $S^3$ remains finite, while the $\psi$--circle collapses.  We again find an asymptotic expansion of the Kahler potential:
\bea
Z=\frac{1}{2}+\frac{y^2}{g^2},\quad
h^{-2}=g
\quad K=-\int dy~y\log y+K_0(z,\zb)+
y^2 K_1(z,\zb)+O(y^4),
\eea
but now it is sufficient to keep only $K_0$ since the metric becomes
\bea
ds_{10}^2&=&g\left[-dt^2+d\Omega_3^2\right]+g^{-1}
\left[2\d_a{\bar\d}_b K_0dz^ad\zb^b+dy^2+y^2d\psi^2\right]
\eea
Clearly this metric describes a regular space, so we arrive at a complement of (\ref{Jan23BC})
\bea\label{Jan23BCa}
y=0:&&Z=\frac{1}{2}.
\eea

To summarize, we saw that the hypersurface $y=0$ splits into two types of regions: $S^3$ collapses to zero size when $Z=-\frac{1}{2}$ and $\psi$--circle shrinks when $Z=\frac{1}{2}$. We also demonstrated that, in a complete analogy with \cite{LLM},  the metric remains regular in the interior points of the regions (although, there is an additional 
requirement: $\d_a\db_b K=0$ inside droplets with $Z=-\frac{1}{2}$).
However, in contrast to the 1/2--BPS case where regions were allowed
to have arbitrary shapes 
\cite{LLM}, in the present situation additional regularity conditions arise 
on the boundary of the droplets (where both $S^3$ and $S^1$ have vanishing warp--factors). Let us discuss the relevant restrictions. 

{\bf Boundaries of the droplets: an example.}

To analyze regularity conditions for the wall separating different regions, it is convenient to begin with example which shows how $S^3$ and $S^1$ are incorporated in a patch of flat space. The simplest 
solution which fits in the ansatz (\ref{DonosSoln}) is 
$AdS_5\times S^5$ and we already presented its local structure in section \ref{SectIIBEx}. Now let us discuss the appropriate boundary conditions. 

The regions with $Z=\frac{1}{2}$ and $Z=-\frac{1}{2}$ are separated by the three sphere $r=1$ (see equation (\ref{StandAdSBdry})) and at each point 
on this sphere the space is locally flat. To see this more explicitly, we make an expansion of the metric in the vicinity of such point. Introducing new coordinates\footnote{Notice that for 
$AdS_5\times S^5$ one has a relation 
$\sqrt{(r^2+y^2-1)^2+4y^2}=\sin^2\theta+\sinh^2\rho=2y\cosh G$, so 
$\zeta$ takes real values.} $v,R,\zeta$:
\bea\label{Dec10E}
v=r^2-1,\quad R=\sqrt{(r^2+y^2-1)^2+4y^2},\quad 
\cos^2\zeta=\frac{ye^G}{R},
\eea
we find the following relations:
\bea\label{Dec10J}
ye^{-G}=R\sin^2\zeta,\quad y=\frac{R}{2}\sin 2\zeta,\quad 
Z=\frac{1}{2}\cos 2\zeta,\quad h^{-2}=R.
\eea
To evaluate the metric, we need two more ingredients: the geometry on the Kahler base and one--form $\omega$. Since Kahler potential 
(\ref{KahlerAdS}) depends on 
$z_a,\zb_a$ only through $v$, the four dimensional metric can be written as
\bea
2\d_a\db_b K dz^ad\zb^b=2\d_v Kdz_ad\zb_a+2\d_v^2 K|\zb_adz_a|^2,\quad v=z_a\zb_a-1.
\eea
The derivatives of (\ref{KahlerAdS}) can be easily evaluated:
\bea
\d_v K&=&\frac{1}{4(1+v)}\left[(v-y^2)+R\right]=
\frac{R}{4}(\cos 2\zeta+1)+O(R^2),\nonumber\\
\d_v^2 K&=&\frac{1}{4}(\cos 2\zeta+1)+O(R),
\eea
and used to compute the leading contribution to the metric on the base:
\bea
2\d_a\db_b K dz^ad\zb^b=(Z+\frac{1}{2})\left[Rdz_ad\zb_a+
|\zb_adz_a|^2\right]+\dots
\eea
Next we simplify the expression for the one--form:
\bea
\omega&=&
\frac{i}{2y}\left[{\bar\d}\d_y K-{\d}\d_y K\right]=
\frac{1}{8(1+v)}
(-2+\frac{4+v+y^2}{R})\eta=\left[\frac{1}{2R}+O(1)\right]\eta,
\nonumber\\
\eta&\equiv&i(z_ad\zb_a-\zb_a dz_a)
\eea
Using this data, we find the leading contribution to the ten--dimensional metric:
\bea\label{AdSmetrReg}
ds_{10}^2&\approx&-R(dt+\frac{\eta}{2R})^2+\frac{1}{R}\left[
R dx_\perp dx_\perp+\frac{1}{4}(dv^2+\eta^2)+dy^2\right]+
R(c_\zeta^2d\Omega_3^2+s_\zeta^2d\psi^2)\nonumber\\
&\approx&
-dt\eta+dx_\perp dx_\perp+\frac{dR^2}{4R}+
R(d\zeta^2+\cos^2\zeta d\Omega_3^2+\sin^2\zeta d\psi^2)
\eea
Since this is a metric of flat space, the solution is regular at the point $R=0$, and regularity cannot be affected by the subleading terms.  Of course this result was expected since we were discussing $AdS_5\times S^5$, but it was important to unravel the precise mechanism which makes the 
geometry regular, since we want to generalize it to other solutions. 

{\bf Shapes of the droplets.}

Let us now use the lessons from $AdS_5\times S^5$ to find the restrictions imposed by regularity. As already discussed, the necessary condition for (\ref{DonosSoln}) to describe a regular geometry is a decomposition of 
$y=0$ hypersurface into droplets with 
$Z=-\frac{1}{2}$. On the boundary separating two types of regions warp factors for both spheres should vanish, so it is convenient to define spherical coordinates by mimicking (\ref{Dec10J}):
\bea
\tan\zeta=e^{-G},\quad R=\frac{2y}{\sin 2\zeta}:\quad Z=\frac{1}{2}\cos 2\zeta,\quad h^{-2}=R.
\eea  
Notice that the relations (\ref{Dec10E}) were specific to $AdS_5\times S^5$ 
case and they will not hold for a general solution.  Since coordinate 
$R$ measures a distance from the wall, we will be interested in the leading terms in $R$--expansion. 

Regularity condition requires that the leading order of 
$v\equiv R\cos 2\zeta$ does not depend on $y$-coordinates, but rather it is a function on the Kahler base. To see this, we rewrite metric in five--dimensional subspace spanned by $y$, $\Omega_3$, $\psi$:
\bea\label{Dec10A}
ds_5^2&\equiv& h^2dy^2+y(e^Gd\Omega_3^2+
e^{-G}d\psi^2)\nonumber\\
&=&\sin^2 2\zeta\frac{dR^2}{4R}+R\cos^2 2\zeta d\zeta^2+
R(\cos^2\zeta d\Omega_3^2+\sin^2\zeta d\psi^2)
\eea
In the ten--dimensional space this metric is combined with contribution 
coming from the Kahler base, and, to describe regular geometry, the
sum should give a metric of the flat space. In other words, in 
$(R,\zeta)$ subspace, the Kahler metric should contribute the difference between flat six--dimensional space and (\ref{Dec10A}):
\bea
ds^2_{flat}-ds_5^2=\frac{1}{R}(dv)^2.
\eea
Thus the one form $dv$ must lie in the Kahler subspace, i.e.  
$\d_y v=0$ at least in the leading order in $R$. Since one can easily invert the relations between $(R,\zeta)$ and $(v,y)$, we conclude that 
the leading contributions to $(R,\zeta)$ depend on the Kahler base only through one real function $v(z_a,\zb_a)$. In the 
$AdS_5\times S^5$ case this statement was true globally, but for a general $1/4$--BPS geometry it holds only in the vicinity of a point on the "wall". 

To proceed we need some additional information about Kahler potential. It can be extracted from the definition  of $Z$:
\bea
\cos 2\zeta=-y\d_y(y^{-1}\d_y K)
\eea
Expressing $\cos 2\zeta$ through $v$ and $y$ 
($\cos 2\zeta=\frac{v}{\sqrt{v^2+4y^2}}$), one can easily integrate 
this equation:
\bea\label{BubblKahl}
K=\frac{v}{8}\sqrt{v^2+4y^2}+\frac{y^2}{2}\left[\log(v+\sqrt{v^2+4y^2})-
\log y\right]+K_0(z,\zb)+y^2 K_1(z,\zb).
\eea
Since various warp factors depend on the four--dimensional base only through $v$, it is clear that this coordinate parameterizes a direction transverse to the wall. We can also introduce three longitudinal 
coordinates, and in the leading order one expects to have translational invariance in those directions. This implies that 
the leading contribution to Kahler potential should be a function of $v$ and $y$ only. In particular, $K_0$ and $K_1$ appearing in 
(\ref{BubblKahl}) depend on their arguments only through $v$. 

Recalling an expression for the one--form $\omega$:
\bea
\omega=\frac{i}{2y}\d_y\d_v K(\d-\db)v
\equiv \frac{1}{y}\d_y\d_v K\eta
\sim \frac{1}{R}\eta,\quad
\eta\equiv\frac{i}{2}(\d-\db)v,
\eea
we can evaluate the leading terms in the metric (\ref{DonosSoln}):
\bea\label{Dec11A}
ds^2&=&-2R\omega dt-R\omega^2+\frac{1}{Rc^2_\zeta}\left[
2\d_v^2 K |\d v|^2+2\d_v K\d\db v+c^2_\zeta dy^2\right]+
R(c_\zeta^2 d\Omega_3^2+s_\zeta^2 d\psi^2)\nonumber\\
&=&-2R\omega dt-R\omega^2+\frac{1}{R\cos^2\zeta}\left[
2\d_v^2 K |\d v|^2+2\d_v K\d\db v-\frac{1}{4}\cos^2\zeta dv^2\right]\\
&&+\frac{dR^2}{4R}+
R(d\zeta^2+\cos^2\zeta d\Omega_3^2+\sin^2\zeta d\psi^2)
\nonumber
\eea
The last line gives a regular metric on $R^6$, so, to avoid singularity, the second line should parameterize $R^{1,3}$ in the vicinity of the wall. Let us analyze this four--dimensional metric in more detail:
\bea\label{Dec11Ba}
ds_4&=&-2R\omega dt-R\omega^2+\frac{1}{R\cos^2\zeta}\left[
2\d_v^2 K |\d v|^2+2\d_v K\d\db v-\frac{1}{4}\cos^2\zeta dv^2\right]\\
&=&-2R\omega dt-
\left(\frac{R}{y^2}(\d_y\d_v K)^2-
\frac{2\d_v^2 K}{R\cos^2\zeta}\right)\eta^2
+\frac{2\d_v K}{R\cos^2\zeta}\d\db v
+\frac{1}{4R}(\frac{2\d_v^2 K}{\cos^2\zeta}-1)dv^2\nonumber
\eea
Since $\d_v K\sim R$ in the vicinity of the wall\footnote{This is obvious
for first two terms in (\ref{BubblKahl}) and for $y^2 K_1$. To argue that 
$\d_v K_0\sim R$, we recall that Kahler potential has vanishing derivatives
in $Z=-\frac{1}{2}$ region (see equation (\ref{Jan23BC})), then continuity
requires that $\d_v K_0\sim R$ on the boundary.}, the $\d\db v$ term in
 the metric remains regular. The contributions proportional to $dv^2$ and $\eta^2$ are independent and naively they both look singular, so the divergences should cancel in both terms. This leads to the following relations\footnote{One should be able to derive 
(\ref{Dec11Bb}) by analyzing the Monge--Ampere equation in the 
vicinity of the wall, but we will not discuss this further.}:
\bea\label{Dec11Bb}
\d_v^2 K=\frac{1}{2}\cos^2\zeta+O(R),\quad 
\d_y\d_v K=\frac{y}{R}+O(R),
\eea
which completely determine the leading contributions to $K$.

Using this information, the four dimensional metric (\ref{Dec11Ba}) can be rewritten in terms of two functions $\la_1$ and $\la_2$ which remain finite in the vicinity of the wall:
\bea\label{Dec11Bc}
ds^2_4&=&-2\eta dt+\d\db v+\la_1\eta^2+\la_2 dv^2
\eea

To avoid singularities in the ten--dimensional metric, the leading contributions to $\la_1$ and $\la_2$ should not depend on $(y,v)$.
Then regularity of the metric (\ref{Dec11Bc}) requires an existence of 
a holomorphic one--form $\xi$, such that
\bea\label{RegDegMetr}
\d{\bar\d}v+\la_1 dv^2+\la_2\eta^2=\xi{\bar\xi}+O(v).
\eea
Indeed, to make the metric (\ref{Dec11Bc}) regular, the matrix appearing in the left--hand side of the last equation should have rank two. This can only be accomplished if (anti)holomorphic terms $dz^adz^b$, $d\zb^ad\zb^b$ cancel out in the lhs of 
(\ref{RegDegMetr}), leading to holomorphicity of  $\xi$ ant to a relation 
$\la_2=4\la_1$. Moreover, on the surface $v=0$, $\xi$ describes a 
one--form in a two--dimensional space, so, by a change 
of coordinates, it can be always be written as $\xi=fdw$. An original  definition of $\xi$ implies that function $w$ is holomorphic.

Now equation (\ref{RegDegMetr}) can be rewritten as a relation 
between hermitean $2\times 2$ matrices, which should be supplemented by requiring $dv$ and $dw$ to be independent:
\bea\label{RegDegMetr1}
\d_a{\bar\d}_b v+\la \d_a v{\bar\d}_b v=g \d_a w \db_b\wb+O(v),\quad
\mbox{det}(\d_a v\d_b w)|_{v=0}\ne 0.
\eea
We conclude that a droplet whose boundary is defined by equation 
$v(z_a,\zb_a)=0$ leads to a smooth metric if and only if function $v$ satisfies (\ref{RegDegMetr1}). Notice that these relations are invariant under holomorphic reparametirizations which are regular on the $v=0$ surface.

To check the conditions (\ref{RegDegMetr1}), it is convenient to start
with evaluating function $\la$. To do so, we compute the determinant on both sides of the first relation in (\ref{RegDegMetr1}):
\bea\label{DetRegCond}
\mbox{det}(\d_a{\bar\d}_b v+\la \d_a v{\bar\d}_b v)=O(v)
\eea
Moreover, since $\d_a w$ and $\d_a v$ are independent,  
equation appearing in (\ref{RegDegMetr1}) implies that
\bea\label{DetRegCondAux}
\left.\mbox{det}(\d_a{\bar\d}_b v)\right|_{v=0}\ne 0.
\eea
Once function $\la$ is determined, one needs to check the remaining differential condition
\bea\label{DiffRegCond}
\d_a{\bar\d}_b v+\la \d_a v{\bar\d}_b v=\xi_a{\bar\xi}_b+O(v),\quad
\xi_a dz^a=f~dw+v\xi'.
\eea

The requirement (\ref{RegDegMetr1})
leads to rather nontrivial restrictions on the surfaces separating the 
droplets, and, to illustrate this fact, we consider few examples.

{\bf Examples of regular and singular droplets.}

The $AdS_5\times S^5$ example has already been discussed before, 
and, as a consistency check, we now demonstrate that conditions 
(\ref{DetRegCond}) and (\ref{DetRegCondAux}) are satisfied for 
that solution. Function $v$ for this case was introduced in 
(\ref{Dec10E}), so we find
\bea
&&v=z_a\zb_a-1,\quad 
M_\la=\| \d_a{\bar\d}_b v+\la \d_a v{\bar\d}_b v\|=
\left(\begin{array}{cc}
1+\la z_1\zb_1&\la z_2\zb_1\\
\la z_1\zb_2&1+\la z_2\zb_2
\end{array}\right),\nonumber\\
&&\mbox{det}M_\la=1+\la z_a\zb_a=1+\la+\la v
\eea
Thus the relation (\ref{DetRegCond}) is satisfied for $\la=-1$, and corresponding one--form is
\bea
\xi=z_2dz_1-z_1dz_2=z_1z_2~d\log\frac{z_1}{z_2},
\eea 
i.e. the requirement (\ref{DiffRegCond}) is also satisfied. 
As expected, we found the the wall located at  $z_a\zb_a=1$ leads 
to a regular solution. 

Inspired by this example, one may consider the most general quadratic function of $z_a$ and $\zb_a$:
\bea\label{Dec10GenQdr}
v=h_{ab}z_a\zb_b+A_{ab}z_az_b+{\bar A}_{ab}\zb_a\zb_b-B
\eea
It is clear that if $\mbox{det}h=0$, then (\ref{DetRegCondAux}) is violated, so such $v$ would lead to a singular solution. 
Assuming that $h_{ab}$ is a non--degenerate matrix, 
we can use linear transformations of $z_a$ 
to diagonalize it\footnote{Since the criteria of regularity 
(\ref{RegDegMetr1}) are invariant under holomorphic reparameterizations, they are not affected by such transformations.}: $h'_{ab}=\delta_{ab}$. The residual $U(2)$ invariance can be used to put $v$ in one of the two canonical forms 
parameterized by real numbers $a$, $b$:
\bea
\mbox{I}:&&v=z_a\zb_a+(az_1z_2+\frac{b}{2}z_1^2+cc)-B\nonumber\\
\mbox{II}:&&v=z_1\zb_1-z_2\zb_2+(az_1z_2+\frac{b}{2}z_1^2+cc)-B
\eea
Equation (\ref{DetRegCond}) can be easily analyzed in each 
case, and, requiring function $\la$ to remain finite at all points on the 
$v=0$ surface, we conclude that $\la$ must be constant:
\bea\label{Dec10Quad1}
\mbox{I}:&&\mbox{det}M_\la=1+\la\d_a v{\db}_a v=
1+\la|\zb_1+az_2+bz_1|^2+\la|\zb_2+az_1|^2\nonumber\\
&&\quad=1+\la(1+a^2)(v+B)+\la b^2z_1\zb_1+
\la\left[(1-a^2)(az_1z_2+\frac{b}{2}z_1^2)+abz_1\zb_2+cc\right]\nonumber\\
&&\mbox{det}M_\la=O(v):\quad b=0,\quad \la=-\frac{1}{B(1+a^2)},\quad a(1-a^2)=0\\
\label{Dec10Quad2}
\mbox{II}:&&-\mbox{det}M_\la=
1+\la|\zb_1+az_2+bz_1|^2-\la|\zb_2-az_1|^2\nonumber\\
&&\quad=1+\la(1-a^2)(v+B)+\la b^2z_1\zb_1+
\la\left[(1+a^2)(az_1z_2+\frac{b}{2}z_1^2)+abz_1\zb_2+cc\right]\nonumber\\
&&\mbox{det}M_\la=O(v):\quad b=0,\quad a=0,\quad 
\la=-\frac{1}{B}.
\eea

The first case gives two possible values of $a$: $a=0$ reduces to the case of the sphere which was discussed before, while $a=1$ leads 
to a surface described by the equation
\bea\label{SingulBndry}
v=|z_1+\zb_2|^2-B= Z\Zb-B,\quad Z\equiv z_1+\zb_2.
\eea
Let us check whether this function satisfies the relation 
(\ref{DiffRegCond}).
We begin with computing the one--form $\xi$ and its differential:
\bea\label{Jan22Xi}
&&\d\db v-\frac{1}{2B}|\d v|^2=\frac{1}{2B}|\Zb dz_1-Zdz_2|^2+O(v):
\nonumber\\
&&\xi=\frac{1}{\sqrt{2B}}(\Zb dz_1-Zdz_2)=
\frac{1}{\sqrt{2B}}(\Zb dz_1+Zd\zb_1-Zd\Zb),\\
&&d\xi=\frac{1}{\sqrt{2B}}(d\Zb\wedge dz_1+dZ\wedge d\zb_1-
dZ\wedge d\Zb).\nonumber
\eea
Equation (\ref{DiffRegCond}) implies that 
$d\xi\wedge\xi=dv\wedge \omega_2+O(v)$, and this relation is not satisfied by (\ref{Jan22Xi}). Indeed, $d\xi\wedge\xi$ has a contribution
\bea
(Zd\Zb-\Zb dZ)dz_1d\zb_1=dvdz_1d\zb_1-2\Zb dZ dz_1d\zb_1
\eea
which is not proportional to $dv$. Then we conclude that 
the surface defined by (\ref{SingulBndry}) does not satisfy 
the relation (\ref{DiffRegCond}), and thus it leads to a singular geometry.

Coming back to the relations (\ref{Dec10Quad1}), 
(\ref{Dec10Quad2}), we arrive at a conclusion any quadratic function
(\ref{Dec10GenQdr}) satisfying the regularity conditions 
(\ref{RegDegMetr1})
can be transformed by holomorphic reparameterizations into one of the following expressions:
\bea\label{RegulHyperb}
v_{{I}}=z_1\zb_1+z_2\zb_2-B,\quad 
v_{{II}}=z_1\zb_1-z_2\zb_2-B
\eea
This demonstrates that the relations (\ref{RegDegMetr1}) are very restrictive. 

Notice that exclusion of the wall (\ref{SingulBndry}) from the set of regular solutions is very important for correspondence between 
regular droplets and probe branes. Indeed, suppose 
(\ref{SingulBndry}) gave an allowed shape of the droplet. Then, sending 
$B$ to zero, one could collapse the droplet to a curve 
$z_1=-\zb_2$ which must carry some amount of D3--brane charge.
However, from the analysis of section 
\ref{SectIIBPrb}, we know that such object does not exist (recall that 
probe branes 
must follow holomorphic curves). This is the key difference between the present situation and the case of $1/2$--BPS bubbles discussed in 
\cite{LLM}: there the branes looked like pointlike sources and there were no restrictions on the shape of the droplets. 

We will now demonstrate that exclusion of surface (\ref{SingulBndry})
is a part of a general pattern: by collapsing a regular droplet in 
$1/4$--BPS case, one always arrives at a holomorphic curve. This fact
 provides a nontrivial correspondence between the regular supergravity solutions and brane probe analysis presented in section \ref{SectIIBPrb}. 

{\bf Collapsing droplets and holomorphic curves.}

Let us consider a family of regular surfaces $v_\eps(z_a,\zb_a)$ parameterized by $\eps$, and assume that at $\eps=0$ equation 
$v_\eps(z_a,\zb_a)=0$ describes a two--dimensional curve rather than 
a three--dimensional surface. Assuming that 
$v_\eps(z_a,\zb_a)$ is a smooth function of $\eps$ and that 
relations (\ref{RegDegMetr1}) are satisfied for any $\eps>0$, we will demonstrate that the two dimensional curve $v_0(z_a,\zb_a)=0$ must be holomorphic.

Near any point on the $v_0(z_a,\zb_a)=0$ curve, one can always make a holomorphic change of coordinates and rewrite the equation $v_0=0$ as
\bea
z_2=f(z_1,\zb_1).\nonumber
\eea
Since surfaces $v_\eps(z_a,\zb_a)=0$ must surround this curve, 
at small $\eps$ one can always write
\bea\label{Jan26Veps}
v_\eps(z_a,\zb_a)=|z_2-f(z_1,\zb_1)|^2-
|\eps{h}(z_a,\zb_a)|^2+
O(\eps^4).
\eea
Moreover, in a vicinity of the surface $v_\eps=0$, 
coordinates $(z_2,\zb_2)$ can 
be eliminated from function $h$.  

Let us first assume that $\mbox{det}(\d_a\db_b v_0)\ne 0$.
Then one can take $\eps\rightarrow 0$ limit in conditions 
(\ref{RegDegMetr1}). While doing this, it is important to introduce a scaling $\la\sim \eps^{-2}$, otherwise the matrix in the lhs of 
(\ref{RegDegMetr1}) would have a non--vanishing determinant. Introducing a finite ${\tilde\la}=\eps^2\la$, we find an equation:
\bea\label{Jan27V0Eqn}
M_{a{\bar b}}\equiv
\d_a\db_b v_0+(\eps^{-2}{\tilde\la}\d_a v\db_b v|_{v=0})_{\eps=0}=
g\d_a w\db_b \wb+O(v_0)
\eea
For small values of $\eps$, the surface $v_\eps=0$ can be parameterized by $z_1,\zb_1$ and a pure phase $\eta$:
\bea
z_2=f(z_1,\zb_1)+\eps \eta~h (z_1,\zb_1),\qquad {\bar\eta}\eta=1,
\eea
so we can compute the derivatives:
\bea\label{Jan27DerV}
\eps^{-1}\db_1 v_\eps |_{v=0}=
-\db_1 f{\bar\eta}{\bar h}-\db_1{\bar f}\eta h+
O(\eps),
\qquad
\eps^{-1}\db_2 v_\eps |_{v=0}=\eta h+O(\eps).
\eea
Substituting these expressions into (\ref{Jan27V0Eqn}) and taking the derivative of the left--hand side, we find an expression for 
${\tilde\la}(z_1,\zb_1,\eta)$. Factorizing (\ref{Jan27V0Eqn}) for this value of ${\tilde\la}$, we find
\bea
g^{1/2}d w=A_a(z_1,\zb_1,\eta)dz^a+O(v_0).
\eea
Since $z_2=f(z_1,\zb_1)$ on the $v_0=0$ curve, the last relation is only possible if eta--dependence factorizes in $A_a$: 
$A_a(z_1,\zb_1,\eta)=F(z_1,\zb_1,\eta){\tilde A}_a(z_1,\zb_1)$. This  means that, up to an overall coefficient, matrix $M_{a{\bar b}}$ is eta--independent.

Substituting the expressions (\ref{Jan27DerV}) into the definition 
(\ref{Jan27V0Eqn}) of $M_{a{\bar b}}$ and keeping track of the $\eta$--dependence and factors of ${\tilde\la}$, one can schematically write 
$M_{a{\bar b}}$ as
\bea\label{Jan27InterMb}
M_{a{\bar b}}=\left(
\begin{array}{cc}
\d_1\db_1 v_0+{\tilde\la}
(a_0+a_1\eta^2)({\bar a}_0+{\bar a}_1{\bar\eta}^2)&
\d_1\db_2 v_0+{\tilde\la}{\bar b}
(a_0+a_1\eta^2)\\
\d_2\db_1 v_0+{\tilde\la}{b}({\bar a}_0+{\bar a}_1{\bar\eta}^2)&
\d_2\db_2 v_0+{\tilde\la}{b}{\bar b}
\end{array}
\right).
\eea
Function ${\tilde\la}$ is determined by solving the equation 
$\mbox{det}(M_{a{\bar b}})=0$. Substituting the resulting value of 
${\tilde\la}$ back into (\ref{Jan27InterMb}), and requiring 
$M_{1{\bar 1}}/M_{1{\bar 2}}$ to be eta--independent, one 
concludes that 
$a_1=0$. This implies that ${\bar \d}_1f=0$, i.e. $f$ is a holomorphic 
function. Such conclusion falsifies our original assumption that 
$\mbox{det}(\d_a\db_b v_0)\ne 0$, so the matrix $\d_a\db_b v_0$ 
has to be degenerate. 

The condition $\mbox{det}(\d_a\db_b v_0)=0$ can be rewritten as
an equation for $f(z_1,\zb_1)$:
\bea
-\d_1\db_1 f(\zb_2-{\bar f})-\d_1\db_1 {\bar f}(z_2-{f})+
\d_1{\bar f}\db_1 f=0
\eea
Restricting this relation to the curve $v_0=0$, we find that 
$f$ must be holomorphic. This means that the droplet collapses to a curve
\bea
z_2=f(z_1),
\eea 

Thus, by utilizing small--$\eps$ analysis, we have shown that the droplets described by equation (\ref{Jan26Veps}) can only be regular 
if function $f(z_1,\zb_1)$ is holomorphic, this implies that regular droplets can only collapse to holomorphic curves. This conclusion is in a perfect agreement with discussion of sections \ref{SectIIBDbr} and 
\ref{SectIIBPrb}, where both probe analysis and consistency of SUGRA were used 
to demonstrate that supersymmetric D3--branes must follow holomorphic profiles. 

\

{\bf Summary}

Let us summarize the results of this long subsection. By requiring the 
geometries (\ref{DonosSoln}) to be regular, we arrived at the following 
picture for the boundary conditions in the $y=0$ hyperplane. This Kahler space is divided into a set of droplets, where Kahler potential satisfies one of the two conditions:
\bea\label{Jan31BndCnd}
y=0:\qquad\begin{array}{l}
Z=-\frac{1}{2},~ 
\d_a \db_b K(z,\zb,y=0)=0,\\
~\\
Z=+\frac{1}{2}
\end{array}
\eea
Notice that the relation $\d_a \db_b K(z,\zb,y=0)=0$ is crucial in enforcing regularity inside the $Z=-\frac{1}{2}$ droplets, and, as we 
will demonstrate in the next subsection, it is also needed to uniquely 
specify the solution of the Monge--Ampere equation. 

Moreover, in contrast to the boundaries of $1/2$--BPS droplets, which can be arbitrary \cite{LLM}, the domain walls in $1/4$--BPS case must obey a 
restriction coming from regularity. In particular, we demonstrated that 
there exists a real function $v(z_a,\zb_a)$ which defines the 
boundaries between droplets (via an equation $v=0$) and satisfies the 
differential relations (\ref{RegDegMetr1}):
\bea\label{RegDgMtr1Cp}
\d_a{\bar\d}_b v+\la \d_a v{\bar\d}_b v=g \d_a w \db_b\wb+O(v),\quad
\mbox{det}(\d_a v\d_b w)|_{v=0}\ne 0.
\eea
Here $w(z_1,z_2)$ is a holomorphic function. Looking at few examples, we showed that (\ref{RegDgMtr1Cp}) gives a nontrivial restriction on the shapes of the droplets. In particular, it was demonstrated that regular droplets can only collapse to holomorphic curves, in a perfect agreement of the D brane analysis presented in sections \ref{SectIIBDbr}, 
\ref{SectIIBPrb}.

%==============================================

\subsection{Asymptotic behavior and perturbative expansion}

\label{SectIIBPert}

%==============================================

In the previous subsection we discussed the behavior of the solutions near $y=0$ hyperplane and found that regularity imposes nontrivial 
boundary conditions on the Kahler potential. However, behavior at $y=0$ cannot fix the solutions on Monge--Ampere equation uniquely: 
one also has to specify Kahler potential at infinity. In this subsection we will discuss examples of large $R$ behavior which lead to metrics with interesting asymptotics, and we will demonstrate that, once the large--distance behavior is fixed, any combination of the boundary conditions 
(\ref{Jan23BC}), (\ref{Jan23BCa}) leads to the unique solution of 
Monge--Ampere equation.

The metrics which are most interesting from the point of view of 
AdS/CFT approach $AdS_5\times S^5$ at large values of $y$. As discussed in section \ref{SectIIBEx}, $AdS_5\times S^5$ can be embedded into the general $1/4$--BPS ansatz in two different 
ways, so, to describe an asymptotically-AdS space, a Kahler 
potential should approach either (\ref{KahlerAdS}) or (\ref{KahlAltAdS}) 
at large values of $y$. It turns out that in both cases a more natural way to impose asymptotic boundary conditions is to formulate them at large values of 
$R=\sqrt{r^2+y^2}$~\footnote{We recall that $r$ was introduced 
as a radial coordinate in the two--dimensional Kahler space, and,  generically being ambiguous, such coordinate is well--defined for spaces whose Kahler potential asymptotes to (\ref{KahlerAdS}) or 
(\ref{KahlAltAdS}).}
rather than $y$:
\bea\label{Jan23AdS}
K_{(\ref{KahlerAdS})}&=&
-\frac{1}{2}y^2\log y+\frac{r^2}{2}+O(\log R),\\
\label{Jan23AltAdS}
K_{(\ref{KahlAltAdS})}&=&
\frac{1}{2}y^2\log y+\frac{y^2}{2}\log(1-x_2^2)+\frac{1}{2}\log R+o(\log R).
\eea
We also recall that $r\rightarrow\infty$ has different geometric 
interpretations for (\ref{Jan23AdS}) and (\ref{Jan23AltAdS}): in the first case one goes to infinity of flat four--dimensional space, while in the second case large values of $r$ correspond to the boundaries and to the ends of the cylinder (see figure \ref{FigAdS5Two}b). It is useful to 
introduce 
special notation for the 
leading terms in (\ref{Jan23AdS}) and (\ref{Jan23AltAdS}):
\bea
K_I\equiv
-\frac{1}{2}y^2\log y+\frac{r^2}{2},\quad
K_{II}\equiv
\frac{1}{2}y^2\log y+\frac{y^2}{2}\log(1-x_2^2)+\frac{1}{2}\log R.
\eea
These Kahler potentials correspond to ${\bf C}^2$ with 
$Z=\frac{1}{2}$ and to a strip with $Z=-\frac{1}{2}$. 
Let us now perturb these solutions.

If some finite region with $Z=-\frac{1}{2}$ is added to the geometry 
described by $K_I$, the asymptotic behavior would remain unchanged, so at large distances 
one can treat the effects of the insertion as perturbation. In other words, we can write
\bea\label{Jan24Pert}
K=K_I+K^{(1)},
\eea
and at large distances (\ref{DonosSoln}) reduces to a linear 
equation for $K^{(1)}$:
\bea\label{Jan24PrtEqn}
\Delta_{z,\zb}K^{(1)}=2y^{-1}\d_y K^{(1)}-
y\d_y(y^{-1}\d_y K^{(1)}).
\eea
While this equation should not be trusted near $y=0$ (where $K_I$ and $K^{(1)}$ become comparable), one can formally extend the perturbation theory in the entire $y\ge 0$ region, keeping in mind that 
the series would converge only for large values of $y$. Such extension will allow us to count the number of degrees of freedom and to show that the boundary conditions (\ref{Jan23BC}), (\ref{Jan23BCa}) specify the 
solution uniquely (at least in perturbation theory).  

We begin with selecting some finite region $D$ of $y=0$ 
hypersurface and 
requiring that
\bea
y\d_y(y^{-1}\d_y K^{(1)})=2
\eea
there (this corresponds to setting $Z=-\frac{1}{2}$ in for 
(\ref{Jan24Pert})). Rewriting equation (\ref{Jan24PrtEqn}) in terms 
of $H=y\d_y(y^{-1}\d_y K^{(1)})$:
\bea
\Delta_{z,\zb}H+y^{-1}\d_y (y\d_y H)+\frac{4}{y^2}H=0,\quad
H|_{y=0}=\left\{\begin{array}{l}
2,\ (z,\zb)\in D\\
0,\ (z,\zb)\notin D,
\end{array}\right.
\eea
one can find a unique solution which vanished at infinity. However, there is still an ambiguity in function $K^{(1)}$: since at infinity we only require $K^{(1)}\ll K_I\sim R^2$, any harmonic function 
${\tilde K}^{(1)}(z,\zb)=o(R^2)$ would lead to a solution with correct asymptotics:
\bea\label{Jan24Amb}
K^{(1)}=\int_\infty^y ydy\int_\infty^y 
\frac{dy}{y}H+{\tilde K}^{(1)}(z,\zb):\qquad
\Delta_{z,\zb}{\tilde K}^{(1)}=0,\quad
\frac{K^{(1)}}{K_I}
\stackrel{{\footnotesize R\rightarrow\infty}}{\longrightarrow} 0.
\eea
Notice that this freedom in choosing ${K}^{(1)}$ is crucial for ensuring 
that the second regularity condition in (\ref{Jan23BC}) can be imposed in region $D$: to cancel a non--zero contribution of function $K_I$ at $y=0$, $(z,\zb)\in D$, one needs some ambiguity in ${\tilde K}^{(1)}$.
Moreover, the harmonic function ${\tilde K}^{(1)}$ and the unwanted 
contribution $K_0$ (which must satisfy the homogeneous 
Monge--Ampere equation (\ref{HomMonAmp})) to the Kahler potential 
have the same amount of freedom\footnote{To demonstrate this, one can 
consider a {\it formal} perturbation theory in (\ref{HomMonAmp}) by 
writing $K_0=\frac{1}{2}z_a\zb_a+\eps {\hat K}(z,\zb)$, 
truncating (\ref{HomMonAmp}) to the first order in $\eps$, and formally setting $\eps=1$. Then one finds that ${\hat K}$ has the same number of free parameters as a harmonic function. Of course, this argument is very heuristic, and one should not take the {\it solution} of the truncated equation seriously, however the {\it number of degrees of freedom} will not be affected by the higher--order terms.}, so it appears 
that the ambiguities in (\ref{Jan24Amb}) and in its higher--order counterparts can be used to remove $K_0$, and, once this is done, the perturbative expansion of function $K$ would be fixed uniquely. 

To fix the ambiguity in (\ref{Jan24Amb}), we will require $K^{(1)}$ to be analytic in the region $D$ and to vanish on all boundaries $\d D$. Due to the maximum principle, this uniquely determines harmonic function 
${\tilde K}^{(1)}(z,\zb)$ in the compact $D$, and we will set 
${\tilde K}^{(1)}=0$ on the complement of this region. Once function 
$K^{(1)}$ is determined, one can repeat the analysis for higher orders in perturbation theory, and again the ambiguity can be fixed order by order:
\bea
K^{(p)}|_{\d D,y=0}=0,\quad p>1.
\eea
While we only expect the perturbation series to converge at large values of $y$, the Kahler potential can be analytically continued to the entire $y>0$ subspace, and the result would satisfy the Monge--Ampere equation (\ref{DonosSoln}) as well as boundary conditions which were 
imposed order by order:
\bea
\left.-\frac{y}{2}\d_y(\frac{\d_y K}{y})\right|_{y=0}=\left\{
\begin{array}{rl}
-\frac{1}{2},&(z,\zb)\in D\\
\frac{1}{2},&(z,\zb)\notin D
\end{array}
\right.,\quad K|_{\d D,y=0}=0,\quad K=K_I+O(\log R).
\eea
Assuming that function $K_0\equiv \lim_{y\rightarrow 0} K$ remains regular inside the region $D$, we arrive at the equation 
(\ref{HomMonAmp}) along with a boundary condition:
\bea\label{Jan24aa}
(z,\zb)\in D:\quad
\d_1\db_1 K_0\d_2\db_2 K_0-\d_1\db_2 K_0\d_2\db_1 K_0=0,
\qquad
K_0|_{\d D,y=0}=0
\eea
We will now demonstrate that function $K_0$ vanishes in the region $D$, so the boundary condition (\ref{Jan23BC}) is satisfied.

As we discussed before, the homogeneous Monge--Ampere 
equation can be easily solved in terms of some holomorphic coordinate $w$: $K_0=K_0(w,\wb)$, then it is convenient to perform a holomorphic reparameterization:
$(z_1,z_2)\rightarrow (w,v)$. Assuming that this change of variables is regular inside $D$, we conclude that the image of $D$ in 
$(w,v,\wb,{\bar v})$ space is compact. Starting with an arbitrary point 
$(w_0,v_0)\in D$, one can consider a complex plane $w=w_0$. Due 
to compactness of $D$, this plane must intersect the boundary $\d D$ 
along some hypersurface, then 
$K_0(w_0,\wb_0)=0$. Since $(w_0,v_0)\in D$ was arbitrary, we conclude that $K_0|_{\d D}=0$ implies $K_0|_D=0$. 
This statement can be interpreted as a "maximum principle" for the homogeneous Monge--Ampere equation. Of course, our arguments were rather heuristic, but they can be made precise. 

To summarize, we showed that starting with solution $K_I$ whose boundary conditions correspond to ${\bf C}^2$ with $Z=\frac{1}{2}$, 
and introducing an arbitrary distribution of compact droplets with 
$Z=-\frac{1}{2}$, one can use perturbation theory to construct a solution which satisfies boundary condition (\ref{Jan23BC}) inside the droplets and condition (\ref{Jan23BCa}) outside. This implies that, for any distribution of 
droplets, perturbation theory leads to a unique regular geometry. Notice that the differential restriction in (\ref{Jan23BC}) was crucial both for enforcing regularity and for ensuring uniqueness of the solution. A similar 
perturbation theory can be developed around solution 
(\ref{Jan23AltAdS}), in this 
case one introduces compact droplets with 
$Z=\frac{1}{2}$ and requires $\d_a\db_b K_0$ to vanish in the exterior
 of the droplets. It would also be interesting to study geometries with more exotic asymptotics: in the 1/2--BPS case, where the system was exactly solvable \cite{LLM}, such solutions were discussed in 
\cite{sheppard}.

%=====================================

\subsection{Topology and charges} 

\label{SectIIBTop}

%=====================================

The bubbling solutions discussed in this section have regular metrics and source--free field strengths, so, to allow non--zero fluxes, the geometries must have nontrivial topology. In this subsection we will explore the topological structure of $1/4$--BPS solutions and show
that they indeed contain some non--contractible 
five--cycles. We will also demonstrate that the integrals of $F_5$ over such cycles give non--zero answers, and discuss some global restrictions on the distributions of the droplets imposed by 
quantization of charge.

%==================

\begin{figure}[tb]
\begin{tabular}{ccc}
\epsfxsize=2.2in \epsffile{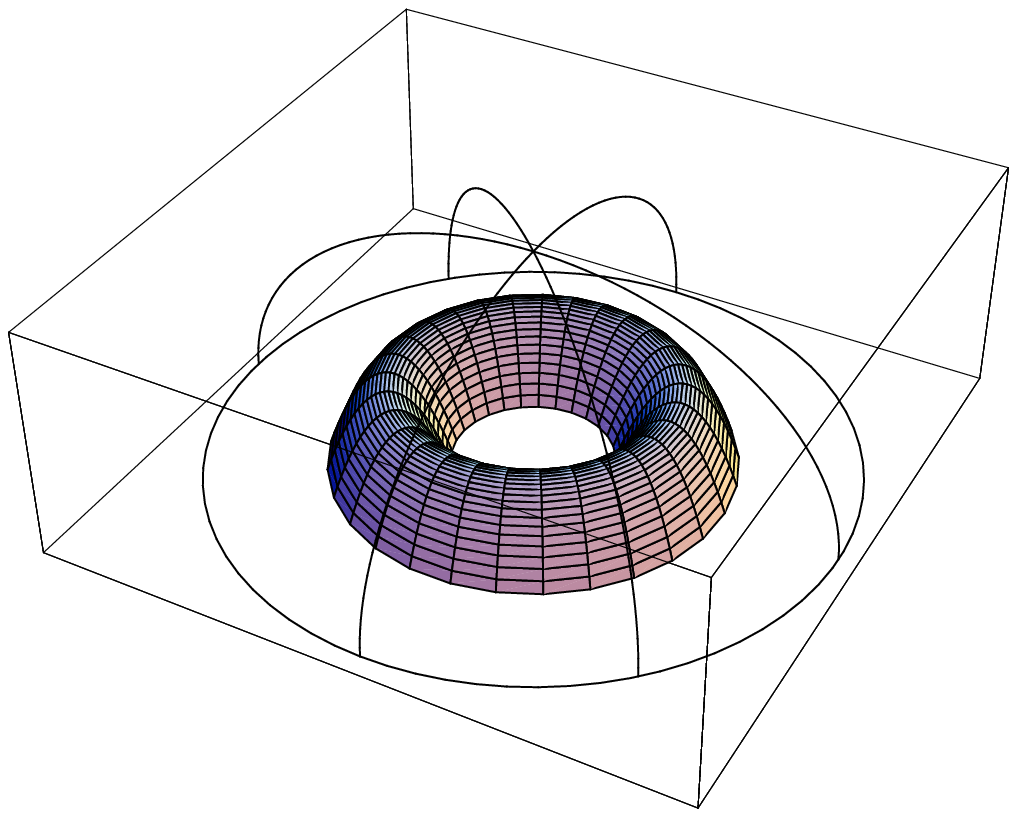}&
\qquad\qquad&
\epsfxsize=2.2in \epsffile{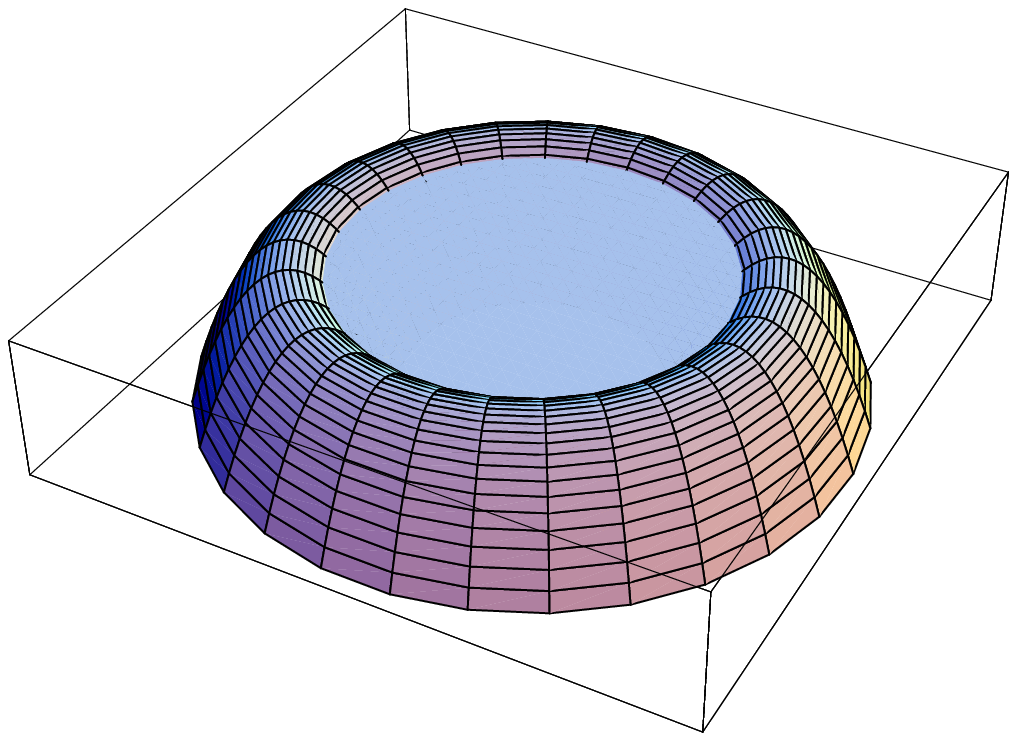}\\
(a)&&(b)
\end{tabular}
\caption{
Topology of the 1/4 geometries. 
To construct a non--contractible cycle, 
one should start with a four--dimensional "cap" which ends on a 
three--dimensional surface surrounding a compact droplet (a), and fiber 
$\psi$ over the cap. The projection of the cap onto the Kahler space fills 
the interior of the sphere depicted in figure (a). 
Alternatively, one can start with a cap ending on a curve inside 
$Z=-\frac{1}{2}$ region, and fiber the three--sphere over it. The projection of the cap onto the Kahler space looks like a "membrane" depicted in 
figure (b).
} \label{FigTop}
\end{figure}

%======================

To analyze the topology of the solution (\ref{DonosSoln}), we recall that  
either $S^3$ or $S^1$ collapses at $y=0$ hyperplane. This leads to  
two simple constructions of non--contractible five--cycles.

\ 

I. If there exists a compact region with $Z=-\frac{1}{2}$ (i.e. with collapsing $S^3$) in $y=0$ hyperplane, then such droplet can be surrounded by a three--dimensional surface ${\cal S}$ which lies entirely in the 
$Z=\frac{1}{2}$ region (see figure \ref{FigTop}a). Constructing a 
four--dimensional "cap", which goes to $y>0$ and has ${\cal S}$ as 
its boundary, and fibering $\psi$ over it, 
one arrives at a five--dimensional surface $\Omega$. Notice that, since radius of 
$\psi$--direction goes to zero on ${\cal S}$, the surface $\Omega$ is compact: restricting the metric (\ref{DonosSoln}) to $\Omega$ in a vicinity 
of ${\cal C}$, one finds a regular geometry without a boundary\footnote{We wrote the four--dimensional Kahler metric appearing in (\ref{DonosSoln}) 
as $2\d_a\db_b K_0dz^ad\zb^b=ds_3^2+dx_\perp^2$, where $x_\perp$ is a direction orthogonal to ${\cal S}$.}:
\bea
ds_5^2=ds^2_{10}|_\Omega=g^{-1}\left[ds_3^2+dy^2+y^2d\psi^2\right].
\eea
Moreover, this construction does not allow the surface ${\cal S}$ to  move into $Z=-\frac{1}{2}$ region ($\psi$--direction does not shrink there), so the five--cycle $\Omega$ is non--contractible. 
An analogous "bubbling" effect was described in \cite{LLM}. 

\

II. A construction of five--cycles containing $S^3$ is 
less--straightforward, since there are no compact droplets with 
$Z=-\frac{1}{2}$~\footnote{One might think that the spherical droplet
(\ref{AltAdS5Dropl}) arising from $AdS_5\times S^5$ has a compact 
interior, but,
since it touches the boundary of the space ($|z_2|=1$), this droplet cannot be surrounded by any three--dimensional surface.}. 

To get some intuition, we consider the $1/2$--BPS geometries of \cite{LLM}. As discussed in section \ref{SectIIBEx}, these solutions can be embedded in the general $1/4$--BPS ansatz, and the boundary between droplets is invariant under phase rotations of $z_2$ (see 
(\ref{Bbl2BcIIB})). This allows a very simple pictorial representation of droplets in terms of coordinates 
$(z_1,\zb_1,|z_2|^2)$, and one can argue that, while there is only one connected region with $Z=-\frac{1}{2}$, generically this region is not simply--connected (an example of map from $(z_1,z_2)$ plane to the boundary of the droplet in $(z_1,\zb_1,|z_2|^2)$ space is presented in figure \ref{Fig3Dbbl}). As was shown in \cite{LLM}, one can get a 
non--trivial five--cycle by taking a non--contractible curve 
${\cal C}$ in ${\tilde Z}=\frac{1}{2}$ region, constructing a two--manifold which ends on ${\cal C}$, and fibering $S^3$ over it. 
To recover $\Omega$ in the context of $1/4$--BPS geometries, we 
select a non--contractible curve ${\cal C}$ in $y=0,z_2=0$ plane, construct a two dimensional surface $\omega$ which ends on ${\cal C}$ and explores $y>0$, $z_2=0$ region, and fiber $S^3$ over 
$\omega$. 
A surface $\omega$ and a curve ${\cal C}$ can be moved to non--zero values of $z_2$, but, as long as ${\cal C}$
stays inside the ${Z}=\frac{1}{2}$ droplet, the resulting five--cycle 
$\Omega$ has no boundary. 
Since the original curve ${\cal C}$ was non--contractible, the projection of $\Omega$ onto Kahler manifold covers a "hole" in the 
${Z}=\frac{1}{2}$ droplet, so $\Omega$ is a non--contractible cycle. An example of two--dimensional surface $\omega$ is presented in figure 
\ref{FigTop}b.

Using the $1/2$--BPS example as a guide, we can propose a general way of getting five--cycles involving the $S^3$. If the region 
${Z}=\frac{1}{2}$ is not simply--connected, it contains a 
non--contractible curve ${\cal C}$. Then one can build a 
two--dimensional surface $\omega$ which stays at positive values of $y$ and has ${\cal C}$ as its boundary. Fibering the 
three--sphere over 
$\omega$, one gets a non--contractible five--cycle $\Omega$. Looking at the metric (\ref{DonosSoln}), one can see that $\Omega$ has no 
boundary at $y=0$, so this five--cycle has a topology of $S^5$.
It appears that the five--spheres I and II are the only 
topologically--nontrivial cycles in the geometries (\ref{DonosSoln}).

\

Since the geometries contain non--contractible five spheres, and 
a non--zero five--form is present in the solution, it is natural to evaluate
the integrals $F_5$ over the cycles. The amount of flux is invariant under small deformations of cycles, so it is convenient to choose these surfaces to make the computation easier. The four--dimensional "cap" surrounding $Z=-\frac{1}{2}$ region can be deformed into a surface located at small values of $y$, then $\Omega$ is parameterized by 
$(z_a,\zb_b)$ and $\psi$.  The metric at small values of $y$ has already been analyzed in section \ref{SectIIBDrpl},
so we find
\bea\label{Jan26Flx1}
ds_{10}^2&=&h^{-2}\left[-(dt+\omega)^2+d\psi^2+
2\d_a\db_b K_1dz^ad\zb^b\right]+h^2(dy^2+y^2d\Omega_3^2)
\nonumber\\
F_5&=&-d[y^4h^4(dt+\omega)]\wedge d\Omega_3+
h^{-4}d\psi\wedge \mbox{det}(\d_a\db_b K_1)d^2 zd^2\zb+\dots
\eea
The field strength contains additional terms at the same order, but 
they will be irrelevant for out computation. Integrating $F_5$ over the 
$Z=-\frac{1}{2}$ region, one finds a very simple expression for the
flux in terms of the volume of the droplet:
\bea\label{FluxIIB}
ds_{10}^2&=&h^{-2}[-(dt+\omega)^2+d\psi^2]+
g^{(4)}_{ij}dx^idx^j+h^2(dy^2+y^2d\Omega_3^2)\nonumber\\
\int_{\Omega_5}F_5&=&2\pi V_4\equiv 2\pi \int \sqrt{g^{(4)}}d^4 x.
\eea
A counterpart of this formula for $1/2$--BPS geometries was encountered in \cite{LLM}, where the agreement with field theory picture \cite{beren} 
was also demonstrated. Unfortunately the matrix model in the 1/4--BPS case is more complicated, but it would be nice to interpret (\ref{FluxIIB}) 
in terms of eigenvalues of holomorphic matrices $X$ and $Y$.

Let us now turn to the type II cycles. Again, they can be moved to 
small values of $y$, but now one also has freedom in deforming the
contour ${\cal C}$ inside $Z=-\frac{1}{2}$ bubble. 
Integrating the expression for $F_5$, we find the flux:
\bea\label{Jan26Flx2}
\int_{\Omega_5} F_5=
\left.4i\pi^2\int_\omega \d_a\db_b Kdz^ad\zb^b\right|_{y=0}:\qquad
\d \omega={\cal C}.
\eea
Notice that the contribution to the integral comes only from the part of $\omega$ which lies inside $Z=\frac{1}{2}$ region: $\d_a\db_b K=0$ otherwise. Moreover, it is clear that the integral in (\ref{Jan26Flx2}) 
is invariant under small deformations of $\omega$.

\

To summarize, we found that the $1/4$--BPS bubbling solutions have very interesting topological structure: they can contain two types of non--contractible five cycles, and the fluxes through such spheres have very simple geometrical meaning. Since the charges computed in (\ref{Jan26Flx1}), (\ref{Jan26Flx2}) must be quantized, one finds a 
set of global restrictions on the "volumes" of the droplets and  two--cycles $\omega$. 
It would be very nice to connect this data with field theory picture developed in \cite{beren4}.

%==================================

\subsection{Relation to brane webs}

\label{SectIIBCmpWb}

%==================================

By looking at the explicit form of the Killing spinor for the geometry 
(\ref{DonosSoln}), one can see that the isometry $\d_t$ is 
rotational.
Since the timelike Killing vector in flat space generates translational isometry, the system (\ref{DonosSoln}) cannot describe asymptotically--flat solutions. However, the flat asymptotics can arise as a result of certain singular limits and in this subsection we will discuss two interesting possibilities. The first one leads to standard D3 branes with flat worldvolume, while the second limit reproduces the geometry 
(\ref{WebD3Metr}) produced by the webs of D3 branes. 

The Killing spinor has a very simple time dependence
($\eps\sim e^{it/2}$), which can be removed by rescaling the 
$t$-coordinate: $t=\la^{-1}{\tilde t}$, and sending $\la$ to infinity while keeping ${\tilde t}$ fixed. To keep $g_{\tilde t\tilde t}$ finite, we also need to rescale $h^{-2}=\la^2 {\tilde h}^{-2}$ and there are two natural ways to do it.

{\bf 1. The limit of flat branes.}

Let us consider a shift $e^G=\la e^{{\tilde G}}$ and send 
$\la$ to infinity, while keeping ${\tilde G}$ finite. To retain the canonical periodicity of $\psi$, we will also rescale $y=\la{\tilde y}$. Then the leading terms in $\frac{1}{\la}$ expansion become
\bea
h^{-2}=\la^2 {\tilde y}e^{\tilde G},\quad 
Z=\frac{1}{2}-\la^{-2}e^{-2{\tilde G}},\quad 
K=\la^2 K_{-1}(z)+K_0(y)+\la^{-2}K_1.
\eea
Equation relating Kahler potential and $Z$ implies that
\bea
y^{-1}\d_y K_0=-\log {\tilde y}+C
\eea
To make the metric finite, we rescale the coordinates on $S^3$, then, dropping tildes, we arrive at the solution:
\bea
ds_{10}^2&=&ye^Gdx_{1,3}^2+(ye^G)^{-1}\left[2
\d_a{\bar\d}_b K_{-1}dz^ad\zb^b+dy^2+y^2d\psi^2\right],
\nonumber\\
&&\mbox{det} h_{a{\bar b}}=\frac{1}{4}.
\eea
This geometry describes a two--dimensional Calabi-Yau manifold 
and a set of branes whose distribution is governed by function $H=(ye^G)^{-2}$.
In this order the equation for the metric decouples and we have to 
look at the equation of motion for $F_5$ to conclude that $H$ is harmonic. The branes are occupying points in the transverse space.

{\bf 2. D3 branes wrapping holomorphic cycles}

An alternative flat limit can be obtained by a rescaling which keeps 
the radius of the sphere finite: $e^G=\la^{-1} e^{{\tilde G}}$, 
$y=\la {\tilde y}$. Then we find
\bea
h^{-2}&=&\la^2 ye^{-G},\quad Z=-\frac{1}{2}+\la^{-2}e^{2G},\quad
\psi=\la^{-1}{\tilde\psi},\quad t=\la^{-1}{\tilde t}
\nonumber\\
ds^2&=&ye^{-G}(-dt^2+d\psi^2)+y^{-1}e^G\left[dy^2+y^2 d\Omega_3^2\right]+2y^{-1}e^{-G}\d_a{\bar\d}_b K dz^ad\zb^b
\eea
The Kahler potential is not rescaled and it satisfies the following equations:
\bea
&&K=\la^2 K_{-1}(y)+K_0,\quad \d_y K_{-1}=y\log y,\quad 
e^{2G}=-\frac{1}{2}y\d_y(y^{-1}\d_y K_0),\nonumber\\
&&\mbox{det}~h_{a{\bar b}}=\la^{-1}y^2 e^{2G}W{\bar W}\quad
\rightarrow\quad \mbox{det}~h_{a{\bar b}}=\frac{1}{4}y^2 e^{2G}
\eea
The gauge $W=\frac{1}{2}\la^{1/2}$ was chosen to avoid a singularity in the metric. To remove an explicit $y$--dependence of the metric, we define $H=y^{-1}e^G$ and rescale the Kahler potential: 
$K_0=y^2 {\tilde K}$. Then we arrive to the following geometry:
\bea
ds^2&=&H^{-1}(-dt^2+d\psi^2)+H\left[dy^2+y^2 d\Omega_3^2\right]
+2H^{-1}\d_a{\bar\d}_b {\tilde K} dz^ad\zb^b\nonumber\\
H^2&=&-\frac{1}{2}y^{-3}\d_y(y^ 3\d_y {\tilde K}),\qquad 
\mbox{det}~h_{a{\bar b}}=\frac{1}{4} H^2
\eea
This solution goes over to (\ref{WebD3Metr}), (\ref{MembrMAH}), 
(\ref{MembrMAN}) if one identifies $H$ with 
$e^{-3A/2}$. The expressions for the RR five--form also agree. 

%=====================================

\subsection{Summary}

\label{SectIIBSumm}

%=====================================

Let us summarize the results of this section. While the local description of 1/4--BPS bubbling geometries (\ref{DonosSoln}) has been discussed 
before 
\cite{donos}, the allowed boundary conditions were not known. In particular, 
a recent proposal by \cite{vaman} appeared to be incomplete since it 
clearly disagreed with expectations from the probe analysis (by shrinking 
the droplets of \cite{vaman}, one could arrive at sources which are not 
allowed in string theory). By requiring the geometries of \cite{donos} to be 
regular at the hyperplane $y=0$, we found an improved version of the 
picture proposed in \cite{vaman}: the hyperplane is divided into regions 
where one of the conditions (\ref{Jan31BndCnd}) should be satisfied.
Moreover, in contrast to 1/2--BPS case, the shapes of the droplets 
cannot be arbitrary, but rather they are described by an equation 
$v(z_a,\zb_a)=0$, and function $v$ must satisfy the relations 
(\ref{RegDgMtr1Cp}). We demonstrated that these restrictions on $v$ are consistent with results of probe analysis which requires the brane profiles to be holomorphic. We also showed that the allowed locations of D3 branes can be determined either from open string computations (which reduces to a DBI analysis) or from consistency of supergravity equations (this 
amounts to a description in terms of close strings), and there is a perfect agreement between these independent results. As a by-product of the probe analysis, we gave a clear geometric interpretation of Mikhailov's description of giant gravitons \cite{mikhail}. 

Once the distribution of droplets in $y=0$ plane is specified, one can try to solve a nonlinear Monge--Ampere equation to construct the corresponding 
geometry. While we were not able to find new explicit solutions, we used perturbation theory to demonstrate that, for a fixed asymptotic behavior,  conditions (\ref{Jan31BndCnd}) specify the solution uniquely, and any allowed distribution of droplets leads to a regular geometry without 
sources. All fluxes are geometric: we showed that solutions contain 
non--contractible five--cycles and evaluated fluxes through them (see equations (\ref{FluxIIB}), (\ref{Jan26Flx2})). 

We also considered some explicit examples of 1/4--BPS geometries, in particular we showed that one can embed $AdS_5\times S^5$ into an ansatz (\ref{DonosSoln}) in two different ways, and we also embedded all 1/2--BPS solutions constructed in \cite{LLM}. Finally, by taking a limit of bubbling solution (\ref{DonosSoln}), we recovered the web of D3 branes which was discussed in section \ref{Sect35Brn}.

%=====================================

\section{1/4--BPS geometries in M theory.}

\label{SectM4geom}

\renewcommand{\theequation}{7.\arabic{equation}}
\setcounter{equation}{0}

%=====================================

In the previous section we discussed 1/4--BPS geometries with 
$AdS_5\times S^5$ asymptotics. The interest in such metrics is driven by their relation to supersymmetric states in four dimensional field theory. However, there are other important cases of AdS/CFT correspondence and understanding of eleven dimensional bulk configurations might shed some light on the dynamics of $(2,0)$ six--dimensional CFT and of conformal 
theory on M2 brane.  In this section 
we will discuss $1/4$--BPS geometries in M theory. As we will see, 
they share many properties of their ten--dimensional counterparts, in particular, the discussion of droplets and their boundaries would essentially mimic the arguments presented in the previous section. It is interesting to notice that, while in the $1/2$--BPS case the equation governing ten--dimensional geometries was much easier than its M theory counterpart \cite{LLM}, the Monge--Ampere equations describing the 1/4--BPS geometries appear to have the same degree of difficulty in ten and 
eleven dimensions. 

We will look at a particular set of geometries which preserve $SO(6)$ symmetry. They could correspond either to supersymmetric states in (2,0) theory on $R\times S^5$, or to some space--dependent configurations in three--dimensional CFT. It might be useful to recall that 1/2--BPS solutions of \cite{LLM} had $SO(6)\times SO(3)$ symmetry, so now we are breaking half of the supercharges which were used to produce "translations" along $SO(3)$. 

%=====================================

\subsection{Local structure of the solution}

\label{SectM4Loc}

%=====================================

As already mentioned, we want to study eleven dimensional geometries which preserve $8$ supercharges as well as bosonic  
$SO(6)$ symmetry. Fortunately, the local structure of the solution can be easily extracted from the results of \cite{GMSW}. 
 
We recall that authors of \cite{GMSW} constructed the most general solution of eleven dimensional supergravity which contains $AdS_5$ factor\footnote{To avoid unnecessary complications and to make contact with notation introduced in \cite{LLM}, we set $m=\frac{1}{2}$ 
in the formulas of \cite{GMSW} and rewrite the solution in terms of AdS space with unit radius.}:
\bea\label{GMSWmetr}
ds^2&=&4e^{2\la}ds_{AdS}^2+e^{-4\la}\left(
h_{ij}dx^idx^j+\frac{dy^2}{\cos^2\zeta}\right)+
\frac{4e^{2\la}}{9}\cos^2\zeta(d\psi+\rho)^2\\
F_4&=&-(\d_y e^{-6\la})V_4+
\frac{1}{\cos^2\zeta}(*_4d_4 e^{-6\la})dy-\frac{4}{9}\cos^4\zeta 
(*_4\d_y\rho)(d\psi+\rho)\nonumber\\
&&+\left[\frac{4}{9}\cos^2\zeta *_4 d_4\rho-\frac{4}{3}e^{-6\la}J\right]
dy(d\psi+\rho).\nonumber
\eea
As shown in \cite{GMSW}, the four dimensional metric $h_{ij}$ is Kahler and there are various differential relations between the metric components and the Kahler form $J$:
\bea
&&y=e^{3\la}\sin\zeta,\quad \rho=
{J}\cdot d_4\left[\frac{1}{2}\log(\cos^2\zeta\sqrt{h})\right],\quad
\d_y J=-\frac{2}{3}yd_4\rho\nonumber\\
&&\d_y\log\sqrt{h}=-3y^{-1}\tan^2\zeta-2\d_y\log\cos\zeta
\eea
Notice that, as a consequence of supersymmetry, $\d_\psi$ turns 
out to be a Killing vector for the geometry (\ref{GMSWmetr}). 
It is convenient to introduce complex coordinates $z_a$, $\zb_a$ 
then the relations between metric, Kahler form and Kahler potential become especially simple:
\bea
h_{ij}dx^idx^j=2h_{a{\bar b}}dz^a d\zb^b,\quad 
J=ih_{a{\bar b}}dz^a\wedge d\zb^b,\quad h_{a{\bar b}}=\d_a\db_b K
\eea 

The geometries with $SO(6)$ symmetries can be obtained from 
(\ref{GMSWmetr}) by performing the following analytic continuation:
\bea
ds^2_{AdS}\rightarrow-d\Omega_5^2,\quad 
e^{\la}\rightarrow ie^{\la},\quad \zeta\rightarrow i\zeta,\quad 
\psi\rightarrow t,
\eea
and the resulting solution reads:
\bea\label{Mbubble1}
ds^2&=&-\frac{4e^{2\la}}{9}\cosh^2\zeta(dt+\rho)^2+
4e^{2\la}d\Omega_5^2+e^{-4\la}\left(
2h_{a{\bar b}}dz^ad\zb^b+\frac{dy^2}{\cosh^2\zeta}\right),\nonumber\\
F_4&=&(\d_y e^{-6\la})V_4-
\frac{1}{\cosh^2\zeta}(*_4d_4 e^{-6\la})dy-\frac{4}{9}\cosh^4\zeta 
(*_4\d_y\rho)(dt+\rho)\\
&&+\left[\frac{4}{9}\cosh^2\zeta *_4 d_4\rho+\frac{4}{3}e^{-6\la}J\right]
dy(dt+\rho).\nonumber
\eea
Various functions appearing in this solution satisfy a system of differential equations\footnote{We introduced a 
convenient function $D$ whose normalization is chosen to agree with 
results of \cite{LLM} for the 1/2--BPS case.}:
\bea\label{GMSWanCont}
&&\sinh\zeta=ye^{-3\la},\quad \rho=\frac{3i}{2}(\d D-\db D),\quad
y^{-1}\d_y J=2i\d\db D,\nonumber\\
&&\d_y D=y^{-1}\tanh^2\zeta=\frac{ye^{-6\la}}{1+y^2 e^{-6\la}},
\quad e^{3D}=4\cosh^2\zeta~\mbox{det} h_{a\bar b}.
\eea
Considering the equation for the $y$--dependence of the Kahler potential $K$, we find a relation between $K$ and $D$:
\bea
\d\db\left[y^{-1}\d_y K-2D\right]=0.
\eea
Since metric is not affected by addition of an (anti)holomorphic function to Kahler potential, we can choose a gauge where
\bea\label{Jan31DtK}
D=\frac{1}{2y}\d_y K.
\eea 
At this point the entire solution is completely specified by the 
Kahler potential (although some expressions are simpler in 
terms of $D$):
\bea\label{Mbubble2}
e^{-6\la}=\frac{y^{-1}\d_y D}{1-y\d_y D},\quad \tanh^2\zeta=y\d_y D,\quad \rho=\frac{3i}{2}(\d-\db)D,\quad D=\frac{1}{2y}\d_y K.
\eea
The last remaining equation relates $D$ and 
$\mbox{det}h_{a{\bar b}}$:
\bea\label{Mbubble3}
4\mbox{det}h_{a{\bar b}}=(1-y\d_y D)e^{3D}=
\left[1-\frac{1}{2}y\d_y(y^{-1}\d_y K)\right]
\exp\left(\frac{3}{2y}\d_y K\right).
\eea
This is an M theory counterpart of the Monge--Ampere equation 
(\ref{DonosSoln}). 

To summarize, the most general eleven--dimensional solution preserving $8$ supercharges and $SO(6)$ isometry is given by 
(\ref{Mbubble1}), it is completely specified by the Kahler potential $K$ 
(see (\ref{Mbubble2})) which satisfies a Monge--Ampere equation 
(\ref{Mbubble3}). In the remaining part of this section we will study the system (\ref{Mbubble1}), (\ref{Mbubble2}) (\ref{Mbubble3}) in more detail. We begin with discussing some known solutions which are covered by the general ansatz (\ref{Mbubble1}). 

%=====================================

\subsection{Examples}

\label{SectM4Ex}

%=====================================

The simplest supersymmetric geometries with $SO(6)$ symmetry are 
$AdS_4\times S^7$ and $AdS_7\times S^4$, so it seems natural to 
discuss them first. However, to avoid repetition, we will begin with 
embedding a more general class of solutions preserving only $16$ supercharges, and then come back to $AdS_p\times S^q$. 

\subsubsection{1/2--BPS bubbling solutions.}

The most general $1/2$--BPS solution of M theory was constructed in 
\cite{LLM}, where it was shown that the metric and fluxes are uniquely 
determined in terms of one function ${\tilde D}(z,\zb,x)$ which 
satisfies a continual Toda equation:
\bea\label{TodaEqn}
\d_x^2 e^{\tilde D}+4\d_z\d_{\zb}{\tilde D}=0.
\eea
The explicit form of the solution is reviewed in the Appendix 
\ref{SectApEmbM} (see equation (\ref{LLMthry})), here we will only need 
to recall that the 
$1/2$--BPS metrics have $SO(6)\times SO(3)$ isometries and coordinate $x$ is related to the radii of five-- and three--spheres in a very simple way: $x=\frac{1}{2}R_2R_5^2$. Moreover, 
regularity restricts the allowed boundary conditions for the normal
derivative of ${\tilde D}$ at $x=0$:
\bea\label{TodaBCond}
x=0:\qquad
\begin{array}{l}\d_x{\tilde D}=0,~ {\tilde D}-\mbox{finite}\\
\d_x{\tilde D}=\frac{1}{x}+O(1)
\end{array}
\eea
Thus the entire $x=0$ plane is divided into two types of regions with arbitrary curves separating them (see figure \ref{Fig2Bbl}). 

In the Appendix \ref{SectApEmbM} we embed the $1/2$--BPS solutions into the more general geometry (\ref{Mbubble1}), (\ref{Mbubble2}), 
(\ref{Mbubble3}) which is parameterized by one function 
$D(z_a;\zb_a;y)$. The relation between 1/2--BPS and 1/4--BPS variables is given by (\ref{MthryMap}):
\bea\label{MainMthryMap}
D(z,We^{-i\phi};\zb,We^{i\phi};y)={\tilde D}(z,\zb;x),\quad 
y=x\cos\theta,\quad 
W=x\sin\theta e^{-{\tilde D}}.
\eea
Here coordinates $\theta$ and $\phi$ parameterize the two--sphere in the $1/2$--BPS solution:
\bea
d\Omega_2^2=d\theta^2+\sin^2\theta \left[d(\phi-\frac{2t}{3})\right]^2.
\eea
To find interpretation of the boundary conditions (\ref{TodaBCond}),
we recall the relation (\ref{RelXYMder}):
\bea
\d_y {D}=\frac{\cos\theta\d_x {\tilde D}}{1-\sin^2\theta x\d_x{\tilde D}},
\qquad x=0:\quad
\begin{array}{l}
\d_x{\tilde D}=0\rightarrow \d_y D=0\\
\d_x{\tilde D}=\frac{1}{x}\rightarrow \d_y D=\frac{1}{y}
\end{array}
\eea
Now it is natural to impose boundary conditions at the hypersurface where $y=x\cos\theta=0$: it consists of the regions described by the last equation as well as points where $\cos\theta=0$ and $\d_y D=0$.
Thus we see that the lift of 1/2--BPS geometries gives the solutions 
of (\ref{Mbubble1}), (\ref{Mbubble2}), (\ref{Mbubble3}), and for regularity the $y=0$ surface is divided into regions with two types of 
boundary conditions:
\bea\label{MBblBCond}
y=0:\qquad
\begin{array}{cl}
\mbox{I}:&\d_y{D}=\frac{1}{y}+O(1)\\
\mbox{II}:&\d_y{D}=0,~ {D}-\mbox{finite}
\end{array}
\eea
As we will see in section \ref{SectM4BC}, regularity also imposes some restrictions on the derivatives of Kahler potential in the region I.

The boundary between the regions can be found by repeating the logic 
which led to (\ref{Bbl2BcIIB}):
\bea
z_2\zb_2=e^{-2{\hat D}(z,\zb)},\quad {\hat D}={\tilde D}-\log x.
\eea

While there are many similarities in the description of $1/2$--BPS solutions in IIB string theory and in the eleven--dimensional supergravity, on the technical level Toda equation (\ref{TodaEqn}) is much more complicated than its IIB counterpart (\ref{LaplLLM}). In particular, 
for the Laplace equation one can easily write down an explicit solution corresponding to the most general boundary condition 
(\ref{12BPSoldBC}), while it is not clear how to do so for the system (\ref{TodaEqn}), 
(\ref{TodaBCond}). While we believe that the appropriate solution does exist for any distribution of droplets, only few explicit solutions are know, and now we will discuss their embedding into the $1/4$--BPS 
ansatz in more detail.  

\subsubsection{pp--wave and $AdS_p\times S^q$.}

The simplest solution of the Toda equation corresponds to the 
eleven--dimensional pp-wave \cite{LLM}:
\bea
x=\frac{1}{4}u^2v,\quad z-\zb=i\left(\frac{u^2}{2}-v^2\right),\quad
e^{\tilde D}=\frac{u^2}{4}
\eea
Using (\ref{MainMthryMap}), we can introduce the variables appropriate 
for the 1/4--BPS case:
\bea
y=\frac{1}{4}u^2 v\cos\theta,\quad z_2=v\sin\theta e^{-i\phi},\quad
z_1-\zb_1=i\left(\frac{u^2}{2}-v^2\right),\quad
e^{D}=\frac{u^2}{4}.
\eea
It is easy to see that the boundary conditions (\ref{MBblBCond}) are satisfied on the surface $y=0$, and one can extract the equations for the surface separating regions I and II. To do so, we observe that region I corresponds to $u=0$, where
\bea
i(z_1-\zb_1)=v^2\ge v^2\sin^2\theta =z_2\wb\nonumber
\eea
Thus the boundary between two regions is a three--dimensional surface
\bea\label{MplnWave}
\mbox{Im}~z_1=-\frac{1}{2}z_2\zb_2.
\eea
Once function $D$ is known, one can easily recover the Kahler potential integrating the last equation in (\ref{Mbubble2}). In the case of the pp-wave the simplest way to proceed is to eliminate $v,\theta$ from the expression for $y$:
\bea
y=\frac{u^2}{4}\sqrt{\frac{u^2}{2}-Z-W^2},\quad W=|z_2|,\quad 
Z=-i(z_1-\zb_1)
\nonumber
\eea
This leads to the Kahler potential
\bea
K&=&\int^{u^2/4} d{\tilde u}\log{\tilde u}[6{\tilde u}^2-2{\tilde u}(Z+W^2)]
\nonumber\\
&=&\left.\frac{{\tilde u}^2}{6}
(-4{\tilde u}+3(Z+W^2)+6(2{\tilde u}-Z-W^2)\log{\tilde u})
\right|_{{\tilde u}=\frac{u^2}{4}}.\nonumber
\eea
This expression is not very illuminating, we presented its derivation just to illustrate the general procedure. 

\

Next we look at AdS$_7\times$S$^4$. In this case the solution of Toda equation is parameterized by the radial coordinate $r$ of AdS and one of the coordinates $\alpha$ of the sphere \cite{LLM}:
\bea
e^D=\frac{r^2 L^{-6}}{4+r^2},\quad 
z=\left(1+\frac{r^2}{4}\right)\cos\alpha e^{i\psi},\quad 
4x=L^{-3}r^2\sin\alpha.
\eea
The coordinates for 1/4--BPS embedding are
\bea
y=\frac{r^2}{4L^3}\sin\alpha\cos\theta,\quad 
z_1=\left(1+\frac{r^2}{4}\right)\cos\alpha e^{i\psi},\quad
z_2=L^3\left(1+\frac{r^2}{4}\right)\sin\alpha\sin\theta e^{-i\phi},
\eea
and the region I corresponds to $r=0$ where
\bea
z_1\zb_1=\cos^2\alpha\le 1-L^{-6}z_2\zb_2.\nonumber
\eea
Thus the boundary between regions corresponds to an ellipsoid
\bea\label{MAdS}
z_1\zb_1+L^{-6}z_2\zb_2=1.
\eea
This equation reduces to (\ref{MplnWave}) if one writes 
$z_1=i+L^{-6}{\tilde z}_1$ and takes $L$ to infinity. 

\

The AdS$_4\times$S$^7$ solution works in a similar way: 
\bea
&&e^D=4L^{-6}\sqrt{1+\frac{r^2}{4}}\sin^2\alpha,\quad 
z_1=e^{i\psi}\left(1+\frac{r^2}{4}\right)^{1/4}\cos\alpha,\quad 
2x=L^{-3}r\sin^2\alpha,\nonumber\\
&&y=\frac{1}{2}L^{-3}r\sin^2\alpha\cos\theta,\quad 
z_2=\frac{L^3}{4}\frac{r}{\sqrt{r^2+4}}\sin\theta e^{-i\phi},
\eea
the region I now corresponds to $\alpha=0$:
\bea
16L^{-6}z_2\zb_2\le \frac{r^2}{r^2+4}=1-\frac{1}{(z_1\zb_1)^2}.
\nonumber
\eea
and the boundary between regions is given by the surface
\bea
16L^{-6}z_2\zb_2=1-\frac{1}{(z_1\zb_1)^2}
\eea
This reduces to (\ref{MAdS}) if we make a holomorphic reparameterization $z_1\rightarrow z_1^{-1/2}$ and a constant rescaling of $W$. Notice that this conformal map exchanges the interior and exterior of the circle $|z_1|=1$. This observation is consistent with the fact that AdS$_7\times$S$^4$ and  
AdS$_4\times$S$^7$ had complementary boundary conditions in $x=0$ plane.

To summarize, we have shown that 1/2--BPS geometries of \cite{LLM} fit nicely into more general solution (\ref{Mbubble1}), (\ref{Mbubble2}) (\ref{Mbubble3}), and regular geometries must satisfy the boundary conditions (\ref{MBblBCond})\footnote{As we will see in the next subsection, regularity also imposes an extra restriction on the values of Kahler potential in $y=0$ hyperplane.  It can be viewed as a requirement on the "integration constant' in equation (\ref{Jan31DtK}).} with some additional restrictions on the 
surfaces separating I and II regions. We also presented explicit examples of such surfaces which came from the AdS$_p\times$S$^q$
and plane wave geometries. In the next subsection we will demonstrate that conditions (\ref{MBblBCond}) are required for regularity even for the most general 1/4--BPS solution (which cannot be embedded into the ansatz of \cite{LLM}) and we will also derive the restrictions on the shapes of the droplets. 

%=====================================

\subsection{Boundary conditions.}

\label{SectM4BC}

%=====================================

As in the IIB case, local equations (\ref{Mbubble1}), (\ref{Mbubble2}) (\ref{Mbubble3}) have to be supplemented by some boundary conditions. The five--sphere contracts along the hypersurface where $e^\la=0$ and, since we want to keep 
$g_{tt}$ finite, this implies that $y=0$. However, the $y=0$ hypersurface has another region where $\zeta=0$ and $e^\la$ does not vanish. Let us consider regularity conditions which should be imposed on these two subsets.

I. If $e^\la\rightarrow 0$, then to keep nonzero $g_{tt}$, one should send $\zeta$ to infinity. In the vicinity of such points it is convenient 
to parameterize the leading contribution to the metric in terms of a 
new function $f=ye^{-2\la}$ which remains finite:
\bea
ds^2&=&-\frac{4f^2}{9}(dt+\rho)^2+
\frac{4}{f}\left[y d\Omega_5^2+\frac{dy^2}{4y}\right]+
2\frac{f^2}{y^2}h_{a{\bar b}}dz^ad\zb^b,\nonumber\\
e^D&=&ye^{\tilde D},\quad \rho=\frac{3i}{2}(\d-\db){\tilde D},\quad
K=\int 2y\log y~dy+\int dy^2 {\tilde D}+K_0(z,\zb),\nonumber\\
e^{-6\la}&=&-\frac{1}{y^3\d_y{\tilde D}},\qquad 
f=|\d_y {\tilde D}|^{-1/3}.
\eea
This metric describes regular geometry if function ${\tilde D}$ 
remains finite as $y$ goes to zero and
\bea
\d_a\db_b K_0=0,\qquad \mbox{det}(\d_a\db_b {\tilde D})|_{y=0}\ne 0,\quad
\d_y {\tilde D}<0.
\eea
It is easy to see that the last two requirements are satisfied, so in this case the boundary condition reduces to 
\bea\label{MBdry1}
y=0:\quad\d_y D=\frac{1}{y}+O(1),\quad \d_a\db_b K=0.
\eea

II. Assuming that $e^{2\la}$ remains finite, we find the following 
expansions:
\bea
&&\sinh\zeta=ye^{-3\la},\quad 
\d_y D=ye^{-6\la},\quad 
K=K_0+y^2 K_1+y^4 K_2,\\
ds^2&=&-\frac{4e^{2\la}}{9}(dt+\rho)^2+
4e^{2\la}d\Omega_5^2+e^{-4\la}\left(
2h_{a{\bar b}}dz^ad\zb^b+\frac{dy^2}{\cosh^2\zeta}\right).\nonumber
\eea
This metric is regular as long as $K_0$ is a non--singular Kahler potential. Looking at the right--hand side of equation (\ref{Mbubble3}), we observe that this is indeed the case as long as $D$ remains finite. 
Thus we arrive at the boundary condition:
\bea\label{MBdry2}
y=0:\quad \d_y D=0,\quad D-\mbox{finite}.
\eea

III. The regularity conditions are slightly more involved near the hypersurfaces which separate regions I and II. A similar problem for 
the IIB geometries was analyzed 
in section \ref{SectIIBDrpl}, and now the intuition developed there will be applied to the eleven--dimensional case. 

{\bf Shapes of the droplets.}

As we saw, the hyperplane $y=0$ is divided into regions where either 
$e^{2\la}$ or $y^2e^{-4\la}$ go to zero, so both factors should vanish 
near the wall separating different domains. This suggests natural coordinates in a vicinity of a point on the wall:
\bea
X=e^{2\la},\quad Y=ye^{-2\la}.
\eea
In particular, this change of variables leads to the following relations:
\bea
y=X^2Y,\quad e^{2\la}\cosh^2\zeta=X^2+Y^2\equiv R^2.
\eea
As one approaches a wall, function $R$ goes to zero, and we  
require that the metric (\ref{Mbubble1}):
\bea\label{Jan17PreMetr}
ds^2&=&-\frac{4e^{2\la}}{9}\cosh^2\zeta(dt+\rho)^2+
4e^{2\la}d\Omega_5^2+e^{-4\la}\left(
h_{ij}dx^idx^j+\frac{dy^2}{\cosh^2\zeta}\right)
\eea
remains regular at the points where $R=0$. 
Starting with relation between $y$ and $X,Y$, one can always 
parameterize the Kahler metric by $v(X,Y)$ and three more coordinates ${\tilde x}_a$ which are orthogonal to it. As we will see, regularity conditions impose certain restrictions on function 
$v(z_a,\zb_b)$.

Near a point where $R=0$, the metric (\ref{Jan17PreMetr}) becomes 
singular unless the five--dimensional sphere combines with radial coordinate $X$ to give a flat six--dimensional space ($ds_6^2=4dX^2+4X^2d\Omega_5^2$), and the metric in remaining five directions describes a flat space $R^{1,4}$. 
Subtracting $ds_6^2$ from (\ref{Jan17PreMetr}) and keeping only 
the leading terms in powers of $R$, we find: 
\bea\label{Jan17Mtr5d}
ds_5^2&=&-\frac{4}{9}(X^2+Y^2)(dt+\rho)^2+
\left[-4dX^2+\frac{dy^2}{X^2(X^2+Y^2)}\right]
+X^{-4}h_{ij}dx^idx^j\nonumber\\
&=&-\frac{4}{9}(X^2+Y^2)(dt+\rho)^2+
\left[dY^2-\frac{(2XdX-YdY)^2}{(X^2+Y^2)}\right]
+X^{-4}h_{ij}dx^idx^j
\eea
This five--dimensional space should be orthogonal to $ds_6^2$, so, after the Kahler metric is rewritten in terms of $(v[X,Y],{\tilde x}_a)$, 
the contributions proportional to $dX$ and to $dX^2$ should disappears from the last expression. This can only happen if $v$ is a function of 
one variable $2X^2-Y^2$, and, without loss of generality, we can set
\bea
v=2X^2-Y^2=2e^{2\la}-y^2e^{-4\la}.
\eea

Treating this relation as a cubic equation for $e^{2\la}$, one concludes that this warp factor depends on the Kahler base only through $v$ 
(i.e. $e^{2\la}=f(v,y)$), then, integrating equations (\ref{GMSWanCont}), 
(\ref{Jan31DtK}) 
for $D$ and $K$, one finds that Kahler potential is also a function of 
$v$ and $y$ only\footnote{There is also an "integration constant" 
$K_0(z_a,\zb_a)$, and, just as in IIB case, but we will not discuss 
this further.}. Of course, 
just as in the IIB case, this reduction of happens only in the vicinity of the wall. Starting with $K(v,y)$, we deduce 
Kahler metric and one--form $\rho$:
\bea
h_{ij}dx^idx^j&=&2\d_v^2 K|\d v|^2+2\d_v K\d\db v=2\d_v K\d\db v+
\frac{\d_v^2 K}{2}\left(dv^2+|(\d-\db)v|^2\right)\nonumber\\
\rho&=&\frac{3i}{4y}\d_y\d_v K(\d-\db)v\equiv 
\frac{3}{2y}\d_y\d_v K\eta,\qquad \eta=\frac{i}{2}(\d-\db)v.
\eea
Substituting these expressions into (\ref{Jan17Mtr5d}) and requiring the resulting space to be regular, we conclude that\footnote{To arrive at 
(\ref{Jan17AAA}), one should follow the steps which led to (\ref{Dec11Bb}), 
(\ref{RegDegMetr}).}
\bea\label{Jan17AAA}
&&\d_v^2 K=\frac{X^4}{2R^2}+O(R^3),\quad
\d_y\d_v K=\frac{X^2 Y}{R^2}+O(R^2),\quad R^2\equiv X^2+Y^2,
\nonumber\\
&&ds_5^2=-\frac{2}{3}\eta dt+dY^2+\frac{2}{X^4}\d_v K\d\db v+
\la_1 dv^2+\la_2\eta^2.
\eea
In the leading order in $R$, coefficients $\la_1$ and $\la_2$ must be 
$(y,v)$--independent. Moreover, extracting the contribution to $\d_v K$ 
from equations for $\d_v^2 K$ and $\d_y\d_v K$, we conclude that 
the leading contribution to $X^{-4}\d_v K=\frac{1}{2}+O(R)$ is a constant. 
To describe regular geometry, the metric (\ref{Jan17AAA}) 
should correspond to a flat five--dimensional space:
\bea
ds_5^2=-\frac{2}{3}\eta dt+dY^2+\d\db v+
\la_1 dv^2+\la_2\eta^2=dY^2+dS_{R^{1.3}}^2,
\eea
then, repeating the steps which led to (\ref{RegDegMetr1}), we find a restriction on function $v$:
\bea\label{Jan17RegBnd}
\d_a\d_b v+\la\d_a v\db_b v=g\d_a w\db_b \wb+O(v),\quad 
\mbox{det}(\d_a v\d_b w)|_{v=0}\ne 0
\eea
We conclude that a decomposition of $y=0$ hyperplane into regions I and II produces a regular geometry if and only if the function $v$ defining the boundaries of the droplets\footnote{We recall that 
the domain walls separating the regions are defined by 
$v(z_a,\zb_a)=0$.} obeys the relations listed in (\ref{Jan17RegBnd}). 
The same differential conditions restrict the boundary of 1/4--BPS droplets in IIB supergravity (see equation (\ref{RegDegMetr1})), and their consequences were discussed in great detail in section \ref{SectIIBDrpl}. 
Here we only mention that some necessary (but not sufficient) conditions can be formulated as {\it algebraic} relations 
(\ref{DetRegCond}), (\ref{DetRegCondAux}) between $\la$ and 
derivatives of $v$:
\bea\label{Jan31AAA}
\mbox{det}(\d_a{\bar\d}_b v+\la \d_a v{\bar\d}_b v)=O(v),\qquad
\left.\mbox{det}(\d_a{\bar\d}_b v)\right|_{v=0}\ne 0.
\eea

\

As in section \ref{SectIIBPert}, one can show that introducing the droplets 
whose shapes satisfy (\ref{Jan17RegBnd}), imposing the boundary conditions (\ref{MBdry1}), (\ref{MBdry2}), and specifying asymptotic behavior of Kahler potential, one arrives at the unique solution of the Monge--Ampere equation, which leads to a regular geometry. 

One can also discuss probe membranes and M5 branes on the 1/4--BPS 
geometries (\ref{Mbubble1}), and, following the arguments presented in 
sections \ref{SectIIBDbr}, \ref{SectIIBPrb}, it can be shown that the 
probe analysis 
and consistency of supergravity lead to the same requirements for 
the brane profiles. Moreover, the restrictions (\ref{Jan31AAA}) prevent droplets from shrinking to branes which are excluded by the probe analysis
(see similar discussion in section \ref{SectIIBDrpl}). 

To analyze topology of the solutions (\ref{Mbubble1}), one needs to make minor 
modifications in the discussion of section \ref{SectIIBTop}: geometry 
(\ref{Mbubble1}) 
contains non--contractible four--cycles, which surround the regions with collapsing $S^5$, and seven--cycles, which are constructed by fibering 
$S^5$ over two--cycle in $\d_y D=0$ region (see figure \ref{FigTop}). It 
is easy to see that such four-- and sever--spheres carry nontrivial fluxes, 
but 
explicit expressions for $\int F_4$ and $\int F_7$ are not very 
illuminating.

%============================================

\section{Unified description of bubbling solutions}

\label{SectUnif}

\renewcommand{\theequation}{8.\arabic{equation}}
\setcounter{equation}{0}

%============================================

Looking back at discussions in the last two sections, one observes 
that, while there are striking similarities in the descriptions of 
1/4--BPS states in ten and eleven dimensions, the boundary 
conditions look somewhat different: going from IIB to M theory, 
one interchanges Dirichlet and Neumann boundary conditions.  
A similar difference was also encountered in \cite{LLM} for the 
1/2-BPS solutions. In this section we will introduce an alternative parameterization of IIB solutions which makes the boundary 
conditions identical to those encountered in eleven dimensions.
The analysis of this section is inspired by the matching 
1/2--BPS solutions with 1/4--BPS ansatz: as discussed in the Appendix 
\ref{SectApEmbIIB}, the "improved" variables for IIB SUGRA arise 
naturally in the process of embedding. We will begin with discussion of 
$1/2$--BPS solution.

{\bf 1/2--BPS case.}

We begin with recalling the bubbling geometries in M theory \cite{LLM}:
\bea\label{LLMthryOne}
ds^2&=&-4e^{2\la}\cosh^2\xi (d\tau+V)^2+\frac{e^{-4\la}}{\cosh^2\xi}
\left[dx^2+e^{D}dzd{\bar z}\right]+4e^{2\la}d\Omega_5^2+
x^2e^{-4\la}d\Omega_2^2\nonumber\\
F_4&=&d\left[-4x^3 e^{-6\la}(d\tau+V)+2~^*_3\left\{e^{-{D}}
x^2\d_x\left(\frac{\d_x e^{D}}{x}\right)dx+x\d_x d_2 {D}
\right\}\right]\wedge d^2\Omega\\
&&e^{-6\la}=\frac{\d_x {D}}{x(1-x\d_x {D})},\quad 
V=\frac{i}{2}(dz\d_z-d\zb\d_{\bar z}){D},\quad \sinh\xi=xe^{-3\la}.
\nonumber
\eea
The solutions are parameterized by a function ${D}$ satisfying 
Toda equation 
(\ref{TodaEqn}) and Neumann boundary conditions (\ref{TodaBCond}):
\bea\label{Jan30Mbndr}
x=0:\quad
\left\{\begin{array}{l}\d_x{D}=0,~ {D}-\mbox{finite},\\
\d_x{D}=\frac{1}{x}+O(1).
\end{array}\right.
\eea

This should be contrasted with situation in IIB SUGRA, where 1/2--BPS 
solutions are parameterized by a 
harmonic function ${\tilde Z}(z,\zb,x)$ (see (\ref{LLMIIBdy}), 
(\ref{LaplLLM})), which satisfies the 
Dirichlet boundary conditions \cite{LLM}:
\bea
{\tilde Z}(z,\zb,x=0)=\pm\frac{1}{2}.
\eea
However, as we saw in section \ref{SectIIBEx}, one can also describe ten--dimensional solutions in terms of function $D$ which has Neumann boundary conditions (\ref{Jan30Mbndr}) (see equations (\ref{Jan7Eqn1}), (\ref{AltIIBcnd})):
\bea\label{LLMthryTwo}
ds^2&=&-e^{2\la}\cosh^2\xi(dt+V)^2+\frac{1}{e^{2\la}\cosh^2\xi}
(dx^2+dzd\zb)+
e^{2\la}d\Omega_3^2+x^2e^{-2\la}d{\tilde\Omega}_3^2,\nonumber\\
F_5&=&-\frac{1}{4}d\left[x^4 e^{-4\la}(dt+V)-*_3\left\{x^2\d_x\left(
\frac{\d_x D}{x}\right)+x\d_x d_2 {D}\right\}\right]\wedge 
d{\tilde\Omega}_3+dual,\\
&&e^{-4\la}=\frac{\d_x {D}}{x(1-x\d_x {D})},\quad 
V=-i(dz\d_z-d\zb\d_{\zb}){D},\quad
\sinh\xi=xe^{-2\la}.\nonumber
\eea
The systems (\ref{LLMthryOne}) and (\ref{LLMthryTwo}) look strikingly similar, and the boundary conditions (\ref{Jan30Mbndr}) are identical in both cases. The differences between (\ref{LLMthryOne}) and (\ref{LLMthryTwo}) stem from the fact that in IIB case the time--like Killing vector is translational, while in M theory it is rotational, so function $D$ obeys different equations (compare (\ref{Jan7Eqn2}) and (\ref{TodaEqn})).
In the systems without fluxes, the relation between translational (rotational) Killing vector and Laplace (Toda) equation has been extensively discussed in the past \cite{gibHawk,bachas}. 

{\bf 1/4--BPS case.}

Given the similarities between 1/2--BPS geometries in ten and eleven dimensions, it is interesting to see whether they persist for the 
1/4--BPS solutions as well. To determine this, we rewrite the IIB solutions 
(\ref{DonosSoln}) in terms of a new function $D$, and compare the results with (\ref{Mbubble1}), (\ref{Mbubble2}), (\ref{Mbubble3}). Motivated by  
the discussion of section \ref{SectIIBEx}, we extend the relation 
(\ref{Jan7Eqn4}) to arbitrary 1/4--BPS geometries by making a definition:
\bea\label{Jan30In}
D=\frac{1}{2}(y^{-1}\d_y K+\log y).
\eea
Then equation for $Z$ (\ref{DonosSoln}) implies that
\bea
Z=-y\d_y D+\frac{1}{2}.
\eea
Substituting this into (\ref{DonosSoln}) and introducing $e^{2\la}=ye^G$, one finds the following metric:
\bea
ds_{10}^2&=&-e^{2\la}\cosh^2\zeta(dt+\omega)^2+
e^{-2\la}\left[2
\d_a{\bar\d}_b Kdz^ad\zb^b+\frac{dy^2}{\cosh^2\zeta}+y^2d\psi^2\right]+
e^{2\la}d\Omega_3^2,\nonumber\\
&&\qquad \omega=i(\db-\d)K,\quad 
e^{-4\la}=\frac{y^{-1}\d_y D}{1-y\d_y D},\quad
\sinh\zeta=ye^{-2\la}.
\eea
The Monge--Ampere equation also simplifies in terms of $D$:
\bea\label{Jan30Out}
\mbox{det} h_{a{\bar b}}=\frac{1}{4}(1-y\d_yD)e^{2D}. 
\eea
We observe that similarities between the type IIB system 
(\ref{Jan30In})--(\ref{Jan30Out}) and its M theory counterpart
((\ref{Mbubble1}), (\ref{Mbubble2}), (\ref{Mbubble3})) are even more striking than in the 1/2--BPS case. Perhaps this is related to the fact 
that now Killing vectors are rotational in both cases. 

%============================================

\section{Discussion}

\label{SectDisc}

\renewcommand{\theequation}{9.\arabic{equation}}
\setcounter{equation}{0}

%============================================

Since the results of this paper have already been summarized in the introduction, in this section we will discuss some applications and open problems. 

While D branes can be described in terms of either open or closed strings, 
traditionally one utilizes open string picture to find the positions of the branes, and then uses this information as an input for the SUGRA analysis. Along with earlier work \cite{myCM}, this article provides an evidence for 
plausibility of an alternative approach, where brane profiles are 
found directly in the closed--string picture. Then an agreement with open string analysis serves as a nontrivial check of the open/closed string 
duality. So far, such agreement was demonstrated only for configurations preserving eight or more supercharges, and it would be nice to see 
whether it persists for branes with lower supersymmetry. It would also be interesting to understand the nature of the agreement: since DBI and SUGRA descriptions are valid in different corners of the parameter space, one may unravel some new non--renormalization theorems. 

While discussing bubbling geometries, we encountered a very interesting restriction (\ref{IntroV}) on the shape of the droplets, and it would be nice to acquire a more geometrical understanding of this conditions. Moreover, since the 
geometries discussed in section \ref{SectIIBbl} correspond to 1/4--BPS 
states in a ${\cal N}=4$ SYM, one should be able to find a field--theoretic counterpart of (\ref{IntroV}). We recall that in the 1/2--BPS case, 
boundary conditions in supergravity \cite{LLM} had a direct interpretation 
in terms of a matrix model on the field theory side 
\cite{beren}. Using this correspondence as a guide, one expects to see the 
condition (\ref{IntroV}) in a matrix model introduced in \cite{beren4}, but, 
to do so, a better understanding of the matrix model is required. In fact, the quantum mechanical system introduced in \cite{beren4} describes 
1/8--BPS states as well, and the most general gravity solution with this amount of supersymmetry was discussed in \cite{kim}. It would be very nice to understand regularity condition in this case. It is also very interesting to find the geometries preserving less than eight supercharges, since they might  viewed as microscopic states contributing to an entropy of the 
black hole  constructed in \cite{16BH}. 

There are also open problems in the 1/4--BPS case. While the geometries (\ref{DonosSoln}) give the bulk description of local states in field theory, the metrics corresponding to 1/4--BPS non--local states are still not known. In the 1/2--BPS case, the geometries corresponding to various defects in IIB string \cite{myWils} and M \cite{myMdef} theories turned out to be more complicated than the solutions of \cite{LLM}, and this trend is expected to continue for configurations with lower supersymmetry. However, the boundary conditions encountered in \cite{myWils,myMdef} are as transparent as ones found in \cite{LLM}, so it would be interesting to find their 1/4--BPS counterparts. An additional motivation for such investigation comes from the fact that the brane probe analysis (i.e. an analog of 
\cite{mikhail}) has been already performed in \cite{MaldDruk}. 

Finally, the Monge--Ampere equations encountered in this article present an interesting technical challenge. From the probe analysis, we know that the sources can be freely superposed, this indicates that Monge--Ampere equations might have some hidden linear structure. If this is indeed the case, it would be very interesting to find the right variables which make 
the equations linear and lead to explicit solutions. 

\section*{Acknowledgments}

It is a pleasure to thank Samir Mathur and Ilarion Melnikov for useful discussions. 
This work is supported by DOE grant DE-FG02-90ER40560.

\appendix

%============================================

\section{Derivation of metrics produced by brane webs}

\label{SectAp1}

\renewcommand{\theequation}{A.\arabic{equation}}
\setcounter{equation}{0}

%============================================

To construct a geometry produced by a connected string web, one needs to 
find a supersymmetric background which fits into the following ansatz:
\bea
ds_{IIB}^2&=&-f_1 dt^2+g_{ij}dx^idx^j+f_2 d\Omega_6^2\\
B+iC^{(2)}&=&dt\wedge V_i dx^i,\qquad 
\tau\equiv C^{(0)}+ie^{-\Phi}=\tau(x_i).\nonumber
\eea
Rather than solving the equations for Killing spinors on this geometry, 
we will choose an alternative path and construct the metrics describing a U--dual system. To be more precise, we begin with smearing the web in one of the transverse directions and performing a T duality in this direction. The lift of the resulting IIA solution to M theory gives a configuration with 
$SO(6)\times U(1)^2\times U(1)_t$ symmetry:
\bea\label{EqnAppIso}
ds^2&=&-e^{2A}dt^2+e^{2C}(dw_1+\chi dw_2)^2+e^{2D}dw_2^2+g_{MN}dX^MdX^N+
e^{2B}d\Omega_5^2\nonumber\\
G_4&=&dt\wedge W_\alpha\wedge dw^\alpha
\eea
To find supersymmetric geometries, we should solve the equations for the Killing spinor and it turns out that, while an assumption of translational invariance in $w_1$ and 
$w_2$ leads to certain simplifications in these equations, the first few steps towards solving them rely only on $SO(6)\times U(1)_t$ isometries. Thus we begin with searching for the the most general solution consistent with latter symmetry:
\bea\label{Eqn1107App}
ds^2&=&-e^{2A}dt^2+g_{\mu\nu}dX^\mu dX^\nu+e^{2B}d\Omega_5^2\nonumber\\
G_4&=&dt\wedge F_3
\eea
and in subsection \ref{SectApIsom} we will analyze the additional 
restrictions imposed by the ansatz (\ref{EqnAppIso}). Our first goal is to solve the equations for the Killing spinor:
\bea\label{11DSpinorEqn}
\nabla_m\eta+\frac{1}{288}\left[
-\frac{1}{2}{\gamma_m}{\not G}+\frac{3}{2}{\not G}\gamma_m\right]\eta=0.
\eea
Looking at the components of this equation along the isometry directions, one can produce the projectors which do not contain derivatives of the Killing spinor. We begin with projectors which correspond to time and sphere directions:
\bea\label{TimeKillEqn}
&&\nabla_t\eta-\frac{1}{144}{\gamma_t}{\not G}\eta=0:\qquad
\frac{1}{2}{\not\d}A\eta-\frac{1}{144}{\not G}\eta=0\\
\label{SphKillEqn}
&&\nabla_a\eta+\frac{1}{288}{\gamma_a}{\not G}\eta=0:\qquad
-\frac{i}{2}e^{-B}\Gamma_S\eta+\frac{1}{2}{\not\d}B\eta+
\frac{1}{288}{\not G}\eta=0
\eea
To arrive at the second relation we introduced the standard invariant fermions on the odd--dimensional sphere with unit radius \cite{ramf}:
\bea
{\tilde\nabla}_a\eta=-\frac{i}{2}{\tilde\gamma}_a\Gamma_S\eta
\eea
The projectors (\ref{TimeKillEqn}) and (\ref{SphKillEqn}) can be combined to produce a relation which does not contain fluxes:
\bea\label{GeomProjIn}
-i\Gamma_S\eta+{\not\d}(B+\frac{A}{2})\eta=0.
\eea
We can use the diffeomorphisms in five--dimensional space spanned by $X^M$ to choose $y=e^{B+A/2}$ to be one of the coordinates and to parameterize the orthogonal subspace by $x_1,\dots x_4$. Then the "geometric" projector (\ref{GeomProjIn}) determines the $y$--component of the metric:
\bea\label{3DYmetric}
y=e^{B+A/2},\qquad g_{\mu\nu}dX^\mu dX^\nu\equiv e^{-A} dy^2+
g_{ij}dx^idx^j.
\eea
In this frame the projector (\ref{GeomProjIn}) reduces to a very simple relation:
\bea\label{GeomProj}
i\Gamma_S\eta-\Gamma_y\eta=0.
\eea

So far we only discussed the time and sphere components of (\ref{11DSpinorEqn}), let us not look at the remaining seven equations. Using the relation (\ref{TimeKillEqn}), we can rewrite those components as
\bea\label{eqn1107a}
&&\nabla_\mu\eta-\frac{1}{8}\gamma_\mu{\not\d}A\eta+
\frac{3}{4}\frac{1}{144}{\not G}\gamma_\mu\eta=0:\nonumber\\
&&e^{A/8}\nabla_\mu(e^{-A/8}\eta)+\frac{1}{8}\d^\nu A\gamma_{\nu\mu}\eta-
\frac{e^{-A}}{48}\Gamma_0{\not F}\gamma_\mu\eta=0.
\eea
Convenient rescalings
\bea
\eta=e^{A/8}{\tilde\eta},\quad e_\mu^{\bf a}={\tilde e}_\mu^{\bf a} e^{A/4}
\eea
lead to simplifications in the equation (\ref{eqn1107a}):
\bea\label{eqn1107b}
{\tilde\nabla}_\mu{\tilde\eta}-\frac{e^{-3A/2}}{48}\Gamma_0
{\not {\tilde F}}{\tilde\gamma}_\mu{\tilde\eta}=0.
\eea
Notice that this relation, as well as two remaining projectors 
(\ref{TimeKillEqn}), (\ref{SphKillEqn}), does not 
mix $\eta_+=(1+i\Gamma_0)\eta$ and $\eta_-=(1-i\Gamma_0)\eta$, so, without loss of generality, we can impose a projection
\bea
\Gamma_0\eta=i\eta.
\eea

Let us summarize what we have learned so far. Assuming only $SO(6)\times U(1)_t$ symmetry and introducing convenient coordinates, we showed that the eleven--dimensional geometry must have the form
\bea\label{ReducMetr}
ds^2&=&-e^{2A}dt^2+e^{A/2}\left[e^{-3A/2}dy^2+h_{ij}dx^idx^j\right]+y^2 e^{-A}d\Omega_5^2\\
G_4&=&dt\wedge F\nonumber
\eea
and the Killing spinor must satisfy the following relations 
\bea\label{SpinSystAPr}
&&{\not\d}A\eta-\frac{ie^{-3A/2}}{18}{\not F}\eta=0,\\
\label{SpinSystDfE}
&&{\nabla}_\mu{\eta}-\frac{ie^{-3A/2}}{48}
{\not {F}}{\gamma}_\mu{\eta}=0,\\
\label{SpinSystOrtPr}
&&\Gamma_0\eta=i\eta,\qquad i\Gamma_S\eta-\Gamma_y\eta=0.
\eea
in the {\it reduced} five--dimensional space\footnote{From now on we will only use the metric which appears in the square brackets in (\ref{ReducMetr}), so we drop tildas in the equation (\ref{eqn1107b}). We will also use Greek letters to denote five--dimensional indices and Latin letters for the four-dimensional ones.}:
\bea\label{5Dmetric}
g_{\mu\nu}dX^\mu dX^\nu\equiv e^{-3A/2} dy^2+h_{ij}dx^idx^j
\eea

Before we proceed with analysis of these equations, let us count the preserved supersymmetries. Eleven--dimensional spinor $\eta$ has $32$ real components and two
independent projections (\ref{SpinSystOrtPr}) reduce the number of components to $8$.
The projector (\ref{SpinSystAPr}) breaks one half of the remaining supersymmetries, 
so it appears that we are dealing with $1/8$--BPS configuration. However a closer inspection of the system (\ref{SpinSystAPr})--(\ref{SpinSystOrtPr}) demonstrates that 
the same bosonic background preserves spinors with both signs in 
$\Gamma_0\eta=\pm i\eta$ (while other signs in 
(\ref{SpinSystAPr})--(\ref{SpinSystOrtPr}) being adjusted accordingly), so, even though 
we will only look for a four--component spinor satisfying 
(\ref{SpinSystAPr})--(\ref{SpinSystOrtPr}), the resulting geometries will be $1/4$--BPS.

The system (\ref{SpinSystAPr})--(\ref{SpinSystOrtPr}) can be viewed as a set of relations for a spinor in seven dimensions spanned by matrices 
$(\gamma_\mu,\gamma_t,\Gamma_S)$. These seven objects are not independent since 
the product of gamma matrices in eleven dimensions is equal to one:
\bea\label{ProdGam7}
\Gamma_0\Gamma_{1234}\Gamma_y\Gamma_S=1.
\eea
Multiplying this relation by $\eta$ and using projections (\ref{SpinSystOrtPr}), 
we conclude that
\bea\label{4DProj}
\Gamma_{1234}\eta=\eta.
\eea

The seven dimensional spinor appearing in (\ref{SpinSystAPr})--(\ref{SpinSystOrtPr}) came from the reduction on a five--sphere, so it can have at most $8$ independent components. The projectors (\ref{SpinSystOrtPr}) truncate the number of components to 
$2$, and after enforcing (\ref{SpinSystAPr}) one should end up with a spinor which is parameterized by one real number. In the remaining part of this section we will discuss
the properties of this one--component spinor in more detail. 

%=====================================

\subsection{Equations for the spinor bilinears}

\label{SectApBlnr}

%=====================================

To find the restrictions on the five--dimensional metric (\ref{5Dmetric}), we will study the equations for the spinor bilinears. Since equations 
(\ref{SpinSystAPr}), (\ref{SpinSystDfE}) and their hermitean conjugates 
will be used extensively, we rewrite them here for future 
reference: 
\bea\label{Ap1DifEqnZZ}
&&{\nabla}_\mu\eta-\frac{ie^{-3A/2}}{48}
{\not {F_3}}{\gamma}_\mu\eta=0,\quad
{\nabla}_\mu\eta^\dagger-\frac{ie^{-3A/2}}{48}
\eta^\dagger\gamma_\mu{\not {F_3}}=0\\
\label{Ap1ProjZZ}
&&{\not\d}A\eta-\frac{ie^{-3A/2}}{18}{\not F}_3\eta=0,\quad
\eta^\dagger{\not\d}A-\frac{ie^{-3A/2}}{18}\eta^\dagger{\not F}_3=0
\eea
We begin with solving the equation for the scalar bilinear $\eta^\dagger\eta$ (the choice of integration constant fixes the normalization of the spinor):
\bea
\nabla_\mu(\eta^\dagger\eta)-\frac{18}{48}2\d_\mu A~\eta^\dagger\eta=0:
\qquad
\eta^\dagger\eta=e^{3A/4}.
\eea
Next we look at tensor bilinear:
\bea
J_{\mu\nu}=\eta^\dagger\gamma_{\mu\nu}\eta.
\eea
The second projector in (\ref{SpinSystOrtPr}) implies that this tensor 
cannot have legs in $y$ direction:
\bea
J_{\mu\nu}e^\mu_{\bf y}=0:\qquad J\equiv 
\frac{1}{2}J_{\mu\nu}dX^\mu\wedge dX^\nu=\frac{1}{2}J_{mn}dx^m\wedge dx^n.
\eea
Let us compute the derivative of the tensor and the exterior derivative of the two--form:
\bea\label{Nov8DerJ}
\nabla_{\mu} J_{\nu\la}&=&\frac{ie^{-3A/2}}{48}
\eta^\dagger \left(\gamma_{\nu\la}{\not {F_3}}{\gamma}_\mu+
\gamma_m{\not {F_3}}{\gamma}_{kl}\right)\eta\\
&=&-\frac{3}{8}
\eta^\dagger(\gamma_{\nu\la}\gamma_\mu{\not\d}A+
{\not\d}A\gamma_\mu\gamma_{\nu\la})\eta+
\frac{ie^{-3A/2}}{8}
\eta^\dagger(\gamma_{\nu\la}F_{\mu\rho\sigma}\gamma^{\rho\sigma}+
6F_{\mu\rho\sigma}\gamma^{\rho\sigma}\gamma_{\nu\la})\eta\nonumber\\
\nabla_{[\mu} J_{\nu\la]}&=&-\frac{9}{4}\d_{[\mu}A J_{\nu\la]}+
\frac{ie^{-3A/2}}{8}
\eta^\dagger(\gamma_{[\nu\la}F_{\mu]\rho\sigma}\gamma^{\rho\sigma}+
F_{\rho\sigma[\mu}\gamma^{\rho\sigma}\gamma_{\nu\la]})\eta\nonumber
\eea
We begin with simplifying the $y$ component of the last relation:
\bea\label{PreFyy}
\frac{1}{3}\d_y J_{kl}&=&-\frac{3}{4}\d_y A J_{kl}+
\frac{ie^{-3A/2}}{24}
\eta^\dagger\{\gamma_{kl},\gamma^{pq}\}F_{ypq}\eta+
\frac{ie^{-3A/2}}{12}4F_{lky}\eta^\dagger\eta\nonumber\\
&=&-\frac{1}{4}\d_y A J_{kl}+
\frac{ie^{-3A/4}}{3}F_{lky}
\eea
To eliminate the term with anticommutator we used the relations which can be obtained by combining the second projection in (\ref{SpinSystOrtPr}) 
with (\ref{SpinSystAPr}):
\bea
{\d}_yA\eta-\frac{ie^{-3A/2}}{6}{F}_{ypq}\gamma^{pq}\eta=0,\quad
\eta^\dagger{\d}_yA-\frac{ie^{-3A/2}}{6}\eta^\dagger
{F}_{ypq}\gamma^{pq}=0
\eea
Equation (\ref{PreFyy}) can be rewritten as a simple expression for the flux:
\bea\label{EqnFyy}
F_{kly}=i\d_y(e^{3A/4}J_{kl}).
\eea
Motivated by this relation, we compute the four--dimensional components of the 3--form 
$d(e^{3A/4}J)$:
\bea
\nabla_{[m}(e^{3A/4}J_{kl]})
&=&-\frac{1}{4}e^{3A/4}\eta^\dagger\{{\not\d}A,\gamma_{mkl}\}\eta+
\frac{ie^{-3A/4}}{8}
\eta^\dagger(\gamma_{[kl}F_{m]pq}\gamma^{pq}+
F_{pq[m}\gamma^{pq}\gamma_{kl]})\eta\nonumber\\
&=&-\frac{ie^{-3A/4}}{12}\left(F_{pq[k}\eta^\dagger\gamma^{pq}\gamma_{lm]}\eta-
F_{pq[l}\eta^\dagger\gamma_k\gamma^{pq}\gamma_{m]}\eta+
F_{pq[m}\eta^\dagger\gamma_{kl]}\gamma^{pq}\eta
\right)\nonumber\\
&&+\frac{ie^{-3A/4}}{8}
\eta^\dagger(\gamma_{[kl}F_{m]pq}\gamma^{pq}+
F_{pq[m}\gamma^{pq}\gamma_{kl]})\eta\nonumber\\
&=&\frac{ie^{-3A/4}}{24}\left(F_{pq[m}\eta^\dagger\gamma^{pq}\gamma_{kl]}\eta+
2F_{pq[m}\eta^\dagger\gamma_l\gamma^{pq}\gamma_{k]}\eta+
F_{pq[m}\eta^\dagger\gamma_{kl]}\gamma^{pq}\eta
\right)\nonumber\\
&=&\frac{ie^{-3A/4}}{6}\left(F_{p[km}\eta^\dagger\gamma^{p}
\gamma_{l]}\eta-
F_{p[lm}\eta^\dagger\gamma_{k]}\gamma^{p}\eta
\right)\nonumber\\
&=&\frac{i}{3}F_{lkm}\nonumber
\eea
Combining this with (\ref{EqnFyy}), we arrive at the final expression for the flux:
\bea\label{Nov8Flux}
F=id(e^{3A/4}J)
\eea
Alternatively, we can extract the flux from looking at a bilinear built out of the projectors (\ref{Ap1ProjZZ}): 
\bea
&&\eta^\dagger\{\gamma_{\mu\nu\la},{\not\d}A\}\eta-\frac{i}{18}e^{-3A/2}
\eta^\dagger\{\gamma_{\mu\nu\la},{\not F}\}\eta=0\nonumber\\
&&6\d_{[\mu}A J_{\nu\la]}-\frac{i}{18}e^{-3A/2}(12F_{\mu\nu\la}e^{3A/4}+
18F_{\rho\sigma[\mu}\eta^\dagger{\gamma_{\nu\la]}}^{\rho\sigma}\eta)=0.\nonumber
\eea
Different components of the last equation give\footnote{Notice that the projector 
(\ref{4DProj}) implies that 
$\eta^\dagger\gamma_{klmp}\eta=\eps_{klmp}\eta^\dagger\eta$.}:
\bea\label{Nov8dAJ}
&&\d_{y}A J_{kl}=\frac{i}{3}e^{-3A/4}(F_{ylk}+\frac{1}{2}
F_{pqy}{\eps_{kl}}^{pq}),\\
&&6\d_{[m}A J_{kl]}=\frac{2i}{3}e^{-3A/4}(F_{mlk}+
F_{pq[m}{\eps_{kl]}}^{pq}).
\eea
To simplify the second equation, we observe that
\bea
\eps^{klms}F_{pq[m}{\eps_{kl]}}^{pq}=\frac{1}{3}\eps^{klms}
{\eps_{kl}}^{pq}F_{pqm}=\frac{2}{3}(g^{mp}g^{sq}-g^{mq}g^{sp})
F_{pqm}=0.
\eea
This leads to a very simple expression for the four dimensional components of the flux:
\bea\label{FluxWdg}
F_{mkl}=9ie^{3A/4}\d_{[m}AJ_{kl]}=3ie^{3A/4}(dA\wedge J)_{klm},
\eea
and, combining it with (\ref{Nov8Flux}), we arrive at the equation for $J$:
\bea\label{PreKahlStr}
0&=&\left[d(e^{3A/4}J)-3e^{3A/4}dA\wedge J\right]_{klm}=
e^{3A}\left[d(e^{-9A/4}J)\right]_{klm}
\eea
Let us go back to the equation (\ref{Nov8dAJ}). First of all, it implies that $J_{kl}$ is an anti--self--dual tensor. Further, by substituting the expression (\ref{EqnFyy}) for the flux
into the right--hand side of that equation, we arrive at a useful relation between the $y$ derivatives of the tensor bilinear:
\bea\label{DualJYder}
\d_y(e^{-3A/2}{\eps_{kl}}^{pq}J_{pq})+e^{-3A/2}\d_y(J_{pq}){\eps_{kl}}^{pq}=0.
\eea
The two--form $J$ is especially useful since it is related to an almost complex structure.
Indeed, we can use the Fierz identities to show that 
\bea
J_{mp}J^{pn}=e^{3A/2}\delta_m^n
\eea
This implies that 
\bea
{\tilde J}_m^{~n}=ie^{-3A/4}{J}_m^{~n}
\eea
is an almost complex structure on the four--dimensional manifold parameterized by $x_m$. In the next subsection we will demonstrate that this almost complex structure is integrable and we will also 
discuss a holomorphic two--form. 

%=====================================

\subsection{Complex structure and holomorphic two--form}

\label{SectApCmplS}

%=====================================

To demonstrate that ${\tilde J}_m^{~n}$ is a complex structure, we need to show that the Nijehuis tensor 
\bea
{N_{mn}}^p={\tilde J}_m^{\ \ q}{\tilde J}_{[n\ \ ,q]}^{\ \ p}-
{\tilde J}_n^{\ \ q}{\tilde J}_{[m\ \ ,q]}^{\ \ p}
\eea
vanishes. We begin with recalling the four--dimensional components of 
the equation (\ref{Nov8DerJ}):
\bea\label{SpcDerOfJ}
\nabla_m J_{kl}&=&-\frac{3}{8}\eta^\dagger(
\{\gamma_{klm},{\not\d}A\}+2g_{m[l}[\gamma_{k]},{\not\d}A])\eta+
\frac{i}{8}e^{-3A/4}(4F_{mlk}+2{\eps_{kl}}^{pq}F_{mpq})\nonumber\\
&=&-\frac{9}{4}J_{[kl}\d_{m]}A-\frac{3}{2}g_{m[l}J_{k]p}\d^p A+
\frac{i}{4}e^{-3A/4}(2F_{mlk}+{\eps_{kl}}^{pq}F_{mpq}).
\eea
Taking antisymmetric part, we find
\bea\label{Nov8DerJAnti}
\nabla_{[m} J_{k]l}
&=&-\frac{9}{4}J_{[kl}\d_{m]}A-\frac{3}{4}g_{l[m}J_{k]p}\d^p A+
\frac{i}{4}e^{-3A/4}(2F_{mlk}-{\eps_{l[k}}^{pq}F_{m]pq}).\nonumber
\eea
Notce that 
\bea
{\eps_{l[k}}^{pq}F_{m]pq}=-\frac{1}{3}g_{lm}{\eps_k}^{rpq}F_{rpq},
\eea
so we need to find the four--dimensional dual of the three--form. To this end 
we construct a bilinear using projectors (\ref{Ap1ProjZZ}):
\bea
&&0=\eta^\dagger[\gamma_k,{\not\d}A]\eta+\frac{i}{9}e^{-3A/2}
F^{pqr}\eta^\dagger\gamma_{pqrk}\eta=
2\d^p AJ_{kp}+\frac{i}{9}e^{-3A/4}F^{pqr}\eps_{pqrk}
\nonumber\\
&&i{\eps_{l[k}}^{pq}F_{m]pq}=
-\frac{i}{3}g_{l[m}{\eps_{k]}}^{rpq}F_{rpq}=
-6g_{l[m}\d^p AJ_{k]p}e^{3A/4}
\eea
Substituting this expression and the one for $F_{klm}$ into (\ref{Nov8DerJAnti}), 
we find
\bea
\nabla_{[m} J_{k]l}
&=&-\frac{9}{4}J_{[kl}\d_{m]}A-\frac{3}{4}g_{l[m}J_{k]p}\d^p A
+\frac{9}{2}\d_{[m}AJ_{kl]}
+\frac{3}{2}g_{l[m}\d^p AJ_{k]p}\nonumber\\
&=&\frac{3}{4}\left[3J_{[kl}\d_{m]}A+g_{l[m}J_{k]p}\d^p A\right].
\eea
Let us rewrite this in terms of an almost complex structure 
${\tilde J}=ie^{-3A/4}J$:
\bea
\nabla_{[m} {\tilde J}_{k]l}
&=&\frac{3}{4}\left[3{\tilde J}_{[kl}\d_{m]}A+
g_{l[m}{\tilde J}_{k]p}\d^p A\right]-\frac{3}{4}\d_{[m}A {\tilde J}_{k]l}
\nonumber\\
&=&\frac{3}{8}\left[{\tilde J}_{kl}\d_{m}A-{\tilde J}_{ml}\d_{k}A+
2{\tilde J}_{mk}\d_l A+g_{lm}{\tilde J}_{kp}\d^p A-g_{lk}{\tilde J}_{mp}\d^p A
\right]\nonumber
\eea
To evaluate the Nijehuis tensor, we need to compute
\bea
{\tilde J}_k^{\ \ q}\nabla_{[m}{\tilde J}_{q]l}&=&
\frac{3}{8}\left[-{g}_{kl}\d_{m}A-{\tilde J}_{ml}{\tilde J}_k^{\ q}\d_{q}A+
2g_{mk}\d_l A-g_{lm}{g}_{kp}\d^p A-{\tilde J}_{kl}{\tilde J}_{mp}\d^p A
\right]\nonumber\\
&=&\frac{3}{8}\left[-{g}_{kl}\d_{m}A-{\tilde J}_{ml}{\tilde J}_k^{\ \ q}\d_{q}A+
2g_{mk}\d_l A-g_{lm}\d_k A-{\tilde J}_{kl}{\tilde J}_m^{\ \ p}\d_p A
\right]\nonumber
\eea
The right-hand side of this expression is symmetric under interchange of $k$ and 
$m$, so we conclude that 
Nijehuis tensor vanishes. This implies that ${\tilde J}$ is an integrable complex structure and we can choose complex coordinates:
\bea\label{ComplCoord}
ds^2=2g_{a{\bar b}}dz^a d\zb^b,\quad 
J=\frac{1}{2}J_{mn}dx^m\wedge dx^n=e^{3A/4}g_{a{\bar b}}dz^a\wedge d\zb^b.
\eea
Notice that not only the four--dimensional space is complex, but it is also related to a Kahler space by a very simple rescaling. To see this we recall the equation 
(\ref{PreKahlStr}) for the two--form and rewrite it in terms of the metric $g_{a{\bar b}}$:
\bea
d\left[e^{-3A/2}g_{a{\bar b}}dz^a\wedge d\zb^b\right]
\eea
This implies that $e^{-3A/2}g_{a{\bar b}}$ is a Kahler metric and it can be written in terms of the potential $K(z,{\bar z},y)$. In particular, we find:
\bea\label{KahlMetr}
g_{a{\bar b}}=e^{3A/2}\d_a{\bar \d}_b K.
\eea

At this point the solution is completely specified in terms of a real function $A$ and Kahler potential $K$ and the rest of this subsection will be devoted to finding a relation between them. To extract such relation, we define a new two--form 
\bea\label{OmegaForm}
\Omega=\frac{1}{2}\eta^\dagger\gamma_{mn}\Gamma_{\overline{12}}\eta~dx^{mn}
\eea
and compute its derivatives: 
\bea\label{Nov9DerOmega}
\nabla_\mu(\eta^\dagger\gamma_{\nu\la}\Gamma_{\overline{12}}\eta)-
\frac{i}{48}e^{-3A/2}
\eta^\dagger\left(\gamma_{\nu\la}\Gamma_{\overline{12}}{\not F}\gamma_\mu+
\gamma_\mu{\not F}\gamma_{\nu\la}\Gamma_{\overline{12}}\right)\eta=0.
\eea
To proceed it is convenient to introduce a holomorphic veilbein and 
flat gamma matrices:
\bea
g_{a{\bar b}}=\frac{1}{2}e_a^{A}{e}_{\bar b}^{\bar B}\delta_{A{\bar B}},\qquad
\{\Gamma_A,\Gamma_{\bar B}\}=\delta_{A{\bar B}},\quad 
\{\Gamma_A,\Gamma_{B}\}=0.
\eea
In particular, looking at various components of (\ref{ComplCoord}), we observe that
\bea\label{Nov8Cmpl}
\eta^\dagger\Gamma_{AB}\eta=0,\qquad 
\eta^\dagger[\Gamma_A,\Gamma_{\bar B}]\eta=\delta_{A{\bar B}}.
\eea
As we mentioned before, $\eta$ should be viewed as a eight--component spinor in seven dimensional space and gamma matrices acting on this spinor are constrained 
by the relation (\ref{ProdGam7}). To proceed it is convenient to choose an explicit set of seven gamma matrices:
\bea
&&\Gamma_0=i\sigma_3\otimes\sigma_3\otimes\sigma_3,\quad 
\Gamma_y=\sigma_1\otimes\sigma_3\otimes\sigma_3,\quad
\Gamma_S=\sigma_2\otimes\sigma_3\otimes\sigma_3,\nonumber\\
&&\begin{array}{c}\Gamma_{1}\\
\Gamma_{\bar 1}\end{array}=1\otimes 1\otimes 
\begin{array}{c}\sigma_-\\
\sigma_+\end{array},\qquad
\begin{array}{c}\Gamma_{2}\\
\Gamma_{\bar 2}\end{array}=1\otimes 
\begin{array}{c}\sigma_-\\
\sigma_+\end{array}\otimes \sigma_3,
\eea
with the following actions of sigmas:
\bea
\sigma_3|\uparrow\rangle=|\uparrow\rangle,\quad 
\sigma_3|\downarrow\rangle=-|\downarrow\rangle,\quad
\sigma_+|\downarrow\rangle=|\uparrow\rangle,\quad
\sigma_-|\uparrow\rangle=|\downarrow\rangle.
\eea
The projector involving $\Gamma_0$ leads to the decomposition of the Killing spinor:
\bea
\eta=e_1|\uparrow\uparrow\uparrow\rangle+
e_2|\downarrow\downarrow\uparrow\rangle+
e_3|\downarrow\uparrow\downarrow\rangle+
e_4|\uparrow\downarrow\downarrow\rangle.
\eea
Substituting this into (\ref{Nov8Cmpl}), we 
arrive at the relations
\bea
e_1e_4=e_2 e_3=0,\quad 
-|e_1^2|-|e_2^2|+|e_3^2|+|e_4^2|=\sum |e_i^2|=
-|e_1^2|-|e_3^2|+|e_2^2|+|e_4^2|.
\eea
Combining these relations, we conclude that $e_1=e_2=e_3=0$. In other words, the spinor has only one independent component $|\uparrow\downarrow\downarrow\rangle$ and thus it satisfies the projections
\bea\label{HolonProj}
\Gamma_A\eta=0.
\eea
This relation implies that the two--form (\ref{OmegaForm}) has only holomorphic components\footnote{We use normalization 
$\eps_{\bf 12}=\eps_{\overline{\bf 12}}=1$. In our conventions the determinant of the metric has the following expression in terms of 
{\it curved} epsilon tensor: 
$\sqrt{g}=\frac{1}{4}\eps_{12}\eps_{\overline{12}}$.}:
\bea\label{HolonForm}
\Omega_{ab}=-e^{3A/4}e_a^{\bf A} e_b^{\bf B}\eps_{AB}=-e^{3A/4}\eps_{ab},\quad
\Omega_{{\bar a}b}=\Omega_{{\bar a}{\bar b}}=\Omega_{ym}=0.
\eea
Using this information as well as projector (\ref{Ap1ProjZZ}), we 
can simplify the antisymmetric 
part of (\ref{Nov9DerOmega}):
\bea
&&\nabla_{[\mu}(\eta^\dagger\gamma_{\nu\la]}\Gamma_{\overline{12}}\eta)-
\frac{i}{8}e^{-3A/2}F_{\rho\sigma[\mu}
\eta^\dagger\left(\gamma_{\nu\la]}\Gamma_{\overline{12}}\gamma^{\rho\sigma}+
\gamma^{\rho\sigma}\gamma_{\nu\la]}\Gamma_{\overline{12}}\right)\eta\nonumber\\
&&\qquad+\frac{18}{48}
\eta^\dagger\left(\gamma_{[\nu\la}\Gamma_{\overline{12}}\gamma_{\mu]}{\not\d}A+
{\not\d}A\gamma_{\mu\nu\la}\Gamma_{\overline{12}}\right)\eta=0\nonumber\\
&&\d_{[\mu}\Omega_{\nu\la]}-
\frac{i}{8}e^{-3A/2}F_{\rho\sigma[\mu}~
\eta^\dagger\gamma_{\nu\la]}
[\Gamma_{\overline{12}},\gamma^{\rho\sigma}]\eta
%\nonumber\\
%&&\qquad
+\frac{3}{8}
\eta^\dagger\left(\{\Gamma_{\overline{12}},\gamma_{[\nu\la}\}\d_{\mu]}A+
3{\d}_{[\mu}A\gamma_{\nu\la]}\Gamma_{\overline{12}}\right)\eta=0\nonumber
\eea
We are interested in the situation where two of the indices $(\mu,\nu,\la)$ are holomorphic, then the last equation simplifies:
\bea\label{Nov9DerOmg}
&&\frac{1}{3}\d_{\mu}\Omega_{ab}+\frac{1}{2}\d_\mu A~\Omega_{ab}=
\frac{i}{4}e^{-3A/2}\eps^{\overline{AB}}F_{{\bf\bar B}\sigma[\mu}~
\eta^\dagger\gamma_{ab]}
[\Gamma_{\overline{A}},\gamma^{\sigma}]\eta\nonumber\\
&&\qquad=
\frac{i}{4}e^{-3A/2}\left[2F_{{\bf\bar B}\sigma[\mu}
\Omega_{ab]}e^{\sigma{\bf\bar B}}-
2\eps^{\overline{AB}}F_{{\bf\bar B\bar A}[\mu}
\eta^\dagger\gamma_{ab]}\eta+
2\eps^{\overline{AB}}F_{{\bf\bar B}y[\mu}~
\eta^\dagger\gamma_{ab]}\Gamma_{\overline{A}}\gamma^{y}\eta\right]\nonumber\\
&&\qquad=
\frac{i}{2}e^{-3A/2}\left[F_{\bf\bar B\ \ [\mu}^{\ \ \bf\bar B}
\Omega_{ab]}+
2F_{{\bf\bar 1\bar 2}[\mu}
\eta^\dagger\gamma_{ab]}\eta-\frac{1}{3}\delta^y_\mu
\eps^{\overline{AB}}F_{{\bf\bar B}y[a}~
\eta^\dagger\gamma_{b]}\Gamma_{\overline{A}}\eta\right]
\eea
Here index $\mu$ takes values ${\bar a}$ and $y$. It is convenient to consider these 
two cases separately. We begin with $y$--component:
\bea
&&\frac{1}{3}\d_{y}\Omega_{ab}+\frac{1}{2}\d_y A~\Omega_{ab}=
\frac{i}{6}e^{-3A/2}\left[F_{{\bf\bar B}\ \ y}^{\ \ \bf\bar B}
\Omega_{ab}-\frac{1}{2}e^{3A/4}\eps_{ab}\eps^{cd}
\eps^{\overline{AB}}F_{{\bf\bar B}yc}e_{d}^{\bf A}\delta_{A{\bar A}}\right]\nonumber\\
&&\qquad=\frac{i}{6}e^{-3A/2}\left[F_{{\bf\bar B}\ \ y}^{\ \ \bf\bar B}+
\frac{1}{2}F_{{\bf\bar B}yc}(-e^{c {\bf\bar B}})\right]\Omega_{ab}=
\frac{i}{4}e^{-3A/2}g^{{\bar c}e}F_{{\bar c}ey}\Omega_{ab}\nonumber\\
&&\qquad=-\frac{e^{-3A/2}}{4}g^{{\bar c}e}\d_y(e^{3A/4}J_{{\bar c}e})
\Omega_{ab}=\left[\frac{3}{4}\d_y A+\frac{1}{4}\d_y\log\sqrt{\mbox{det}~g}\right]
\Omega_{ab}
\nonumber
\eea
At the last stage we used the relation (\ref{ComplCoord}): 
$J_{a{\bar b}}=e^{3A/4}g_{a{\bar b}}$.

Simplifying the last equation and using the expression for the holomorphic form (\ref{HolonForm}), we arrive at a relation
\bea\label{PreYderEps}
\frac{1}{3}\d_y \eps_{ab}=\frac{1}{4}\eps_{ab}\d_y\log\sqrt{\mbox{det}~g}
\eea
To determine the $y$--dependence of ${\mbox{det}~g}$, we multiply 
the last equation by 
$\eps^{ab}$:
\bea
\frac{1}{3}\eps^{ab}\d_y \eps_{ab}=\frac{1}{2}\d_y\log\sqrt{\mbox{det}~g},\nonumber
\eea 
and add this relation to its conjugate:
\bea
\frac{2}{3}\d_y \log\sqrt{\mbox{det}~g}=\d_y\log\sqrt{\mbox{det}~g}.\nonumber
\eea
This leads to the conclusion that $\d_y(\mbox{det}~g)=0$ and to simplification in 
equation (\ref{PreYderEps}):
\bea\label{YderEps}
\d_y \eps_{ab}=0.
\eea
Let us now demonstrate that the anti--holomorphic 
derivatives $\d_{\bar c}\eps_{ab}$ vanish as well. 
To do so, we go back to equation (\ref{Nov9DerOmg}):
\bea\label{PreAhDerEps}
&&\frac{1}{3}\d_{\bar c}\Omega_{ab}+\frac{1}{2}\d_{\bar c} A~\Omega_{ab}=
\frac{i}{6}e^{-3A/2}\left[F_{{\bf\bar B}\ \ \ {\bar c}}^{\  \ \bf\bar B}~
\Omega_{ab}+4F_{{\bf\bar 1\bar 2}[a}\eta^\dagger\gamma_{b]{\bar c}}\eta\right]
\nonumber\\
&&=\frac{ie^{-3A/2}}{12}\eps^{\overline{ef}}F_{\overline{ef}d}
\left[-\eps_{\overline{hc}}g^{{\bar h}d}\Omega_{ab}+2\eps_{ab}\eps^{de}J_{e{\bar c}}
\right]=
\frac{ie^{-3A/2}}{12}\eps^{\overline{ef}}F_{\overline{ef}d}
\left[4g_{{\bar c}h}\eps^{hd}-2\eps^{de}g_{e{\bar c}}\right]\Omega_{ab}\nonumber\\
&&=3\eps^{\overline{ef}}\d_{\bar e}A~g_{{\bar f}d}~
g_{{\bar c}h}~\eps^{hd}~\Omega_{ab}=\frac{3}{4}\d_{\bar c}A~\Omega_{ab}.
\eea
At the last stage we used the relation 
$\eps_{ab}g^{b\bar c}=4g_{a\bar b}\eps^{\overline{bc}}$ as well as an expression for the field strength: 
\bea
\eps^{\overline{ef}}F_{\overline{ef}d}=6ie^{3A/4}\eps^{\overline{ef}}
\d_{\bar e}A J_{{\bar f}d}=-6ie^{3A/2}\eps^{\overline{ef}}\d_{\bar e}A g_{{\bar f}d},
\nonumber
\eea
which can be easily extracted from (\ref{FluxWdg}).

Combining equations (\ref{YderEps}) and (\ref{PreAhDerEps}), we conclude that 
\bea\label{DerOmega}
\d_y\eps_{ab}=\d_{\bar c}\eps_{ab}=0:\qquad d\left[e^{-3A/4}\Omega\right]=0.
\eea
Since $\eps_{ab}$ is function of $z_2$ and $z_2$, we can make a holomorphic change of coordinates to set
\bea\label{DefFRame}
\eps_{12}=1,\quad \sqrt{g}=\frac{1}{4}
\eps_{12}\eps_{\overline{12}}=\frac{1}{4}.
\eea
Recalling the expression (\ref{KahlMetr}) for the metric, we arrive for the relation between $e^A$ and Kahler potential:
\bea\label{WarpThKahl}
\d_1{\bar \d}_1K~\d_2{\bar \d}_2K-\d_2{\bar \d}_1K~\d_2{\bar \d}_1K=\frac{1}{4}e^{-3A}
\eea
Notice that this relation holds only in a particular coordinate frame defined by 
(\ref{DefFRame}). 

This completes our discussion of the equations for Killing spinors, let us summarize the results. 
We began with an assumption that eleven dimensional geometry was static and had $SO(6)$ 
symmetry. Since we were interested in the geometries produced by membranes, we also assumed that the flux was electric. Then, solving the equations for the spinors, we arrived at the geometry:
\bea\label{IntermGeom}
ds^2&=&-e^{2A}dt^2+e^{A/2}\left[e^{-3A/2}dy^2+2g_{a{\bar b}}dz^ad{\bar z}^b\right]+
y^2 e^{-A}d\Omega_5^2\\
G_4&=&idt\wedge d(e^{3A/2}g_{a{\bar b}}dz^a\wedge d{\bar z}^b)
\eea 
Moreover, we demonstrated that the metric $h$ has a simple expression in terms of a Kahler 
potential $K(z,{\bar z},y)$:
\bea
g_{a{\bar b}}=e^{3A/2}\d_a{\bar \d}_b K.
\eea
and the warp factor $e^A$ is determined by (\ref{WarpThKahl}). Thus the solution is uniquely parameterized by one real function $K$ and in the next subsection we will use the equations fluxes 
to find the restrictions on this function. 

%=====================================

\subsection{Equations for the flux.}

\label{SectApFlx}

%=====================================

Looking at the solution (\ref{IntermGeom}), we can easily write down the equation of motion for 
the flux $G_4$:
\bea\label{G4EOM10}
d[e^{-7A/2-A/4}y^5*_5d(e^{3A/4}{J})]=0
\eea
Here five--dimensional Hodge duality is taken with respect to the metric which appears in the square brackets in (\ref{IntermGeom}) and the two--form $J$ has been introduced before (see 
(\ref{ComplCoord})):
\bea
J=e^{3A/4}g_{a{\bar b}}dz^a\wedge d\zb^b
\eea
We begin the study of (\ref{G4EOM10}) with analysis of the terms 
which do not contain $dy$:
\bea
d[e^{-7A/2-A/4+3A/4}y^5*_4\d_y(e^{3A/2}{\hat J})]=0=
d_4\left[e^{-3A}*_4\d_y(e^{3A/4}{J})\right]
\eea
Using the relation (\ref{DualJYder}) and anti--self--duality of $J$, we can rewrite the last equation:
\bea
0&=&d_4\left[e^{-9A/4}\left(-\d_y(*J)+\frac{9}{4}\d_y A*J\right)\right]=
d_4\d_y(e^{-9A/4}J)\nonumber
\eea
This is an integrability condition for the equation which has been encountered before. 

Next we look at the $y$--component of the field equation (\ref{G4EOM10}):
\bea\label{PreYComp10}
&&\d_y[e^{-3A}y^5*_4\d_y(e^{3A/4}{J})]-d_4[e^{-9A/2}y^5*_4d_4(e^{3A/4}{J})]=0
\eea
To rewrite the first term we again use equation (\ref{DualJYder}) and anti--self--duality of 
$J$:
\bea
e^{-3A}*_4\d_y(e^{3A/4}{J})=\d_y(e^{-9A/4}J)=\d_y\d_a\db_b K 
dz^a\wedge d\zb^b,
\eea
and to simplify the second term, we use equation (\ref{PreKahlStr})
as well as anti--self--duality of $J$:
\bea
&&d_4[e^{-9A/2}*_4d_4(e^{3A/4}{J})]=3d_4[e^{-15A/4}*_4\left(d_4 A\wedge J\right)]=
-3d_4[e^{-15A/4}{J_m}^p\d_p A dx^m]\nonumber\\
&&=-3d_4[e^{-3A}(\d_a A dz^a-\db_a Ad\zb^a)]=
-2\d_a{\bar\d}_b e^{-3A}dz^a\wedge d{\bar z}^b
\nonumber
\eea
To go to the second line we used the relations
\bea
{J_a}^b=e^{3A/4}\delta_a^b,\quad 
{J_{\bar a}}^{\bar b}=-e^{3A/4}\delta_{\bar a}^{\bar b}.
\eea
Using all this information, equation (\ref{PreYComp10}) can be rewritten as
\bea\label{Nov12PreGauge}
\Delta_y \d_a{\bar \d}_b K+2\d_a{\bar \d}_b e^{-3A}=0
\eea
Since (anti)holomorphic functions can be added to the Kahler potential, we can choose the gauge where
\bea\label{Nov12Gauge}
\Delta_y K+2e^{-3A}=0
\eea

%=====================================

\subsection{Solution with $SO(6)\times U(1)^2\times U(1)_t$ isometry}

\label{SectApIsom}

%=====================================

In this appendix we constructed the most general $SO(6)\times U(1)_t$--invariant geometry 
which is produced by supersymmetric membranes. However, in order to describe the string webs in 
IIB string theory, we are interested in the solutions with two extra isometries (see ansatz 
(\ref{EqnAppIso})). In this subsection we will discuss the additional restrictions which are imposed by these symmetries.  First of all, it is clear that the coordinates $w_1$ and $w_2$ are orthogonal to the 
$y$--direction, so one can repeat the earlier arguments to arrive at the geometry (\ref{ReducMetr}), 
but now the four--dimensional metric is given by
\bea
g_{ij}dx^idx^j={\hat g}_{\alpha\beta}dw^\alpha dw^\beta+{\tilde g}_{MN}dr^M dr^N
\eea
and all functions are invariant under translations in $w_1$, $w_2$. Our goal is to establish the connection between $w_\alpha$ and the complex coordinates $z_a$. To simplify the 
$w$--components of the equation (\ref{SpinSystDfE}) we observe that the ansatz (\ref{EqnAppIso}) implies that
\bea
{\not F}=3\gamma^\alpha{\not W}_\alpha.
\eea
Since we are planning to perform a reduction and T duality in the isometry directions, the Killing spinor should not depend on $w_\alpha$, then the differential equation (\ref{SpinSystDfE}) simplifies:
\bea
\frac{1}{4}\omega_\alpha\eta-\frac{ie^{-3A/2}}{16}
\gamma^\beta{\not W}_\beta{\gamma}_\alpha{\eta}=0
\eea
To extract an expression for $W_\alpha\eta$, we compare the last relation with equation (\ref{Ap1ProjZZ}):
\bea
{\not\d}A\eta-\frac{ie^{-3A/2}}{6}\gamma^\beta{\not W}_\beta\eta=0.
\eea
Since $\gamma_\alpha$ commutes with ${\not W}$, the last two equations can be combined to yield
\bea\label{Nov11}
\omega_\alpha\eta+\frac{3}{2}\gamma_\alpha{\not\d}A\eta-\frac{ie^{-3A/2}}{2}
{\not W}_\alpha{\eta}=0
\eea
Applying an operator $\gamma_y(1-i\Gamma_y\Gamma_S)$ to this relation and using projector 
(\ref{SpinSystOrtPr}), we arrive at the relation which does not contain  
$\Gamma_y$~\footnote{To arrive at this equation we used an expression for the spin connection: 
$\omega_\alpha=-({\not\d}g_{\alpha\beta})\gamma^\beta$.}:
\bea
-\d_yg_{\alpha\beta}\gamma^\beta\eta-
\frac{3}{2}\gamma_\alpha{\d}_yA\eta
-ie^{-3A/2}({W}_\alpha)_{yM}\gamma^M{\eta}=0
\eea
Assuming that a tensor $({W}_\alpha)_{yM}$ has at least one nontrivial 
component\footnote{If all components of $({W}_\alpha)_{yM}$ vanish, then, by applying an 
operator $(1-i\Gamma_y\Gamma_S)$ to the relation (\ref{Nov11}), we can again arrive at 
(\ref{Nov11Proj}) as long as ${\not F}$ is not identically equal to zero. The latter case is 
degenerate and we will not discuss it further.}, we find a projector involving $\gamma_\alpha$ and 
$\gamma_M$:
\bea\label{Nov11Proj}
(a^\alpha\gamma_\alpha+b^M\gamma_M)\eta=0
\eea
Due to projectors (\ref{SpinSystOrtPr}) and (\ref{ProdGam7}), the product of four gamma matrices acts on the Killing spinor in a very simple way, so we can find another relation which is "dual" to 
(\ref{Nov11Proj}). These two relations can be combined to give independent projectors:
\bea\label{Nov11Proj1}
\gamma_M\eta=ia_M^\alpha\gamma_\alpha\eta=i(a_M^\alpha e_\alpha^{\bf a})
{\hat\Gamma}_{\bf a}\eta.
\eea 
Notice that $E_M^{\bf a}=a_M^\alpha e_\alpha^{\bf a}$ can be viewed as a veilbein in the 
$r_1$, $r_2$ directions:
\bea
2{\tilde g}_{MN}\eta=\{\gamma_M,\gamma_N\}\eta=E_M^{\bf a}E_N^{\bf b}
\{{\hat\Gamma}_{\bf a},{\hat\Gamma}_{\bf b}\}\eta=2\delta_{\bf ab}E_M^{\bf a}E_N^{\bf b}\eta.
\eea
We can also use reparameterizations in $r_1$, $r_2$ subspace to set $a_M^\alpha=\delta_M^\alpha$.
Assuming that the coordinates and the vielbein are chosen in this fashion, equation 
(\ref{Nov11Proj1}) can be rewritten as
\bea
({\tilde\gamma}_M-i\delta_M^\alpha{\hat\gamma}_\alpha)\eta=0.
\eea
These relations must be consistent with holomorphic projectors (\ref{HolonProj}), so, going to 
complex coordinates and using (\ref{HolonProj}), we can rewrite the last relation as
\bea
\left(\frac{\d{\bar z}^a}{\d r^M}-i\delta_M^\alpha\frac{\d{\bar z}^a}{\d w^\alpha}\right)
\gamma_{\bar a}\eta=0.
\eea
Since spinor $\eta$ is constrained by (\ref{HolonProj}), the last relation can be satisfied only if the coefficients in front of $\gamma_{\bar 1}$ and $\gamma_{\bar 2}$ vanish separately, so we find
\bea
\frac{\d{\bar z}^a}{\d r^m}-i\frac{\d{\bar z}^a}{\d w^m}=0:\qquad 
{\bar z}^a={\bar z}^a(r_1-iw_1,r_2-iw_2).
\eea
By making a holomorphic reparameterization, we can choose convenient coordinates coordinates:
\bea\label{IsoZcord}
z_1=r_1+iw_1,\qquad z_2=r_2+iw_2.
\eea
Notice that apriori the new coordinates are not consistent with the gauge choice which led to 
(\ref{Nov12Gauge}), so we have to go back to a more general equation (\ref{Nov12PreGauge}). 
In the present case the Kahler potential does not depend on $w$, so both holomorphic and 
anti--holomorphic derivatives should be replaced by the variations with respect to corresponding $r$
and equation (\ref{Nov12PreGauge}) becomes
\bea
\d_M\d_N(\Delta_y K+2e^{-3A})=0:\qquad \Delta_y K+2e^{-3A}=h(y)
\eea
Since Kahler potential can be shifted by an arbitrary function of $y$ without affecting the metric, we can again impose the gauge (\ref{Nov12Gauge}):
\bea
\Delta_y K+2e^{-3A}=0.
\eea
To summarize, we showed that solutions which have two translational isometries, in addition to 
$SO(6)\times U(1)_t$, are still described by the system 
(\ref{IntermGeom}), (\ref{WarpThKahl}), (\ref{Nov12Gauge}), 
and the isometry directions 
must be related with complex coordinates in a very simple way (\ref{IsoZcord}). 

%=====================================

\subsection{Summary of the solution.}

\label{SectApSmr}

%=====================================

Let us collect the results derived in this appendix. 
We have found the most general eleven--dimensional geometry which 
preserved eight supercharges along with $SO(6)\times U(1)_t$ bosonic isometries:
\bea\label{FinalGeom}
ds^2&=&-e^{2A}dt^2+2e^{2A}h_{a{\bar b}}dz^ad{\bar z}^b+
e^{-A}(dy^2+y^2d\Omega_5^2)\\
G_4&=&idt\wedge d(e^{3A}h_{a{\bar b}}dz^a\wedge d{\bar z}^b),\quad
h_{a{\bar b}}=\d_a{\bar \d}_b K.\nonumber
\eea 
The solution is parameterized in terms of the  Kahler 
potential $K(z,{\bar z},y)$ which should satisfy two differential equations (\ref{WarpThKahl}), (\ref{Nov12Gauge}): 
\bea
\d_1{\bar \d}_1K~\d_2{\bar \d}_2K-\d_2{\bar \d}_1K~\d_2{\bar \d}_1K=
-\frac{1}{8}\Delta_y K,\quad
\Delta_y K=-2e^{-3A}.
\eea
It is easy to guess a generalization of the solution (\ref{FinalGeom}) to 
the situations without $SO(6)$ isometries:
\bea
ds^2&=&-e^{2A}dt^2+2e^{2A}h_{a{\bar b}}dz^ad{\bar z}^b+
e^{-A}d{\bf y}_6\\
G_4&=&idt\wedge d(e^{3A}h_{a{\bar b}}dz^a\wedge d{\bar z}^b),\quad
h_{a{\bar b}}=\d_a{\bar \d}_b K.
\eea
Starting with this ansatz, we can explicitly check that conditions for supersymmetry and equations of motion are satisfied as long as 
Kahler potential obeys the differential equations 
(\ref{WarpThKahl}), (\ref{Nov12Gauge})\footnote{Similar checks for 
other brane intersections were 
performed in \cite{myCM}.}.
 
To find the geometries produced by the string webs, one needs to look 
at  eleven dimensional geometries which have two extra translational isometries in addition to $U(1)_t\times SO(6)$. The restrictions coming from this requirement were discussed in the last subsection where we found that the complex coordinates must have the form 
\bea
z_1=r_1+iw_1,\qquad z_2=r_2+iw_2,
\eea
where $w_1$ and $w_2$ are the isometry directions. To find the IIB solution describing the string web, we write the eleven dimensional metric in a slightly more explicit form:
\bea
ds_{M}^2&=&-e^{2A}dt^2+e^{2A}h_{ab}
(dw^adw^b+dr^adr^b)+e^{-A}d{\bf y}_6\\
G_4&=&dt\wedge d(e^{3A}h_{a{b}}~dr^a\wedge d{w}^b),
\quad
h_{a{b}}=\frac{1}{2}\d_a{\d}_b K
\eea
Reducing this geometry along $w_1$ and T dualizing along $w_2$, 
we find the solution in IIB SUGRA:
\bea
ds_{IIB}^2&=&\sqrt{h_{11}}\left[-e^{3A}dt^2+
e^{3A}h_{ab}dr^adr^b+\frac{e^{-3A}}{\mbox{det}h}dw_2^2+
d{\bf y}_6^2\right]\\
e^{2\Phi}=\frac{h_{11}^2}{\mbox{det}h},&&
C^{(0)}=-\frac{h_{12}}{h_{11}},\quad 
B=e^{3A}h_{1a}~dt\wedge dr^a,\quad
C^{(2)}=e^{3A}h_{2a}~dt\wedge dr^a\nonumber
\eea
Recalling that equation (\ref{WarpThKahl}) implies a very simple 
expression for the determinant ($\mbox{det}h=e^{-3A}$), we arrive at 
the metric in 
the Einstein frame:
\bea
ds_{E}^2&=&e^{-3A/4}\left[-e^{3A}dt^2+
e^{3A}h_{ab}dr^adr^b+dw_2^2+d{\bf y}_6^2\right].
\eea

%=====================================

\section{Embedding of 1/2--BPS bubbling geometries}

\label{SectApEmb}

\renewcommand{\theequation}{B.\arabic{equation}}
\setcounter{equation}{0}

%=====================================

In section \ref{SectIIBbl} we discussed bubbling solutions preserving eight supercharges, in particular, 1/2--BPS geometries constructed in 
\cite{LLM} fall into this category. However, embedding of 1/2--BPS geometries into the general ansatz (\ref{DonosSoln}) requires some algebraic manipulations, and we present them in this Appendix. We will also embed the 1/2--BPS solutions of M theory \cite{LLM} into the ansatz (\ref{Mbubble1}). 

%=====================================

\subsection{IIB supergravity}

\label{SectApEmbIIB}

%=====================================

Ten--dimensional case was discussed in section \ref{SectIIBEx}, where
we wrote down an embedding of 1/2--BPS solutions of \cite{LLM} into the general 1/4--BPS geometry 
(\ref{DonosSoln}). Here we will derive the relations 
(\ref{Jan7Eqn1})--(\ref{Jan7Eqn4}). 

Half--BPS geometries constructed in \cite{LLM} were parameterized by one harmonic function ${\tilde Z}(z,\zb;x)$, and, with slight change of notation, the metric can be written as
\bea\label{LLMIIB}
ds^2&=&-{\tilde h}^{-2}(dt+V)^2+{\tilde h}^2(dx^2+dzd\zb)+
xe^{H}d\Omega_3^2+xe^{-H}d{\tilde\Omega}_3^2,\\
{\tilde h}^{-2}&=&2x\cosh H,\quad (dV)_{z\zb}=\frac{i}{2x}\d_x {\tilde Z},
\quad x\d_x V=i(dz\d_z-d\zb\d_{\zb}){\tilde Z},
\quad {\tilde Z}=\frac{1}{2}\tanh H,\nonumber\\
&&\qquad
4\d_z\d_{\zb}{\tilde Z}+x\d_x\left(\frac{\d_x {\tilde Z}}{x}\right)=0.
\nonumber
\eea
To compare this with (\ref{DonosSoln}), we identify the three dimensional sphere appearing in (\ref{DonosSoln}) with $S^3$ in the metric (\ref{LLMIIB}), while embedding the Killing direction $\psi$ from 
(\ref{DonosSoln}) into ${\tilde S^3}$:
\bea\label{ThrSphrA}
d{\tilde\Omega}_3^2=d\theta^2+\cos^2\theta d\psi^2+
\sin^2\theta d{\tilde\phi}^2.
\eea
As discussed in section \ref{SectIIBEx}, this identification follows naturally from the analysis of the R symmetry group. Once the embedding of 
$S^3$ and $\psi$ is specified, we can compare the appropriate warp factors in (\ref{DonosSoln}) and (\ref{LLMIIB}), this leads to the following relations:
\bea
y=x\cos\theta,\quad e^G=\frac{e^H}{\cos\theta},\quad
Z=\frac{1}{2}\tanh G=
\frac{1}{2}\frac{e^H-e^{-H}c_\theta^2}{e^H+e^{-H}c_\theta^2}.
\eea 
By comparing the coordinate dependence of the Killing spinor in 1/2--and 1/4--BPS cases, one concludes that an appropriate coordinate on the Kahler base in (\ref{DonosSoln}) is $\phi={\tilde\phi}+t$ rather 
than ${\tilde\phi}$. Implementing this shift in 
(\ref{LLMIIB})--(\ref{ThrSphrA}) and comparing with (\ref{DonosSoln}), we reproduce the correct $g_{tt}$ and find the expression for 
one--form $\omega$:
\bea\label{Jan5OmForm}
\omega=\frac{\cosh H}{\cos\theta\cosh G}V+
\frac{e^{-G}\tan^2\theta}{e^G+e^{-G}} d\phi.
\eea
Let us now extract the metric of the four--dimensional Kahler space appearing in (\ref{DonosSoln}). We begin with looking at the line element in the $(x,\theta)$ subspace of (\ref{LLMIIB}) and subtracting the $dy^2$ term from (\ref{DonosSoln}):
\bea\label{TempJan5}
ds^2_{x,\theta,\perp}&=&\frac{dx^2}{2x\cosh H}+xe^{-H}d\theta^2-
\frac{dy^2}{2y\cosh G}=\frac{y\cosh H}{e^H\cosh G}\left[
\frac{e^H\sin\theta dx}{2y\cosh H}+d\theta\right]^2\nonumber\\
&=&\frac{e^{-H}\cosh H}{y\cosh G}\left[d(x\sin\theta)+
({\tilde Z}-\frac{1}{2})\sin\theta dx\right]^2
\eea
To proceed its is convenient to introduce a new function\footnote{
This definition was inspired by the discussion of the M theory case, where function ${D}$ arises in a more natural way (see next subsection).} ${D}$:
\bea\label{DefDIIB}
{\tilde Z}-\frac{1}{2}\equiv -x\d_x{D}.
\eea
Rewriting the metric (\ref{TempJan5}) in terms of $D$, one finds
\bea\label{Jan5Ap1}
ds^2_{x,\theta,\perp}
&=&\frac{e^{-H}\cosh H}{y\cosh G}\left[e^{D}d(x\sin\theta e^{-D})+
x\sin\theta d_2D\right]^2
\eea
Here $d_2$ denotes a differential along the directions $z,\zb$. The structure of equation (\ref{Jan5Ap1}) suggests that it is convenient to trade variables $(x,\theta)$ for new coordinates
\bea
y=x\cos\theta,\qquad W=x\sin\theta e^{-{D}}.
\eea 
In particular, as we have shown, these two coordinates are 
orthogonal to each other. 

Notice that the relation (\ref{DefDIIB}) does not determine $D$ uniquely: any function of $(z,\zb)$ can be added to it without affecting 
(\ref{DefDIIB}). To fix this ambiguity, we rewrite the expression for 
$x$--derivative of $V$ in terms of $D$ (see (\ref{LLMIIB}):
\bea
x\d_x V=i(dz\d_z-d\zb\d_{\zb})\left[{\tilde Z}-\frac{1}{2}\right]=
-ix\d_x(dz\d_z-d\zb\d_{\zb}){D}.
\eea
This relation implies that $D$ can be determined completely, by requiring that 
\bea\label{DefDIIBa}
V=-i(dz\d_z-d\zb\d_{\zb}){D}
\eea
in addition to (\ref{DefDIIBa}). 

Let us now look at the $t$--$\phi$ subspace. We already extracted the expression for $\omega$, and now we evaluate the rest of the metric:
\bea\label{Jan5Ap2}
ds^2_{t,\phi}&\equiv&-2x\cosh H(dt+V)^2+
xe^{-H}\sin^2\theta (d\phi-dt)^2\nonumber\\
&=&-h^{-2}(dt+\omega)^2+
\frac{x\cosh H}{e^G\cosh G}\tan^2\theta
(V+d\phi)^2\\
&&+\left[y^2h^2 e^{-2G}\tan^4\theta-
\frac{x\cosh H}{e^G\cosh G}\tan^2\theta+
ye^{-G}\tan^2\theta
\right]d\phi^2\nonumber
\eea
Simplifications show that the last line gives vanishing contribution:
\bea
\frac{e^{-G}}{2\cosh G}\tan^2\theta-
\frac{\cosh H}{\cosh G}\frac{1}{\cos\theta}+1=
\frac{e^{-G}}{2\cosh G}\tan^2\theta-
\frac{e^{-H}}{2\cosh G}\frac{\sin^2\theta}{\cos\theta}=0.\nonumber
\eea
Using expressions (\ref{Jan5Ap1}) and (\ref{Jan5Ap2}), we can 
extract the Kahler metric appearing in (\ref{DonosSoln}):
\bea\label{Jan5PreKahl}
2\d_a{\bar\d}_b Kdz^ad\zb^b&=&\frac{Z+\frac{1}{2}}{h^{2}}\left[
\frac{e^{-H}\cosh H}{y\cosh G}\left\{\frac{y^2}{\cot^2\theta}
(V+d\phi)^2+e^{2D}(dW+Wd_2 D)^2\right\}+
\frac{dzd\zb}{{\tilde h}^{-2}}
\right]\nonumber\\
&=&\left[\frac{\cosh H}{\cosh G}\frac{e^{2D}}{\cos\theta}
\left\{W^2
(V+d\phi)^2+(dW+Wd_2 D)^2\right\}+ye^G{\tilde h}^2 dzd\zb\right]
\eea
Let us try to guess the complex structure. We already have a natural 
complex structure in two dimensional space spanned by $z$, 
${\bar z}$, so only the terms inside curly bracket in (\ref{Jan5PreKahl}) need additional analysis. The metric appearing there can be simplified using the expression (\ref{DefDIIBa}) for $V$:
\bea
ds_2^2&=&W^2\left(d\phi-i\{\d_z D dz-{\d}_{\bar z} D d{\bar z}\}\right)^2+
\left(dW+W\{\d_z D dz+{\d}_{\bar z} D d{\bar z}\}\right)^2
\nonumber\\
&=&
dW^2+W^2 d\phi^2+4W^2\d_z D {\d}_{\bar z} D dz d{\bar z}+
\left[2\d_z D dz(WdW-iW^2 d\phi)+c.c.\right].\nonumber
\eea
This relation suggests a natural complex coordinate 
$w=We^{i\phi}$:
\bea
ds_2^2&=&
(dw+2w{\d}_z D dz)(d{\bar w}+2{\bar w}\d_{\bar z} D d{\bar z}).
\eea 
Now the Kahler metric and one--form $\omega$ can be rewritten
in terms of complex coordinates $z$ and $w$:
\bea\label{Jan5Metr}
&&2\d_a{\bar\d}_b Kdz^ad\zb^b
=\left[\frac{\cosh H}{\cosh G}\frac{e^{2D}}{\cos\theta}
\left|dw+2w{\tilde\d}_z D dz\right|^2+ye^G{\tilde h}^2 dzd\zb\right]
\nonumber\\
&&\omega=-i\frac{\cosh H}{\cos\theta\cosh G}({\tilde\d}_z Ddz-
{\tilde\d}_{\zb} Dd\zb)-
\frac{i}{2}\frac{e^{-G}\tan^2\theta}{e^G+e^{-G}} d\log\frac{w}{\wb}
\eea
Notice that the derivatives ${\tilde\d}_z$ and ${\tilde\d}_{\bar z}$ are 
taken at constant $\theta$ and $x$, in contrast to 
${\d}_z$ and ${\d}_{\bar z}$ which are taken at constant $y$ and $W$.

To find the relation between $D$ and Kahler potential parameterizing 
1/4--BPS geometry (\ref{DonosSoln}), we evaluate the determinant of 
the metric (\ref{Jan5Metr}):
\bea
\mbox{det}h_{a{\bar b}}=\frac{1}{4}\frac{e^{G+2D}}{e^G+e^{-G}}=
\frac{1}{4}\left(Z+\frac{1}{2}\right)e^{2D}
\eea
Comparing this with corresponding equation in (\ref{DonosSoln}) (and 
choosing the gauge $W(z)=\frac{1}{2}$ there), we relate $D$ and 
derivative of the Kahler potential:
\bea\label{Jan5KahlD}
y^{-1}\d_y K=2D-\log y.
\eea

To summarize, we showed that 1/2--BPS bubbling solutions constructed in \cite{LLM} can be embedded in the more general 
1/4--BPS ansatz. To construct the appropriate map, one should first rewrite the 1/2--BPS geometries (\ref{LLMIIB}) in terms of function 
$D$ rather than ${\tilde Z}$:
\bea
x\d_x D=\frac{1}{2}-{\tilde Z},\qquad V=-i(dz\d_z-d\zb\d_{\zb}){D}.
\eea
Then, defining a new variable $y$ and holomorphic coordinates $z,w$:
\bea
y=x\cos\theta,\quad w=x\sin\theta e^{-D+i\phi}
\eea
one arrives at the map (\ref{Jan5KahlD}) between 
(\ref{LLMIIB})--(\ref{ThrSphrA}) and (\ref{DonosSoln}).

In section \ref{SectIIBEx} we also needed the expression for $\d_x$ 
(which is taken at constant $z$, $\theta$) in terms of $\d_y$ (which is computed for fixed $z$, $w$). 
To find the desired relation,
we consider various differentials at constant values of $(z,\zb,w,\wb)$:
\bea
dW=e^{-{D}}\left[s_\theta(1-x\d_x {D})dx+c_\theta xd\theta\right]=0: \ 
dy=c_\theta dx-xs_\theta d\theta=\frac{dx}{c_\theta}
(1-xs_\theta^2\d_x {D})\nonumber
\eea
This leads to a general expression for $\d_y$ in terms of $\d_x$, and we will be particularly interested in its implications for the derivatives 
of ${D}$:
\bea\label{RelXYMIIB}
\d_y =\frac{c_\theta}{1-s_\theta^2 x\d_x {D}}\d_x,\quad 
\d_y {D}=\frac{c_\theta\d_x {D}}{1-s_\theta ^2x\d_x{D}}.
\eea

We conclude this subsection by noticing that the linear equation obeyed by ${\tilde Z}$ implies a simple Laplace equation for function $D$:
\bea
4\d_z\d_{\zb} D+x^{-1}\d_x(x\d_x D)=0.
\eea

%=====================================

\subsection{M theory}

\label{SectApEmbM}

%=====================================

Let us now embed the $1/2$--BPS eleven dimensional solutions
constructed in \cite{LLM} into the general $1/4$--BPS ansatz 
(\ref{Mbubble1}). We begin with recalling the metric found in \cite{LLM}:
\bea\label{LLMthry}
ds^2&=&-4e^{2\la}\cosh^2\xi (d\tau+V)^2+\frac{e^{-4\la}}{\cosh^2\xi}
\left[dx^2+e^{\tilde D}dzd{\bar z}\right]+4e^{2\la}d\Omega_5^2+
x^2e^{-4\la}d\Omega_2^2\nonumber\\
&&e^{-6\la}=\frac{\d_x {\tilde D}}{x(1-x\d_x {\tilde D})},\quad 
V=\frac{i}{2}(\d_z-\d_{\bar z}){\tilde D},\quad \sinh\xi=xe^{-3\la}
\eea
Comparing this with (\ref{Mbubble1}): 
\bea\label{Mbubble1App}
ds^2&=&-\frac{4e^{2\la}}{9}\cosh^2\zeta(dt+\rho)^2+
4e^{2\la}d\Omega_5^2+e^{-4\la}\left(
h_{ij}dx^idx^j+\frac{dy^2}{\cosh^2\zeta}\right),
\eea
we observe that function $e^\la$ appearing in both metrics is the 
same. To make further comparison, we write the metric on $S^2$ as
\bea
d\Omega_2^2=d\theta^2+\sin^2\theta d{\tilde\phi}^2
\eea
and introduce a shift and a rescaling\footnote{This change of variables  can be extracted by comparing Killing spinors for $1/2$ and $1/4$--BPS solutions, but we will not present the argument here.}: 
${\tilde\phi}=\phi-2\tau$,\quad $\tau=\frac{t}{3}$. This leads to identifications:
\bea
&&\cosh^2\zeta=\cosh^2\xi-\sinh^2\xi\sin^2\theta=
1+(\sinh\xi\cos\theta)^2:
\quad y=x\cos\theta.\nonumber\\
&&\rho=3V+3\frac{x^2\sin^2\theta e^{-6\la}}{1+y^2 e^{-6\la}}(V+
\frac{1}{2}d\phi)
=3V\frac{\cosh^2\xi}{\cosh^2\zeta}+\frac{3}{2}\tan^2\theta \tanh^2\zeta 
d\phi
\eea
To extract the metric of the four dimensional Kahler corresponding to
(\ref{LLMthry}), we begin with looking at the $x$--$\theta$ subsector:
\bea\label{Dec22Ap1}
ds^2_{x,\theta,\perp}&\equiv& \frac{e^{-4\la}dx^2}{\cosh^2\xi}+
x^2 e^{-4\la}d\theta^2-\frac{e^{-4\la}dy^2}{\cosh^2\zeta}=
\frac{e^{-4\la}}{\cosh^2\zeta}\left(\frac{s_\theta dx}{ch_\xi}+
xc_\theta ch_\xi d\theta \right)^2\\
&=&e^{-4\la}\frac{ch_\xi^2}{ch_\zeta^2}
\left(d(xs_\theta)-xs_\theta\d_x{\tilde D} dx\right)=
e^{-4\la}\frac{ch_\xi^2}{ch_\zeta^2}
\left(e^{\tilde D} d(xe^{-{\tilde D}}s_\theta)+
xs_\theta d_2{\tilde D}\right)^2
\nonumber
\eea 
Here $d_2$ denotes a differential along the directions $z$, ${\bar z}$.
To simplify the last expression the following relation was used:
\bea\label{XderD}
\frac{1}{ch^2\xi}-1=-\frac{x^2 e^{-6\la}}{1+x^2 e^{-6\la}}=
-x\d_x {\tilde D}
\eea
The structure of equation (\ref{Dec22Ap1}) suggests that it is convenient to trade variables $(x,\theta)$ for new coordinates
\bea
y=x\cos\theta,\qquad W=x\sin\theta e^{-{\tilde D}}.
\eea 
In particular, as we have shown, these two coordinates are 
orthogonal to each other. 

Let us now look at the $t$--$\phi$ sector. We already extracted the expressions for $g_{tt}$ and $\rho$, and now we evaluate the rest of the metric:
\bea\label{Dec22Ap2}
ds^2_{t,\phi}&\equiv&-4e^{2\la}\cosh^2\xi(\frac{dt}{3}+V)^2+
x^2e^{-4\la}\sin^2\theta (d\phi-\frac{2}{3}dt)^2\nonumber\\
&=&-\frac{4e^{2\la}}{9}\cosh^2\zeta(dt+\rho)^2+
4e^{2\la}\frac{ch^2_\xi sh^2_\xi}{ch^2_\zeta}s_\theta^2
(V+\frac{1}{2}d\phi)^2\\
&&+x^2e^{-4\la}\sin^2\theta d\phi^2-
e^{2\la}\frac{ch_\xi^2sh_\xi^2}{ch_\zeta^2}s^2_\theta d\phi^2+
e^{2\la}\frac{sh^4_\xi}{ch^2_\zeta}s_\theta^4d\phi^2\nonumber
\eea
Simplifications show that the last line gives vanishing contribution:
\bea
sh^2_\xi -\frac{ch_\xi^2sh_\xi^2}{ch_\zeta^2}+
\frac{sh^4_\xi}{ch^2_\zeta}s_\theta^2=\frac{1}{ch^2_\zeta}(
sh^2_\xi +sh^4_\xi-ch_\xi^2sh_\xi^2)=0
\eea
Using expressions (\ref{Dec22Ap1}) and (\ref{Dec22Ap2}), we can easily extract the Kahler metric appearing in (\ref{Mbubble1App}):
\bea\label{Dec22PreKahl}
h_{ij}dx^idx^j=\frac{ch_\xi^2}{ch_\zeta^2}e^{2{\tilde D}}\left[
\left(dW+W d_2{\tilde D}\right)^2+
4w^2(V+\frac{1}{2}d\phi)^2\right]+
\frac{e^{\tilde D}dzd{\bar z}}{\cosh^2\xi}
\eea
Let us try to guess the complex structure. We already have a natural 
complex structure in two dimensional space spanned by $z$, 
${\bar z}$, so we only the square bracket in (\ref{Dec22PreKahl}) needs additional analysis. The metric appearing there can be simplified using the expression (\ref{LLMthry}) for $V$:
\bea
ds_2^2&=&\left(dW+W\{\d_z D dz+{\d}_{\bar z} D d{\bar z}\}\right)^2+
W^2\left(d\phi+i\{\d_z D dz-{\d}_{\bar z} D d{\bar z}\}\right)^2
\nonumber\\
&=&
dW^2+W^2 d\phi^2+4W^2\d_z D {\d}_{\bar z} D dz d{\bar z}+\left[
2\d_z D dz(WdW+iW^2 d\phi)+c.c.\right].\nonumber
\eea
This relation suggests a natural complex coordinate 
$w=We^{-i\phi}$:
\bea
ds_2^2&=&
(dw+2w\d_z D dz)(d{\bar w}+2{\bar w}\d_{\bar z} D d{\bar z}).
\eea 
Now the Kahler metric and one--form $\rho$ can be rewritten
in terms of $z$ and $w$:
\bea\label{Dec23Metr}
h_{ij} dx^i dx^j&=&
\frac{\cosh^2\xi e^{2{\tilde D}}}{\cosh^2\zeta}
(dw+2w{\tilde \d}_z D dz)(d{\bar w}+2{\bar w}{\tilde\d}_{\bar z} D 
d{\bar z})+
\frac{e^{\tilde D}dz d{\bar z}}{\cosh^2\xi}\nonumber\\
\rho&=&\frac{3\cosh^2 \xi}{2\cosh^2\zeta}
i({\tilde \d}_z {\tilde D} dz-{\tilde \d}_{\bar z} {\tilde D} d{\bar z})+
\frac{3i}{4}\tan^2\theta \tanh^2\zeta d\log\frac{w}{{\bar w}}
\eea
Notice that the derivatives ${\tilde\d}_z$ and ${\tilde\d}_{\bar z}$ are 
taken at 
constant $\theta$ and $x$, in contrast to 
${\d}_z$ and ${\d}_{\bar z}$ which are taken at constant $y$ and $w$.
The relation between these two sets will be found below. Even without 
knowing such map, we can extract a very useful relation by taking a determinant of the metric (\ref{Dec23Metr}):
\bea
\mbox{det}h_{a{\bar b}}=\frac{e^{3{\tilde D}}}{4\cosh^2\zeta}
\eea
Comparing this with equation (\ref{GMSWanCont}), we conclude that 
$D={\tilde D}$. In principle, the relations
\bea\label{MthryMap}
D(We^{-i\phi},z;y)={\tilde D}(z;x),\quad y=x\cos\theta,\quad 
W=x\sin\theta e^{-{\tilde D}}
\eea
provide a complete embedding of $1/2$--BPS states \cite{LLM} into 
the more general ansatz (\ref{Mbubble1}), but to gain a better 
understanding of this map we will now study it in more detail. 

It is very useful to relate various derivatives appearing in the 
description of $1/2$--BPS states with their counterparts for $1/4$--BPS geometries. 
To find the relation between $({\tilde\d}_z,{\tilde\d}_{\bar z})$ and
$({\d}_z,{\d}_{\bar z})$, we consider a variation of $D$ keeping 
$\theta$ and $x$ (and thus $y$) fixed:
\bea
{\tilde\d}_i {\tilde D} dX^i=\d_i {\tilde D}|_{yW} dX^i+
\d_W {\tilde D}|_{Xy} dW=
\d_i {\tilde D}|_{yw} dX^i
-\d_W {\tilde D}|_{Xy} W{\tilde\d}_i {\tilde D} dX^i
\eea
Solving this equation we find the expression for 
${\tilde\d}_z {\tilde D}$:
\bea
{\tilde\d}_z {\tilde D}=\frac{{\d}_z {\tilde D}}{1+W\d_W {\tilde D}}=
\frac{{\d}_z {\tilde D}}{1+w\d_w {\tilde D}+
{\bar w}\d_{\bar w} {\tilde D}}
\eea
and a similar relation for ${\tilde\d}_{\bar z} {\tilde D}$.
To simplify this further, we need to evaluate the derivative
$\d_W {\tilde D}|_{X,y}=\d_x{\tilde D}\cdot \d_W x|_{X,y}$:
\bea
dW|_{y,X}&=&\d_x(xe^{-{\tilde D}})s_\theta dx+xc_\theta e^{-{\tilde D}}
d\theta|_{y,X}=\left[\d_x(xe^{-{\tilde D}})s_\theta+
xc_\theta e^{-{\tilde D}}
\frac{c_\theta^2}{ys_\theta}\right]dx\nonumber\\
&=&e^{-{\tilde D}}\left[-xs_\theta \d_x {\tilde D}+\frac{1}{s_\theta}\right]
dx=e^{-{\tilde D}}\frac{\cosh^2\zeta}{s_\theta\cosh^2\xi}
dx\nonumber\\
\d_W {\tilde D}|_{X,y}&=&e^{\tilde D}\d_x {\tilde D}
\frac{s_\theta\cosh^2\xi}{\cosh^2\zeta}=
\frac{e^{\tilde D}xs_\theta e^{-6\la}}{\cosh^2\zeta}=
\frac{\tanh^2\zeta}{w\cot^2\theta}
,\quad
\frac{1}{1+W\d_W {\tilde D}}=\frac{\cosh^2\zeta}{\cosh^2\xi}.
\nonumber
\eea
Here we used equation (\ref{XderD}) and definitions of $\zeta$ and $\xi$.
Now the expression for $\rho$ (\ref{Dec23Metr}) can be rewritten in terms of derivatives appropriate for the $1/4$--BPS case:
\bea
\rho=\frac{3i}{2}
({\d}_z {\tilde D} dz-{\d}_{\bar z} {\tilde D} d{\bar z})+
\frac{3i}{4}W\d_W{\tilde D} d\log\frac{w}{{\bar w}}=
\frac{3i}{2}(dz\d_z+dw\d_w-cc){\tilde D}
\eea
This equation agrees with (\ref{GMSWanCont}).

To find the relation between $\d_x$ and $\d_y$, we consider various differentials at constant values of $z,{\bar z},W$:
\bea
dW=e^{-{\tilde D}}\left[s_\theta(1-x\d_x {\tilde D})dx+
c_\theta xd\theta\right]=0: \ 
dy=c_\theta dx-xs_\theta d\theta=\frac{dx}{c_\theta}
(1-xs_\theta^2\d_x {\tilde D}).\nonumber
\eea
This leads to a general expression for $\d_y$ in terms of $\d_x$, and we will be interested in its implications for the derivatives of 
${\tilde D}$:
\bea\label{RelXYMder}
\d_y =\frac{c_\theta}{1-s_\theta^2 x\d_x {\tilde D}}\d_x,\quad 
\d_y {\tilde D}=\frac{c_\theta\d_x {\tilde D}}{1-s_\theta ^2x\d_x{\tilde D}}
\eea
Recalling the expression (\ref{XderD}) for the $x$--derivative of 
${\tilde D}$, we can simplify the last relation:
\bea
\d_y {\tilde D}=\frac{c_\theta xe^{-6\la}}{1+c^2_\theta x^2e^{-6\la}}=
\frac{ye^{-6\la}}{1+y^2e^{-6\la}}
\eea
Comparing this with similar relation in (\ref{GMSWanCont}), we 
conclude that 
\bea
\d_y (D-{\tilde D})=0
\eea
This serves as a consistency check of the map (\ref{MthryMap}).


\begin{thebibliography}{99}
%
\bibitem{StromVafa}
A.~Strominger and C.~Vafa,
  %``Microscopic Origin of the Bekenstein-Hawking Entropy,''
  Phys.\ Lett.\  B {\bf 379}, 99 (1996), hep-th/9601029.
  %%CITATION = PHLTA,B379,99;%%
%
\bibitem{HanWitten}
A.~Hanany and E.~Witten,
  %``Type IIB superstrings, BPS monopoles, and three-dimensional gauge
  %dynamics,''
  Nucl.\ Phys.\  B {\bf 492}, 152 (1997), hep-th/9611230;\\
  %%CITATION = NUPHA,B492,152;%%
E.~Witten,
  %``Solutions of four-dimensional field theories via M-theory,''
  Nucl.\ Phys.\  B {\bf 500}, 3 (1997), hep-th/9703166;
%E.~Witten,
  %``Branes and the dynamics of {QCD},''
  Nucl.\ Phys.\  B {\bf 507}, 658 (1997), hep-th/9706109;\\
  %%CITATION = NUPHA,B507,658;%%
  %%CITATION = NUPHA,B500,3;%%
S.~Elitzur, A.~Giveon, D.~Kutasov, E.~Rabinovici and A.~Schwimmer,
  %``Brane dynamics and N = 1 supersymmetric gauge theory,''
  Nucl.\ Phys.\  B {\bf 505}, 202 (1997), hep-th/9704104;\\
  %%CITATION = NUPHA,B505,202;%%
A.~Giveon and D.~Kutasov,
  %``Brane dynamics and gauge theory,''
  Rev.\ Mod.\ Phys.\  {\bf 71}, 983 (1999), hep-th/9802067.
  %%CITATION = RMPHA,71,983;%%
%
\bibitem{BerDougLeih}
M.~Berkooz, M.~R.~Douglas and R.~G.~Leigh,
  %``Branes intersecting at angles,''
  Nucl.\ Phys.\  B {\bf 480}, 265 (1996), hep-th/9606139.
  %%CITATION = NUPHA,B480,265;%%
%
\bibitem{Ohta}
N.~Ohta and P.~K.~Townsend,
  %``Supersymmetry of M-branes at angles,''
  Phys.\ Lett.\  B {\bf 418}, 77 (1998), hep-th/9710129.
  %%CITATION = PHLTA,B418,77;%%
%
\bibitem{uranga}
A.~M.~Uranga,
  %``Chiral four-dimensional string compactifications with intersecting
  %D-branes,''
  Class.\ Quant.\ Grav.\  {\bf 20}, S373 (2003), hep-th/0301032;\\
  %%CITATION = CQGRD,20,S373;%%
R.~Blumenhagen, M.~Cvetic, P.~Langacker and G.~Shiu,
  %``Toward realistic intersecting D-brane models,''
  Ann.\ Rev.\ Nucl.\ Part.\ Sci.\  {\bf 55}, 71 (2005), hep-th/0502005.
  %%CITATION = ARNUA,55,71;%%
%
\bibitem{StrWeb}
K.~Dasgupta and S.~Mukhi,
  %``BPS nature of 3-string junctions,''
  Phys.\ Lett.\  B {\bf 423}, 261 (1998), hep-th/9711094;\\
  %%CITATION = PHLTA,B423,261;%%
A.~Sen,
  %``String network,''
  JHEP {\bf 9803}, 005 (1998), hep-th/9711130;\\
  %%CITATION = JHEPA,9803,005;%%
S.~J.~Rey and J.~T.~Yee,
  %``BPS dynamics of triple (p,q) string junction,''
  Nucl.\ Phys.\  B {\bf 526}, 229 (1998), hep-th/9711202.
  %%CITATION = NUPHA,B526,229;%%
%
\bibitem{D5Webs}
O.~Aharony and A.~Hanany,
  %``Branes, superpotentials and superconformal fixed points,''
  Nucl.\ Phys.\  B {\bf 504}, 239 (1997), hep-th/9704170;\\
  %%CITATION = NUPHA,B504,239;%%
B.~Kol,
  %``5d field theories and M theory,''
  JHEP {\bf 9911}, 026 (1999), hep-th/9705031;\\
  %%CITATION = JHEPA,9911,026;%%
O.~Aharony, A.~Hanany and B.~Kol,
  %``Webs of (p,q) 5-branes, five dimensional field theories and grid
  %diagrams,''
  JHEP {\bf 9801}, 002 (1998), hep-th/9710116.
  %%CITATION = JHEPA,9801,002;%%
%
\bibitem{BHWebs}
R.~Dijkgraaf, E.~P.~Verlinde and H.~L.~Verlinde,
  %``Counting dyons in N = 4 string theory,''
  Nucl.\ Phys.\  B {\bf 484}, 543 (1997), hep-th/9607026;\\
  %%CITATION = NUPHA,B484,543;%%
G.~Lopes Cardoso, B.~de Wit, J.~Kappeli and T.~Mohaupt,
  %``Asymptotic degeneracy of dyonic N = 4 string states and black hole
  %entropy,''
  JHEP {\bf 0412}, 075 (2004), hep-th/0412287;\\
  %%CITATION = JHEPA,0412,075;%%
D.~Shih, A.~Strominger and X.~Yin,
  %``Recounting dyons in N = 4 string theory,''
  JHEP {\bf 0610}, 087 (2006), hep-th/0505094;\\
  %%CITATION = JHEPA,0610,087;%%
D.~P.~Jatkar and A.~Sen,
  %``Dyon spectrum in CHL models,''
  JHEP {\bf 0604}, 018 (2006), hep-th/0510147;\\
  %%CITATION = JHEPA,0604,018;%%
A.~Dabholkar, D.~Gaiotto and S.~Nampuri,
  %``Comments on the spectrum of CHL dyons,''
  JHEP {\bf 0801}, 023 (2008), hep-th/0702150.
  %%CITATION = JHEPA,0801,023;%%
%
\bibitem{FTBion}
O.~Bergman,
  %``Three-pronged strings and 1/4 BPS states in N = 4 super-Yang-Mills
  %theory,''
  Nucl.\ Phys.\  B {\bf 525}, 104 (1998), hep-th/9712211;\\
  %%CITATION = NUPHA,B525,104;%%
K.~Hashimoto, H.~Hata and N.~Sasakura,
  %``3-string junction and BPS saturated solutions in SU(3) supersymmetric
  %Yang-Mills theory,''
  Phys.\ Lett.\  B {\bf 431}, 303 (1998), hep-th/9803127;\\
  %%CITATION = PHLTA,B431,303;%%
O.~Bergman and B.~Kol,
  %``String webs and 1/4 BPS monopoles,''
  Nucl.\ Phys.\  B {\bf 536}, 149 (1998), hep-th/9804160;\\
  %%CITATION = NUPHA,B536,149;%%
K.~M.~Lee and P.~Yi,
  %``Dyons in N = 4 supersymmetric theories and three-pronged strings,''
  Phys.\ Rev.\  D {\bf 58}, 066005 (1998), hep-th/9804174.
  %%CITATION = PHRVA,D58,066005;%%
%
\bibitem{CalMald}
C.~G.~Callan and J.~M.~Maldacena,
  %``Brane dynamics from the Born-Infeld action,''
  Nucl.\ Phys.\  B {\bf 513}, 198 (1998), hep-th/9708147.
  %%CITATION = NUPHA,B513,198;%%
%
\bibitem{Leigh}
J.~Dai, R.~G.~Leigh and J.~Polchinski,
  %``New Connections Between String Theories,''
  Mod.\ Phys.\ Lett.\  A {\bf 4}, 2073 (1989);\\
  %%CITATION = MPLAE,A4,2073;%%
R.~G.~Leigh,
  %``Dirac-Born-Infeld Action from Dirichlet Sigma Model,''
  Mod.\ Phys.\ Lett.\  A {\bf 4}, 2767 (1989);\\
  %%CITATION = MPLAE,A4,2767;%%
P.~Horava,
  %``Strings on World Sheet Orbifolds,''
  Nucl.\ Phys.\  B {\bf 327}, 461 (1989);
  %%CITATION = NUPHA,B327,461;%%
%%P.~Horava,
  %``Background Duality of Open String Models,''
  Phys.\ Lett.\  B {\bf 231}, 251 (1989).
  %%CITATION = PHLTA,B231,251;%%
%
\bibitem{HorowStrom}
G.~T.~Horowitz and A.~Strominger,
  %``Black strings and P-branes,''
  Nucl.\ Phys.\  B {\bf 360}, 197 (1991).
  %%CITATION = NUPHA,B360,197;%%
%
\bibitem{polch}
J.~Polchinski,
  %``Dirichlet-Branes and Ramond-Ramond Charges,''
  Phys.\ Rev.\ Lett.\  {\bf 75}, 4724 (1995), hep-th/9510017.
  %%CITATION = PRLTA,75,4724;%%
%
\bibitem{malda}
J.~M.~Maldacena,
  %``The large N limit of superconformal field theories and supergravity,''
  Adv.\ Theor.\ Math.\ Phys.\  {\bf 2}, 231 (1998),
  Int.\ J.\ Theor.\ Phys.\  {\bf 38}, 1113 (1999), hep-th/9711200.
  %%CITATION = IJTPB,38,1113;%%
%
\bibitem{gkp}
  S.~S.~Gubser, I.~R.~Klebanov and A.~M.~Polyakov,
  %``Gauge theory correlators from non-critical string theory,''
  Phys.\ Lett.\  B {\bf 428}, 105 (1998), hep-th/9802109.
  %%CITATION = PHLTA,B428,105;%%
%
\bibitem{WittAdS}
  E.~Witten,
  %``Anti-de Sitter space and holography,''
  Adv.\ Theor.\ Math.\ Phys.\  {\bf 2}, 253 (1998), hep-th/9802150.
  %%CITATION = 00203,2,253;%%
%
\bibitem{myCM}
O.~Lunin,
  %``Strings ending on branes from supergravity,''
  JHEP {\bf 0709}, 093 (2007), arXiv:0706.3396 [hep-th].
  %%CITATION = JHEPA,0709,093;%%
%
\bibitem{GMSW}
J.~P.~Gauntlett, D.~Martelli, J.~Sparks and D.~Waldram,
  %``Supersymmetric AdS(5) solutions of M-theory,''
  Class.\ Quant.\ Grav.\  {\bf 21}, 4335 (2004), hep-th/0402153.
  %%CITATION = CQGRD,21,4335;%%
%
%
\bibitem{donos}
A.~Donos,
  %``A description of 1/4 BPS configurations in minimal type IIB SUGRA,''
  Phys.\ Rev.\  D {\bf 75}, 025010 (2007), hep-th/0606199.
  %%CITATION = PHRVA,D75,025010;%%
%
\bibitem{LLM}
H.~Lin, O.~Lunin and J.~M.~Maldacena,
  %``Bubbling AdS space and 1/2 BPS geometries,''
  JHEP {\bf 0410}, 025 (2004), hep-th/0409174.
  %%CITATION = JHEPA,0410,025;%%
%
\bibitem{marolf}
A.~Gomberoff, D.~Kastor, D.~Marolf and J.~H.~Traschen,
  %``Fully localized brane intersections: The plot thickens,''
  Phys.\ Rev.\  D {\bf 61}, 024012 (2000), hep-th/9905094.
  %%CITATION = PHRVA,D61,024012;%%
%
\bibitem{myUnp} O.~Lunin, unpublished notes, 2004.
%
\bibitem{vaman}
B.~Chen {\it et al.},
  %``Bubbling AdS and droplet descriptions of BPS geometries in IIB
  %supergravity,''
  JHEP {\bf 0710}, 003 (2007), 0704.2233 [hep-th].
  %%CITATION = JHEPA,0710,003;%%
%
\bibitem{mikhail}
A.~Mikhailov,
  %``Giant gravitons from holomorphic surfaces,''
  JHEP {\bf 0011}, 027 (2000), hep-th/0010206.
  %%CITATION = JHEPA,0011,027;%%
%
\bibitem{ScanRef}
M.~Cederwall, A.~von Gussich, B.~E.~W.~Nilsson and A.~Westerberg,
  %``The Dirichlet super-three-brane in ten-dimensional type IIB
  %supergravity,''
  Nucl.\ Phys.\  B {\bf 490}, 163 (1997), hep-th/9610148;\\
  %%CITATION = NUPHA,B490,163;%%
M.~Aganagic, C.~Popescu and J.~H.~Schwarz,
  %``D-brane actions with local kappa symmetry,''
  Phys.\ Lett.\  B {\bf 393}, 311 (1997), hep-th/9610249;\\
  %%CITATION = PHLTA,B393,311;%%
M.~Cederwall, A.~von Gussich, B.~E.~W.~Nilsson, P.~Sundell and A.~Westerberg,
  %``The Dirichlet super-p-branes in ten-dimensional type IIA and IIB
  %supergravity,''
  Nucl.\ Phys.\  B {\bf 490}, 179 (1997), hep-th/9611159;\\
  %%CITATION = NUPHA,B490,179;%%
E.~Bergshoeff and P.~K.~Townsend,
  %``Super D-branes,''
  Nucl.\ Phys.\  B {\bf 490}, 145 (1997), hep-th/9611173.
  %%CITATION = NUPHA,B490,145;%%
%
\bibitem{smithRev}
D.~J.~Smith,
  %``Intersecting brane solutions in string and M-theory,''
  Class.\ Quant.\ Grav.\  {\bf 20}, R233 (2003), hep-th/0210157.
  %%CITATION = CQGRD,20,R233;%%
%
\bibitem{Kappa}
M.~Aganagic, C.~Popescu and J.~H.~Schwarz,
  %``Gauge-invariant and gauge-fixed D-brane actions,''
  Nucl.\ Phys.\  B {\bf 495}, 99 (1997), hep-th/9612080;\\
  %%CITATION = NUPHA,B495,99;%%
E.~Bergshoeff, R.~Kallosh, T.~Ortin and G.~Papadopoulos,
  %``kappa-symmetry, supersymmetry and intersecting branes,''
  Nucl.\ Phys.\  B {\bf 502}, 149 (1997), hep-th/9705040.
  %%CITATION = NUPHA,B502,149;%%
%
\bibitem{FaySmth}
A.~Fayyazuddin and D.~J.~Smith,
  %``Localized intersections of M5-branes and four-dimensional  superconformal
  %field theories,''
  JHEP {\bf 9904}, 030 (1999), hep-th/9902210.
  %%CITATION = JHEPA,9904,030;%%
%
\bibitem{beren}
S.~Corley, A.~Jevicki and S.~Ramgoolam,
  %``Exact correlators of giant gravitons from dual N = 4 SYM theory,''
  Adv.\ Theor.\ Math.\ Phys.\  {\bf 5}, 809 (2002), hep-th/0111222;\\
  %%CITATION = 00203,5,809;%%
D.~Berenstein,
  %``A toy model for the AdS/CFT correspondence,''
  JHEP {\bf 0407}, 018 (2004), hep-th/0403110.
  %%CITATION = JHEPA,0407,018;%%
%
\bibitem{giant}
J.~McGreevy, L.~Susskind and N.~Toumbas,
  %``Invasion of the giant gravitons from anti-de Sitter space,''
  JHEP {\bf 0006}, 008 (2000), hep-th/0003075;\\
  %%CITATION = JHEPA,0006,008;%%
M.~T.~Grisaru, R.~C.~Myers and O.~Tafjord,
  %``SUSY and Goliath,''
  JHEP {\bf 0008}, 040 (2000), hep-th/0008015.
  %%CITATION = JHEPA,0008,040;%%
%
\bibitem{beren4}
D.~Berenstein,
  %``Large N BPS states and emergent quantum gravity,''
  JHEP {\bf 0601}, 125 (2006), hep-th/0507203;\\
  %%CITATION = JHEPA,0601,125;%%
D.~Berenstein and R.~Cotta,
  %``A Monte-Carlo study of the AdS/CFT correspondence: 
  %An exploration of
  %quantum gravity effects,''
  JHEP {\bf 0704}, 071 (2007), hep-th/0702090.
  %%CITATION = JHEPA,0704,071;%%
%
\bibitem{donos2}
A.~Donos,
  %``BPS states in type IIB SUGRA with SO(4) x SO(2)(gauged) symmetry,''
  JHEP {\bf 0705}, 072 (2007), hep-th/0610259.
  %%CITATION = JHEPA,0705,072;%%
%
\bibitem{sheppard}
P.~Horava and P.~G.~Shepard,
  %``Topology changing transitions in bubbling geometries,''
  JHEP {\bf 0502}, 063 (2005), hep-th/0502127;\\
  %%CITATION = JHEPA,0502,063;%%
A.~E.~Mosaffa and M.~M.~Sheikh-Jabbari,
  %``On classification of the bubbling geometries,''
  JHEP {\bf 0604}, 045 (2006), hep-th/0602270.
  %%CITATION = JHEPA,0604,045;%%
%
\bibitem{gibHawk}
G.~W.~Gibbons and S.~W.~Hawking,
  %``Classification Of Gravitational Instanton Symmetries,''
  Commun.\ Math.\ Phys.\  {\bf 66}, 291 (1979);\\
  %%CITATION = CMPHA,66,291;%%
T.~Eguchi, P.~B.~Gilkey and A.~J.~Hanson,
  %``Gravitation, Gauge Theories And Differential Geometry,''
  Phys.\ Rept.\  {\bf 66}, 213 (1980).
  %%CITATION = PRPLC,66,213;%%
%
\bibitem{bachas}
C.~P.~Boyer and J.~D.~.~Finley,
  %``Killing Vectors In Selfdual, Euclidean Einstein Spaces,''
  J.\ Math.\ Phys.\  {\bf 23}, 1126 (1982);\\
  %%CITATION = JMAPA,23,1126;%%
J. Gegenberg and A. Das, Gen. Rel. Grav. {\bf 16} 817 (1984);\\
I.~Bakas and K.~Sfetsos,
  %``Toda fields of SO(3) hyper-Kahler metrics and free field realizations,''
  Int.\ J.\ Mod.\ Phys.\  A {\bf 12}, 2585 (1997), hep-th/9604003.
  %%CITATION = IMPAE,A12,2585;%%
%
\bibitem{kim}
N.~Kim,
  %``AdS(3) solutions of IIB supergravity from D3-branes,''
  JHEP {\bf 0601}, 094 (2006), hep-th/0511029.
  %%CITATION = JHEPA,0601,094;%%
%
\bibitem{16BH}
J.~B.~Gutowski and H.~S.~Reall,
  %``General supersymmetric AdS(5) black holes,''
  JHEP {\bf 0404}, 048 (2004), hep-th/0401129;\\
  %%CITATION = JHEPA,0404,048;%%
%
H.~K.~Kunduri, J.~Lucietti and H.~S.~Reall,
  %``Supersymmetric multi-charge AdS(5) black holes,''
  JHEP {\bf 0604}, 036 (2006), hep-th/0601156.
  %%CITATION = JHEPA,0604,036;%%
%
\bibitem{myWils}
S.~Yamaguchi,
  %``Bubbling geometries for half BPS Wilson lines,''
  Int.\ J.\ Mod.\ Phys.\  A {\bf 22}, 1353 (2007), hep-th/0601089;\\
  %%CITATION = IMPAE,A22,1353;%%
O.~Lunin,
  %``On gravitational description of Wilson lines,''
  JHEP {\bf 0606}, 026 (2006), hep-th/0604133;\\
  %%CITATION = JHEPA,0606,026;%%
J.~Gomis and C.~Romelsberger,
  %``Bubbling defect CFT's,''
  JHEP {\bf 0608}, 050 (2006), hep-th/0604155;\\
  %%CITATION = JHEPA,0608,050;%%
J.~Gomis and S.~Matsuura,
  %``Bubbling surface operators and S-duality,''
  JHEP {\bf 0706}, 025 (2007), arXiv:0704.1657 [hep-th];\\
  %%CITATION = JHEPA,0706,025;%%
E.~D'Hoker, J.~Estes and M.~Gutperle,
  %``Gravity duals of half-BPS Wilson loops,''
  arXiv:0705.1004 [hep-th].
  %%CITATION = ARXIV:0705.1004;%%
%
\bibitem{myMdef}
O.~Lunin,
  %``1/2-BPS states in M theory and defects in the dual CFTs,''
  JHEP {\bf 0710}, 014 (2007), 0704.3442 [hep-th].
  %%CITATION = JHEPA,0710,014;%%
%
\bibitem{MaldDruk}
A.~Dymarsky, S.~S.~Gubser, Z.~Guralnik and J.~M.~Maldacena,
  %``Calibrated surfaces and supersymmetric Wilson loops,''
  JHEP {\bf 0609}, 057 (2006), hep-th/0604058;\\
  %%CITATION = JHEPA,0609,057;%%
N.~Drukker, S.~Giombi, R.~Ricci and D.~Trancanelli,
  %``Supersymmetric Wilson loops on S^3,''
  arXiv:0711.3226 [hep-th].
  %%CITATION = ARXIV:0711.3226;%%
%
\bibitem{ramf}
H.~Lu, C.~N.~Pope and J.~Rahmfeld,
  %``A construction of Killing spinors on S**n,''
  J.\ Math.\ Phys.\  {\bf 40}, 4518 (1999), hep-th/9805151.
  %%CITATION = JMAPA,40,4518;%%
\end{thebibliography}
\end{document}